\PassOptionsToPackage{unicode}{hyperref}
\PassOptionsToPackage{hyphens}{url}
\documentclass[
]{article}
\usepackage{amsmath,amssymb}
\usepackage{appendix}
\usepackage{iftex}
\ifPDFTeX
  \usepackage[T1]{fontenc}
  \usepackage[utf8]{inputenc}
  \usepackage{textcomp} 
\else 
  \usepackage{unicode-math} 
  \defaultfontfeatures{Scale=MatchLowercase}
  \defaultfontfeatures[\rmfamily]{Ligatures=TeX,Scale=1}
\fi
\usepackage{lmodern}
\ifPDFTeX\else
\fi
\IfFileExists{upquote.sty}{\usepackage{upquote}}{}
\IfFileExists{microtype.sty}{
  \usepackage[]{microtype}
  \UseMicrotypeSet[protrusion]{basicmath} 
}{}
\makeatletter
\@ifundefined{KOMAClassName}{
  \IfFileExists{parskip.sty}{%
    \usepackage{parskip}
  }{
    \setlength{\parindent}{0pt}
    \setlength{\parskip}{6pt plus 2pt minus 1pt}}
}{
  \KOMAoptions{parskip=half}}
\makeatother
\usepackage{xcolor}
\usepackage{color}
\usepackage{fancyvrb}
\usepackage{cite}

\DefineVerbatimEnvironment{Highlighting}{Verbatim}{commandchars=\\\{\}}
\newenvironment{Shaded}{}{}

\newcommand{\BuiltInTok}[1]{\textcolor[rgb]{0.00,0.50,0.00}{#1}}

\newcommand{\CommentTok}[1]{\textcolor[rgb]{0.38,0.63,0.69}{\textit{#1}}}

\newcommand{\ControlFlowTok}[1]{\textcolor[rgb]{0.00,0.44,0.13}{\textbf{#1}}}

\newcommand{\DecValTok}[1]{\textcolor[rgb]{0.25,0.63,0.44}{#1}}

\newcommand{\FloatTok}[1]{\textcolor[rgb]{0.25,0.63,0.44}{#1}}

\newcommand{\KeywordTok}[1]{\textcolor[rgb]{0.00,0.44,0.13}{\textbf{#1}}}
\newcommand{\NormalTok}[1]{#1}
\newcommand{\OperatorTok}[1]{\textcolor[rgb]{0.40,0.40,0.40}{#1}}

\newcommand{\StringTok}[1]{\textcolor[rgb]{0.25,0.44,0.63}{#1}}

\usepackage{longtable,booktabs,array}
\usepackage{calc} 
\usepackage{etoolbox}
\makeatletter
\patchcmd\longtable{\par}{\if@noskipsec\mbox{}\fi\par}{}{}
\makeatother
\IfFileExists{footnotehyper.sty}{\usepackage{footnotehyper}}{\usepackage{footnote}}
\makesavenoteenv{longtable}
\usepackage{graphicx}
\makeatletter
\def\maxwidth{\ifdim\Gin@nat@width>\linewidth\linewidth\else\Gin@nat@width\fi}
\def\maxheight{\ifdim\Gin@nat@height>\textheight\textheight\else\Gin@nat@height\fi}
\makeatother
\setkeys{Gin}{width=\maxwidth,height=\maxheight,keepaspectratio}
\makeatletter
\def\fps@figure{htbp}
\makeatother
\providecommand{\tightlist}{%
  \setlength{\itemsep}{0pt}\setlength{\parskip}{0pt}}
\usepackage[margin=3cm]{geometry}
\usepackage{amssymb}
\ifLuaTeX
  \usepackage{selnolig}  
\fi
\IfFileExists{bookmark.sty}{\usepackage{bookmark}}{\usepackage{hyperref}}
\IfFileExists{xurl.sty}{\usepackage{xurl}}{} 
\hypersetup{
  hidelinks,
  pdfcreator={LaTeX via pandoc}}

\title{Life at the Boundary of Chemical Kinetics and Program Execution}
\author{Thomas Fischbacher\footnote{Google, Brandschenkestrasse
  110, 8002 Zürich, Switzerland -- \texttt{tfish@google.com}}}
\date{}

\begin{document}

\maketitle

\textbf{Abstract}

This work introduces a generic quantitative framework for studying
dynamical processes that involve interactions of polymer
sequences. Possible applications range from quantitative studies of
the reaction kinetics of polymerization processes to explorations of
the behavior of chemical implementations of computational -- including
basic life-like -- processes. This way, we establish a bridge between
thermodynamic and computational aspects of systems that are defined in
terms of sequence interactions. As by-products of these
investigations, we clarify some common confusion around the notion of
``autocatalysis'' and show quantitatively how a chemically implemented
Turing machine can operate close to the Landauer bound.

Using a Markov process model of polymer sequence composition and
dynamical evolution of the Markov process's parameters via an ordinary
differential equation (ODE) that arises when taking the double
``chemical'' many-particle limit as well as ``rarefied interactions''
limit, this approach enables -- for example -- accurate quantitative
explorations of entropy generation in systems where computation is
driven by relaxation to thermodynamic equilibrium. The computational
framework internally utilizes the Scheme programming language's
intrinsic continuation mechanisms to provide nondeterministic
evaluation primitives that allow the user to specify example systems
in straight purely functional code, making exploration of all possible
relevant sequence composition constellations -- which would be
otherwise tedious to write code for -- automatic and hidden from the
user.

As the original motivation for this work came from investigations into
emergent program evolution that arises in computational substrates of
the form discussed in recent work on ``Computational Life''
\cite{alakuijala2024computational}, a major focus of attention is on
giving a deeper explanation of key requirements for the possible
emergence of self-replicators especially in settings whose behavior is
governed by real world physics rather than ad-hoc rules that may be
difficult to implement in a physical system. A collection of fully
worked out examples elucidate how this modeling approach is
quantitatively related to Metropolis Monte Carlo based simulations as
well as exact or approximate analytic approaches, and how it can be
utilized to study a broad range of different systems. These examples
can also serve as starting points for further explorations.

\hypertarget{introduction}{%
\section{Introduction}\label{introduction}}

Computational models of processes that are governed by simple rules
yet give rise to highly nontrivial behavior have a long history. A
very prominent such early example is Conway's ``Game of Life''
\cite{games1970fantastic} in which the state of the world is described
by an infinite square lattice of cells holding one bit of state each,
and evolution follows a simple update rule where the state of each
cell at the next clock cycle is a function of the cell's own state and
the number of direct 1-neighbors in an 8-neighborhood. Despite the
simplicity of its rules, ``Life'' has rich dynamics and allows
patterns that can be used for signal transmission, generation, and
processing, in fact allowing one to embed any Turing
machine~\cite{allen2004winning}. The key observation that complex
systems made of simple components that obey simple but nontrivial
nonlinear dynamical update rules can give rise to surprisingly complex
behavior certainly is intriguing, and having a good theoretical
toolbox that allows one to identify general motifs which allow one to
make predictions about behavior is readily recognized as potentially
highly relevant for understanding the world -- and informing good
decision making. As such, it is not surprising that the behavior of
complex systems with nonlinear dynamics has become a very active field
of research. A useful introduction to key concepts can be found
e.g. in~\cite{thurner2018introduction}.

The main motivation behind the present article is provided by recent
work~\cite{alakuijala2024computational} that used sampling-based
exploration to demonstrate that a broad class of toy models which
share the following common features have a strong tendency to evolve
from random configurations into more ordered configurations that are
dominated by the activity of replicating entities:

\begin{itemize}
\item
  The state of the system is described by a collection of (long)
  sequences (``tapes'') of symbols from a given symbol-alphabet.
\item
  Dynamics arises due to a pair of tape-sequences being chosen to act
  upon one another according to pre-defined rules that resemble using
  one sequence as a ``program'' according to which the content
  on both sequences may get mutated (i.e. programs can be self-modifying).
\end{itemize}

Technically speaking, the systems described in
\cite{alakuijala2024computational} generally involve multiple tapes,
and interaction rules are such that the start (and end of)
fixed-length tapes have a special role. This is only a superficial
difference, since these constructions are readily embedded into a
conceptually simpler framework with only a single long tape where
symbols are taken from an extended alphabet that also includes start-
and end-tokens, and initial tape-composition is not fully random, but
follows some simple rules w.r.t. presence and placement of these
terminal tokens. Also, if program execution can only begin at the
start of a tape, this can be absorbed into a redefinition of
update-rules that allow execution to nominally start using random
tape-positions, but immediately halt the program unless the first
token is a start-of-tape token. Such a simpler and broader framework
is more closely aligned with how one would want to model polymers in
solution and will be the basis of the construction presented
here. Embedding constructions like that of the aforementioned article
into this framework is discussed in
section~\ref{example-6-the-bff-type-system-of-1}.

In contrast to simple processes that would also fit the description of
``transition to an ordered structure over time'' such as
crystallization of a solution (where a seed crystal provides the
pattern for adding more units in an ordered fashion), the dynamics of
these systems is such that one can observe processes that bear some
semblance to a basic form of \emph{evolution}. Specifically,
section~2.1 of~\cite{alakuijala2024computational} describes in detail
one process that can be interpreted as showing the emergence of a
``disease'' in response to which patterns evolve some form of
``immunity''. This observation appears to make these models appealing
candidates for trying to study and understand the phenomenon of
\emph{abiogenesis}, the spontaneous emergence of biological life from
chemical compounds.

Any such effort however immediately runs into the problem that, from a
chemical perspective, the proposed models are rather remote from what
one could consider as having a plausible basis in molecular chemistry.
Key problems are, in likely order of relevance:

\begin{enumerate}
\def\labelenumi{\arabic{enumi}.}
\item While the initial state in general describes an unstructured
  agglomeration of building blocks that provide basic operations, the
  \emph{mechanism} that would have to be in place in order to implement
  evaluation semantics in close alignment with these models would have
  to be incredibly complex. Vice-versa, the basic operations encoded
  by building blocks (``opcodes'') are only simple from the
  computational perspective: If data tapes were modeled as linear
  polymers (which, given the observation that biology for many key
  mechanisms uses such representations of data, looks eminently
  plausible), an operation such as ``scan backwards on the
  program-tape until we encounter a bracketing symbol that matches up
  with the current one if we keep track of opening and closing
  brackets'' as needed to implement the ``\texttt{{]}}''-operation in the
  toy language discussed in~\cite{alakuijala2024computational} would
  be near-impossible to express with a simple chemical machine\footnote{If one
  were to do this, one likely would want to keep track of the number
  of currently open brackets by attaching a chain of monomers to the
  program execution mechanism that grows or shrinks whenever a bracket-symbol
  is encountered.}.
\item
  It is not clear at all what energetic mechanism would lead to a
  thermodynamic inequilibrium in the initial state that then powers the
  program-evaluation machinery.
\item
  Given that ``any symbol can readily be turned into any other symbol by
  program action'', symbols would have to be represented by monomer
  units on the tape either purely in terms of the arrangement of the
  chemical bonds in any such given monomer unit, or, in a solution, by
  removing and attaching components that are freely available in some
  mobile form in the solution.
\end{enumerate}

With respect to the first point, one may argue that ``emergence of
life'' appears to be a somewhat ubiquitous general phenomenon for such
computational substrates that is not too strongly dependent on
details. Optimistically, we may hence hope that, even if any of the
choices of evaluation rules discussed so far suffer from the problem
that they contain ``too complex to be chemically plausible in a
mostly-unstructured initial state'', chemistry as it works in our
universe may well be sufficiently rich and complex to also give rise
to evolution via some not too dissimilar process. If so, we should not
be surprised to find that interpreting the underlying chemical rules
by expressing them in a form understandable to humans that focuses on
data-processing may be difficult.

One observation that appears to be in favor of this interpretation is
that, if emergence of life were a rare and difficult step, one would
naturally expect our planet to have been lifeless for a relevant
fraction of its history since the point when conditions first would
have made life possible in principle. Instead, the geological record
indicates that life appeared very soon after conditions became
suitable!\cite{javaux2019challenges}, and this observation appears to
leave us with only three possible explanations: (a) an incredible
amount of luck, (b) abiogenesis being not a difficult step, and (c) an
extraterrestrial origin of life~\cite{kamminga1982life}, i.e.~the
planet being seeded by life-bearing cosmic debris.

There are good reasons to think that evolution favors efficient use of
resources, which for replicating entities may well mean: efficient use
of the potential to generate entropy for self-replication. In
\cite{england2013statistical}, compelling reasons have been given for
thinking of \emph{Escherichia coli} as an amazingly effective
self-replicator, which under ideal conditions may well be within one
order of magnitude of the theoretical minimal entropic effort required
to produce a copy of itself -- while also having to power other
functions not directly related to replication, such as a basic
bacterial immune system. Clearly, the observed entropic efficiency of
such biological machines alone when it comes to self-repair and
self-reproduction, which is far from what currently is considered
feasible for engineered systems, is in itself reason enough to closely
study nature's underlying design principles. Unfortunately, cruder and
less efficient ancient biochemical processes and also different data
encoding than the near-ubiquitous genetic code shared by practically
all life forms appear to have become extinct, making it difficult for
us to retrace life's origins.

Despite this, some interesting plausible hypotheses have been proposed
for possible precursors of key chemical compounds in near-universal
mechanisms such as the citric acid cycle. While ideas such as
W\"achtershauser's ``Iron-Sulfur World abiogenesis''
\cite{wachtershauser1990case} may or may not be off on many details,
the general idea to look for simple molecules (such as thioacetic acid
at just under 80 Dalton) that chemically can play the role of
precursors of some performance optimized refined compounds used by
current biology (such as acetyl-coenzyme-A at about 800 Dalton) is
likely to be a step in the right direction.

\hypertarget{on-the-role-of-autocatalysis}{%
{\bf On the Role of Autocatalysis}\label{on-the-role-of-autocatalysis}}

Point 2 above, about ``powering the evaluator machinery'' deserves
further attention. One observation is that in constructions such as
those presented in~\cite{alakuijala2024computational}, it is often
possible to ``seed'' the computational substrate with a pattern
capable of self-replication and have it take over, even if such
self-replicators are difficult to emerge on their own. One hence may
be led to believe that just about any system in which an entity can
arise which autocatalytically can create copies of itself will end up
in a replicator-dominated state as soon as a combination of chance and
some not so random precursor processes created the first such
self-replicator -- and that the replicator with highest effective
reproduction rate would drive its competition to
extinction~\cite{lifson1997crucial}.

This interpretation -- especially the idea that a more effective
self-replicator would drive its competitors to extinction, which is
being frequently referred to in the literature (see
e.g.~\cite{hanopolskyi2021autocatalysis, pascal2013towards}) -- is
incorrect in the sense that the underlying mathematical model of
autocatalysis is incompatible with it being a form of (chemical)
catalysis, which is subject to thermodynamic
constraints. Correspondingly, it is easy to come up with
counterexamples.

In~\cite{lifson1997crucial}, the problem is that the mathematical
modeling approach ignores a term that must be retained if one strives
to not violate thermodynamics. Even very basic models that have
self-replication via autocatalytic processes show that one would in
general expect coexistence of species where relative abundances are a
function of their reproduction rate, stability, and removal rate. Two
such examples are discussed in this section (with more detailed
numerical analysis shown in
appendix~\ref{appendix-a-competing-autocatalytic-species-in-a-flow-equilibrium}),
plus a third one is presented in
section~\ref{example-5-a-simple-machine-language}.

The caveat here is that this claim applies to systems for which there
is a notion of thermodynamic equilibrium (which may be a flow
equilibrium) that is both meaningful and attainable, and in
conjunction with this, the set of relevant (chemical) species is
readily enumerable. The claim will not apply to systems that have been
defined in terms of reaction rates chosen \emph{ad-hoc} without paying
attention to whether these would violate the laws of
thermodynamics. Its usefulness might also be limited in situations
where thermodynamic concepts -- while always providing stringent
bounds -- are difficult to apply. Technically, in situations where the
number of available states \(\rho(E)\) in the energy interval
\([E,E+dE]\) as a function of the energy \(E\) grows exponentially or
faster with energy, the system will not be able to distribute its
energy to these states in an entropy-maximizing way according to
$p(E)\propto \rho(E)\exp(-\beta E)$, and the notion of a
``temperature'' loses its meaning~\cite{hagedorn1970thermodynamics} --
somewhat akin to how some probability distributions such as the Cauchy
distribution fail to have a standard deviation. Even in situations
where there is just an extremely large number of relevant species, the
usefulness of thermodynamics might be limited, providing some
constraints on what could happen, whereas even in large physically
realizable systems, finite (system-)size effects (which almost by
definition are ignored in the thermodynamic limit) would play a
prominent role. Also, loss of ergodicity can complicate (but will not
invalidate) thermodynamics based reasoning a lot.

In such situations, the methods offered in the present work may
perhaps be most useful for studying some specific aspect that one
could identify and isolate -- akin to how Lorenz isolated the
``chaotic attractor'' mechanism by extracting a 3-parameter system
that showed the behavior of interest from a larger 12-parameter model
system~\cite{lorenz1963deterministic,shen202350th}.

\hypertarget{autocatalysis-example-1-chiral-tetrominos}{%
\paragraph{Autocatalysis Example 1: Chiral
Tetrominos}\label{autocatalysis-example-1-chiral-tetrominos}}

First, let us consider a toy world of \(2\times 1\times 1\)-block
monomers that dimerize to form either a non-planar ``A-piece'' Soma
Cube \cite{allen2004winning} tetromino, or its mirror-image
``B-piece'' tetromino.  Both tetrominos fill half of a
\(2\times 2\times 2\) cube, and one observes that with respect to this
cube, the complement of an \(A\)-piece is another \(A\)-piece, not its
mirror image (and likewise for the \(B\)-piece). So, both the
\(A\)-piece and \(B\)-piece can be obtained by cutting up a
\(2\times 2\times 2\) cube into two \emph{homochiral} halves --
depending on how one performs the \emph{coupe du roi}
cut~\cite{schwarzenbach2015coupe}. We discuss the quantitative aspects
of this example in detail, for this will also set the scene for
subsequent analysis of other systems.

\begin{figure}
\centering
\includegraphics[width=\textwidth,height=0.25\textheight]{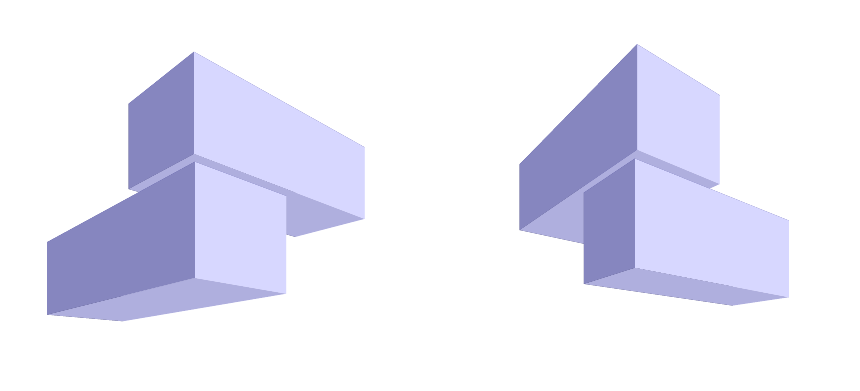}
\caption{The A-piece (left) and B-piece (right) chiral nonplanar
tetrominos}
\end{figure}

One can now plausibly imagine a mechanism by which an \(A\)-piece acts
as a mold for forming another \(A\)-piece from monomers, and so if one
were to seed a solution of the monomer with some \(A\)-piece dimers, one
might expect these to catalyze dimerization to practically-only
\(A\)-piece form. Nevertheless, catalysis only accelerates chemical
processes but does not shift chemical equilibrium. So, if the reverse
process of dissociation is possible at all (i.e.~dimerization does not
come as an effectively-irreversible reaction), some monomers would
spontaneously form \(B\)-piece dimers, and if there is an over-supply of
\(A\)-piece dimers over \(B\)-piece dimers, the rate of dissociation of
\(A\)-piece dimers would be higher than the rate of dissociation of
\(B\)-piece dimers, with the end state then being the highest entropy
state of a racemic mixture (equal concentration) of \(A\)-piece and
\(B\)-piece (in dynamic equilibrium with some dissociated monomers). So,
even if each chiral form can autocatalytically create copies of itself,
the dynamical end state is not dominated by one form.

In general, if no autocatalysis takes place, we will have kinetic
(i.e.~rate) equations for the rate-of-change of concentrations
\(c_{\cdots}\) of the monomer \(M\) as well as the dimers \(A\) and
\(B\) that may take on the a form such as:

\begin{equation}\begin{array}{lcl}
(d/dt)\,c_A&=&K_{A\leftarrow 2M}\cdot c_M\cdot c_M - K_{2M\leftarrow A}\cdot c_A\\
(d/dt)\,c_B&=&K_{B\leftarrow 2M}\cdot c_M\cdot c_M - K_{2M\leftarrow B}\cdot c_B\\
(d/dt)\,c_M&=&-2K_{B\leftarrow 2M}\cdot c_M\cdot c_M -2K_{A\leftarrow 2M} c_M\cdot c_M\\
&&+2 K_{2M\leftarrow A}\cdot c_A +2 K_{2M\leftarrow B}\cdot c_B.
\end{array}\end{equation}

Here, \(K_{\{\text{products}\}\leftarrow\{\text{reagents}\}}\) is the
reaction constant that determines the (concentration-dependent) rate of
the corresponding chemical process. While the reaction kinetics
described above looks simple and reasonable, and is in alignment with
the thinking behind Guldberg and Waage's original derivation of the law
of mass action~\cite{waage1864studier} from the 1860s/70s, one in
general has to be cautious here, since such simplistic reaction kinetics
is often misaligned with reality. A textbook example is the reaction
\(\text{CO}+\text{NO}_2\leftrightharpoons \text{CO}_2+\text{NO}\)
for which the rate of the forward
reaction is, over wide concentration ranges, found to not be
\(\propto c_{\text{CO}}\cdot c_{\text{NO}_2}\) as one might naively expect, but
\(\propto c_{\text{NO}_2}^2\). The reason here is that dynamics is dominated by
an indirect process that is much faster than the direct reaction.
Effectively, most of the \(\text{NO}_2\) does not react directly with \(\text{CO}\)
but via an intermediate product (the nitrate radical \(\text{NO}_3\)) that gets
created at a slow rate (but still faster than the direct \(\text{CO}+\text{NO}_2\)
reaction over wide ranges of concentrations) in the collision of two
\(\text{NO}_2\) molecules which readily reacts with any available \(\text{CO}\) but
then needs to be replenished by this low-rate process.

Despite this possible (actually rather frequent) and occasionally
stark misalignment between the reaction kinetics assumed in secondary
education or college level derivations of the law of mass action and
reality, the law of mass action remains generally valid (apart from
the detail that concentrations have to be replaced with chemical
activities) for a deeper reason: From the reaction kinetics
perspective, equilibrium concentrations are determined by the
condition that the forward and backward reaction rate are equal. There
also is a thermodynamics perspective on this situation, for if we
could shift the balance in equilibrium by tweaking reaction rates -
such as by adding some catalyst or inhibitor -- we could construct a
periodic process via which, by repeated addition and withdrawal of a
catalyst\footnote{In a thought experiment, one might for example
  imagine adding/removing a catalyst by immobilizing the catalyst on
  some surface, and pumping the reactants between containers where
  catalyst is present and other containers where it is not -- the
  energy required for pumping can be made arbitrarily small.}, where
we could in every cycle extract work from the system's drive to relax
to a new equilibrium.  Since this would be a perpetuum mobile of the
2nd kind, we conclude that relative concentrations (or rather,
activities) in equilibrium must be a function of the ``relative
thermodynamic stability'' of the products vs.~reagents given
environmental parameters (such as temperature and pressure). This then
in turn must also equal the ratio of reaction constants for the
forward- and backward-reaction, irrespective of what reaction pathways
are opened or closed. Quantitatively, the equilibrium constant \(K\)
is the ratio of reaction-constants for the forward- and
backward-reaction, and this must be related to the Gibbs free energy
(or ``free enthalpy'') of the reaction\footnote{This follows from
  maximizing total entropy -- effectively the logarithm of the number
  of microstates making up the macrostate -- in a closed system that
  can increase entropy by releasing heat to an ``heat bath''
  environment at constant temperature.} via:
\begin{equation}
  K=\exp\left(-\frac{\Delta G_{R\leftrightharpoons P}}{N_AkT}\right).
\end{equation}

\begin{figure}
\centering
\includegraphics[width=\textwidth,height=0.25\textheight]{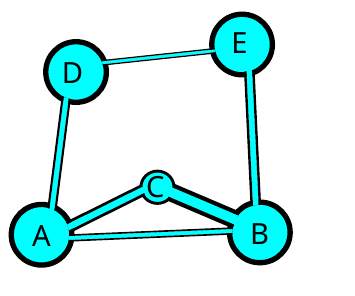}
\caption{Pipe-and-tanks network example. If tank C were removed,
  equilibriation between A and B would only happen due to either
  a slow direct pathway or an alternative slow pathway via D and E.
  In the corresponding chemical picture,
  the species \(\text{Xe}_{(\text{C})}\) would be regarded as a catalyst
  for the \(\text{Xe}_{(\text{A})}\leftrightharpoons \text{Xe}_{(\text{B})}\) reaction.}
\label{fig:pipenetwork}
\end{figure}

An analogy is in order. We can imagine a collection of connected gas
tanks filled with some gas species -- such as Xenon -- at different
altitude with an equilibrium pressure ratio that is set by the
altitude difference, temperature, and gravitational acceleration. No
matter how many pipes we use to connect the tanks, or how we run or
shape them, for every connected component of the pipe network, if the
total amount of gas in all tanks is set, any initial distribution
across the tanks will ultimately equilibriate to a
pressure-distribution where the pressure \(p_\tau\) in tank \(\tau\)
satisfies \(p_T=C\exp(-m_\text{Xe}gh(\tau)/(k_BT))\) where \(C\) is a
tank-independent constant that depends on the total amount of gas, and
$h(\tau)$ is the altitude of tank \(\tau\). The dynamics of how that
equilibrium is reached does depend on the structure of the pipe
network, the lengths and widths of the connecting pipes, and the
initial gas distribution. Notably, if one were to initially only fill
the bottom tank with gas, pressure equilibriation will require lifting
work to be done that is provided by heat entering the system from the
external heat bath that is kept at constant temperature. We are free
to regard each ``Xenon in tank \(\tau\)'' as a different chemical
species \(\text{Xe}_{(\tau)}\) that all differ in Gibbs free energy by
lifting work,
\(\Delta G^\circ(\text{Xe}_{(\tau)})=\Delta
G^\circ(\text{Xe})+N_Am_\text{Xe}gh(\tau)\) and formulate a set of
chemical reactions \(\text{Xe}_{(A)}\leftrightharpoons\text{Xe}_{(B)}\),
\(\text{Xe}_{(A)}\leftrightharpoons\text{Xe}_{(C)}\), etc., where the
equilibrium constants are set by the differences in Gibbs free energy
\(\exp(-\Delta G_{\{\text{reaction}\}}/(k_BT))\), but the
reaction-rates depend on pipe sizes. If we change a pipe network by
adding another (small) tank at low altitude that is connected with big
pipes and thus provides a fast alternate route for equilibriation
between other tanks (or chemical species), this corresponds to adding
a catalyst.  Figure~\ref{fig:pipenetwork} illustrates this.

Correspondingly, even when replacing real world reaction kinetics
(which, as we have seen, can be very subtle and have unexpected
aspects) with some ``cartoon-style'' reaction kinetics where we took
away all the actual reaction pathways and replaced them with entirely
fictitious pathways, such as ones that follow a simplistic
Guldberg-Waage model of reaction kinetics, we still would find the
same thermodynamic equilibrium.

\begin{figure}
\centering
\includegraphics[width=\textwidth,height=0.25\textheight]{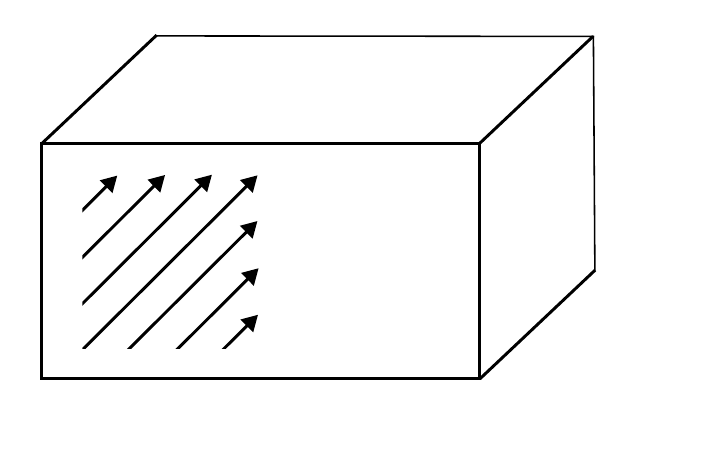}
\caption{Patterned large face on a \(2\times1\times1\) block. If such
  monomers connect in such a way that the directions of the diagonal
  stripes align, it depends on whether arrows prefer to align or to
  anti-align whether we preferentially get an \(A\)-piece dimer or
  \(B\)-piece dimer.}
\label{fig:patterned211}
\end{figure}

Coming back to chiral tetrominos, if \(M\) is the monomer, we have the
chemical reactions \(2M\leftrightharpoons A\) and
\(2M\leftrightharpoons B\). We want to also explore situations where
\(M\) preferentially forms either \(A\) or \(B\). One can imagine
engineering such a situation by providing slightly asymmetric
monomers. To give a mechanical model, if the \(1\times 1\) surface
patch of the \(2\times 1\times 1\) monomer molecule that attaches to
another such patch were sticky and came with grooves oriented as shown
in figure~\ref{fig:patterned211}, where aligning orientation of the
grooves gives a much stronger binding force than having them
misaligned, monomers would tend to form nonplanar dimers,
i.e.~\(A\)-form or \(B\)-form, since only that configuration can make
sticky parts of the surfaces face one another in a way that aligns the
grooves.

It makes sense to start the discussion from the (experimentally in
principle determinable) standard Gibbs free energy of formation
\(\Delta G^\circ\) of \(M\), \(A\), and \(B\), and regard that as given.
From these, one then obtains (by linear combination, and adjusting for
changes to temperature from the thermodynamic reference state via
\(\Delta G(T)=\Delta G^\circ(T_0)+\int_{T_0}^T c_p(\tau)\,d\tau\)) the
Gibbs free energy of for the reactions \(2M\leftrightharpoons A\) and
\(2M\leftrightharpoons B\). These Gibbs free energies can be seen as
quantitatively describing how much maximizing total entropy in a closed
system (at constant pressure) favors having \(A\) over having \(2M\).

If we add chemical pathways that allow \(M\) to form \(A\) (respectively
\(B\)), the concentration-ratios in (entropy-maximizing) chemical
equilibrium do not depend on the number and nature of these pathways (as
long as there are any) -- in analogy to ``adding more pipes that connect
the gas tanks''. Gibbs free energies of formation determine the ratio
\(K_1/K_2\) for reaction constants
\(\{\text{Rate}_{A\leftarrow 2M}=K_1\cdot a_M^2\}\) and
\(\{\text{Rate}_{2M\leftarrow A}=K_2\cdot a_A\}\), where \(a_{\cdots}\) are chemical
activities which for the sake of this discussion we can take to equal
concentrations, and the impossibility of a chemical perpetuum mobile
tells us that if we added other reaction pathways, such as via some
catalyst \(C\): \(C+2M\rightarrow C+A\) and \(C+A\rightarrow C+2M\), we
must have that for the corresponding reactions,
\(\{\text{Rate}_{C+A\leftarrow C+2M}=K_{1,C}\cdot a_M^2\cdot a_C\}\) and
\(\{\text{Rate}_{2M+C\leftarrow A+C}=K_{2,C}\cdot a_A\cdot a_C\}\), we must find
\(K_{1,C}/K_{2,C}=K_1/K_2\). This in particular then also must hold for
autocatalytic reactions, so for \(C=A\).

One obvious consequence of these considerations is that, in a
situation where \(A\) and \(B\) are thermodynamically equally stable
(relative to formation from the elements), and formation (and
dissociation) of either of these is an autocatalytic process, seeding
a pure solution of \(M\) in which equilibriation reactions are
suppressed (such as by cooling) with a small amount of \(A\)-dimer
only and subsequently allowing reactions so that the system can relax
to thermodynamic equilibrium will end up with an equal mixture of
\(A\) and \(B\) in dynamic equilibrium\footnote{This can be
proven by subsequently adding some small amount of \(A\) that was
marked with a radioactive tracer isotope and showing that this tracer
ultimately also shows up in \(B\) and \(M\)} with residual \(M\).
The principle that presence of a catalyst cannot affect the
composition of the entropy-maximizing state also holds for
autocatalysts. While this partly hinges on the question whether some
real chemical system can practically relaxate to the
entropy-maximizing state, thermodynamic theory at least allows us to
give a quantitative answer to the question how much work one could
still extract from a system that has not yet fully equilibriated.

Viewed from the perspective that any reaction kinetic model of
autocatalytic processes can only be in alignment with thermodynamics
if any reaction also includes the backward reaction, with the ratio of
kinetic reaction constants fixed by the thermodynamic stability of
products vs.~reagents, one finds that even oft-cited results from the
literature either ignore or are at odds with thermodynamics. Taking
for example the ``Lenia''~\cite{chan2018lenia} system as a toy model
widely known in the ``artificial life'' research community for having
dynamics reminiscent of emergent ``living organism'' like behavior,
one might wonder if a system like this could spontaneously arise in a
system governed by a chemical flow
equilibrium. In~\cite{kojima2023implementation}, it has been shown
that, while ``Lenia'' is in principle ``asymptotically'' (i.e.~better
than any allowed deviation \(\epsilon>0\)) implementable in terms of a
reaction-diffusion model, there are obstacles to finding a chemical
realization of such a system due to a need to include terms that are
at odds with the law of mass action -- at least in a straightforward
approach.

A more prominent example is given by the claim in
\cite{lifson1997crucial} that if an autocatalytic process were to by
chance create a mutant which is a more effective auto-catalyst in a
system that is in a flow equilibrium rather than in thermodynamic
equilibrium, then that mutant were to drive the original form to
extinction. The underlying mathematical model is expressed in terms of
chemical species and reaction rates, but lacks the terms that would
make autocatalysis respect the principle that no catalyst can shift
chemical equilibrium. If one were to correct this by adding the
reverse reactions with permissible ratios of reaction constants, and
also ensured that the flow equilibrium is implemented in a way that
not merely adds reagents but also removes chemical species in a way
that keeps the total amount of substance constant, one would instead
find that this situation would lead to a flow equilibrium in which the
more stable form is favored over the other, but not extinction. As one
would expect, increasing the flow rate will give both autocatalysts
less opportunity to transform reagent into the corresponding form,
hence reduce the overall amount of product, in such a way that in
terms of relative ratios, a form that can autocatalytically create
more of itself fast is favored over a form with overall smaller
forward- and backward-reaction constants. We give a fully worked out
numerical example in
appendix~\ref{appendix-a-competing-autocatalytic-species-in-a-flow-equilibrium}.
In a wider context, for processes that dynamically produce species
abundance asymmetry, the need for thermodynamic nonequilibrium in
addition to symmetry-violating dynamics, and the relation between
degree of symmetry violation and obtainable asymmetry has been
explored at least since A. Sakharov's 1967 article on
baryogenesis~\cite{sakharov1998violation}.

\hypertarget{autocatalysis-example-2-the-electromagnetic-radiation-field}{%
\paragraph{Autocatalysis Example 2: The Electromagnetic Radiation
Field}\label{autocatalysis-example-2-the-electromagnetic-radiation-field}}

While the ``\(\text{ABM}\)'' toy example is plausible yet
still mildly artificial, given the assumption that a dimer can act as
a mold, the interaction of matter with electromagnetic radiation
provides a very concrete and physically fully realistic second example
for an autocatalytic system that -- despite existence of an
autocatalytic replication mechanism (namely induced emission) -- does
not evolve towards a state where one replicator takes over.

If we imagine box with conductive walls filled with thermal
electromagnetic radiation\footnote{For a
  \([0; L_x]\times [0; L_y]\times [0; L_z]\) box,
  we can get the electromagnetic field modes from a $\vec\nabla\cdot \vec A=0, \phi=0$
  (``Lorenz/Weyl-gauge'') ansatz: \(\vec E=-\partial_{(ct)}\vec A,\;\vec B=\vec\nabla\times\vec A\)
  with \(\vec A(x,y,z,t)=\left({\scriptscriptstyle\begin{array}{l}A_x^{(0)}\cos(k_xx)\sin(k_yy)\sin(k_zz)\\ A_x^{(0)}\sin(k_xx)\cos(k_yy)\sin(k_zz)\\ A_x^{(0)}\sin(k_xx)\sin(k_yy)\cos(k_zz)\end{array}}\right)\cos(\omega t+\varphi)\), where \(\vec A^{(0)}=(A_x^{(0)}, A_y^{(0)}, A_z^{(0)})\), \(\vec k=(k_x, k_y, k_z)\), 
  \(k_{x,y,z}=n_{x,y,z}\pi/L_{x,y,z}\), and \(\omega=c(\vec k\cdot\vec k)^{1/2}\).
  The additional constraint $\vec A^{(0)}\cdot\vec k=0$ removes the unphysical ``longitudinal''
  polarization-direction, requiring us to make a (per-\(\vec k\)) choice
  for the two different physical transversal polarization-directions.
}
(so, effectively, an ``oven'') which also
contains some low-pressure gas species that can absorb and emit
thermal photons at various wavelengths, an oscillation mode of the
electromagnetic field that is thermally excited to contain an expected
number of \(\bar n\) photons\footnote{Each such oscillation mode is
  best thought of as a quantum mechanical harmonic oscillator with
  energy eigenvalues $E_n(n+1/2)hf$, where \(n\) is the ``number of
  photons'' in the oscillation mode. One notes that for quantum states
  with a well-defined number of photons, the expectation value of the
  electrical field strength (given by the ``position'' operator) is
  zero, and as such, oscillator states that correspond to macroscopic
  oscillations, i.e. for which \(\langle \vec E\rangle\propto \vec E_0\cos\omega t\),
  do not have a well-defined number of photons. These can be modeled by
  coherent states -- eigenstates of the non-hermitean lowering operator
  of the Heisenberg algebra that are superpositions of quantum states
  with different numbers of photons.}
can interact with gas atoms (or molecules) not only in such a way that
it transfers energy to an un-excited atom, putting it into an excited
state. Alternatively, a photon can also interact with an excited atom
in such a way that the atom becomes un-excited, and the atom's
excitation energy increases \(\bar n\) by one: the (expected)
\(\bar n\) many photons in the given excitation mode created a new
copy, making another photon in the radiation field have the same
energy, direction, and polarization, as the photons already present in
that oscillation mode.  Schematically, we could express this process
as:

\begin{equation}\{\text{Excited Atom}\}+\gamma_{\omega,\vec k,\pm}\to \{\text{Un-Excited Atom}\} + 2\gamma_{\omega,\vec k,\pm}\end{equation}

The extra photon created has the same frequency \(\omega\), wave-vector
\(\vec k\) (so, direction), and polarization (\(\pm\)) as the incoming
photon.

Still, despite the existence of this ``induced emission''
mechanism~\cite{einstein1916quantentheorie}, the end state of a system
of atoms and radiation modes among which a given amount of energy is
distributed in an entropy-maximizing way will not be dominated by one
kind of photon that turned all (or most) available energy into copies
of itself -- we instead find a blackbody spectrum. Concretely, a
closed cavity with conductive walls will have a discrete infinite
tower - labeled by integer index \(n\) -- of resonant modes at
frequencies \(f_n\). These may be degenerate due to both polarization
and also symmetries related to the shape of the cavity. Each such
oscillation mode \(M_n\) can be in a superposition of quantum states
where it is occupied with \(0, 1, 2, \ldots\) photons, and
thermodynamic equilibrium occupation rates (which only have to take
into account incoherent superpositions) would be of the form
\(p_n(N\,\text{photons})=W_n\exp(-Nh f_n/(k_BT))\). Since these
probabilities have to sum to 1, and using the abbreviations
\(\beta:=1/(k_BT)\) and \(B_n:=\exp(-\beta hf_n)\) we get
\(1/W_n=\sum_N B_n^N=1/(1-B_n)\), and so for the expectation value of
the energy of the mode \(M_n\) with frequency \(f_n\), we have:

\begin{equation}\begin{array}{l}
\langle E_n\rangle=\sum_N p_n(N)\cdot N hf_n
=W_n (-\partial_\beta) \sum_N B_n^N
=-\partial_\beta\ln (1/W_n)=\\
=-(1-B_n)/(1-B_n)^2\partial_{-\beta}(1-B_n)=hf_n\exp(-\beta hf_n)/(1-\exp(-\beta hf_n))=\\
=hf_n/(\exp(\beta hf_n)-1).
\end{array}\end{equation}

Now, if that cavity is filled with some diluted gas that interacts
with electromagnetic radiation by undergoing transitions between
quantum energy levels, and \(c_i\) is the concentration of particles
in energetic state \(i\) (so, ground state or some specific excited
state), we would a priori expect that in thermodynamic equilibrium,
detailed balance is satisfied. That is, for any pair of states, the
total number of \(i\to j\) transitions in the cavity per unit of time
would equal the total number of \(j\to i\) transitions. This is in
alignment with the idea that if we somehow were able to impede or
support other, unrelated transitions that involve some other energy
level(s), any such tweaks would not impact these two rates. We will
here pick a pair of states \((i, j)=(1, 2)\) where \(i=1\) describes
the lower energy state, \(j=2\) the higher energy state, the
difference in energy is \(E_{1\leftarrow 2}=hf_{1\leftarrow 2}\), and
following Einstein~\cite{einstein1916quantentheorie}, we can write the
rate-balance equation as a sum of three terms that are parametrized by
at-first unknown rate coefficients: the rate
\(R^{\rm SE}_{1\leftarrow 2}\) of spontaneous emission of a photon of
energy \(E_{1\leftarrow 2}\) is only proportional to the concentration
\(c_2\) of excited gas particles in quantum state \(2\):
\(R^{\rm SE}_{1\leftarrow 2}=a_{1\leftarrow 2}\cdot c_2\). The rate of
absorption \(R^{\rm Ab}_{2\leftarrow 1}\) of a photon of energy
\(E_{1\leftarrow 2}\) by particles in energy state 1 will be
proportional to the concentration of such particles \(c_1\) as well as
the number of photons of suitable frequency of the given energy in the
cavity. Omitting some irrelevant detail, we here have to pick some
small energy interval \(\Delta E\) (and all subsequent claims will
become exact in the limit where \(\Delta E\) goes to zero while the
size of the cavity is taken to go to infinity faster than
\(L=hc/\Delta E\)) and write the relevant
suitable-photons-per-unit-volume density as the product of the number
of available modes \(\rho_M(f)\) per energy interval \(\Delta E\) and
the expected number of photons in a mode \(M\) with suitable
frequency. As we have seen, this is a function of mode frequency and
temperature. The mode-density function \(\rho_M(f)\) will in general
depend on the geometry of the cavity -- this is obvious for frequencies
for which the associated wavelengths are comparable to the size of the
cavity (``oven''), since we will not be able to fit a standing wave of
half-wavelength \(\lambda/2\gg L\) into an oven of diameter \(L\). For
increasingly large frequencies, the mode-density \(\rho_M\) will
increasingly lose information about the actual shape of the cavity and
asymptote towards some common form which can be worked out but will
have no relevance for our considerations. Overall, the rate of
absorption then is:
\begin{equation}R^{\rm Ab}_{2\leftarrow 1}=\frac{b_{2\leftarrow 1}\cdot c_1\cdot\rho_M(f)\Delta E}{\exp(\beta hf_{1\leftarrow 2})-1}.\end{equation}

Finally, we want to include a (perhaps at first speculative) process via
which a photon that would have energy suitable to excite a gas particle
in energy state~1 to energy state~2 interacts with a particle in the
excited state 2(!), inducing a transition to the lower-energy state
while increasing the number of photons in the oscillation mode of the
original photon by~\(+1\) -- i.e.~``autocatalytically creating an exact copy
of that photon''. If such a process were to not exist, we would expect
deeper analysis to show that its coefficient has to be zero, but on
general grounds of microscopic reversibility, it is very reasonable to
include this term in an ansatz. With reasoning as above, we would find
that this Induced Emission rate \(R^{IE}_{1\leftarrow 2}\) would have to
be of the form
\begin{equation}R^{\rm IE}_{1\leftarrow 2}=\frac{b_{1\leftarrow 2}\cdot c_2\cdot\rho_M(f)\Delta E}{\exp(\beta hf_{1\leftarrow 2})-1}.\end{equation}

Detailed balance then requires that in thermodynamic equilibrium, the
total contribution of \(1\leftrightarrow 2\) transitions to the rate of
change of concentration \(c_2\) is zero: \begin{equation}
R^{\rm SE}_{1\leftarrow 2}+R^{\rm IE}_{1\leftarrow 2}-R^{\rm Ab}_{2\leftarrow 1}=0.
\end{equation}

Using the above expressions, we can write this as:

\begin{equation}
a_{1\leftarrow 2}c_2
+b_{1\leftarrow 2}\cdot c_2\cdot\rho_M(f)\Delta E\cdot\left(\exp(\beta hf_{1\leftarrow 2})-1\right)^{-1}
=b_{2\leftarrow 1}\cdot c_1\cdot\rho_M(f)\Delta E\cdot\left(\exp(\beta hf_{1\leftarrow 2})-1\right)^{-1}.
\end{equation}

In thermodynamic equilibrium, we will have
\(c_2=c_1\exp(-\beta hf_{1\leftarrow 2})=:c_1 B_f\), and so we get:

\begin{equation}
a_{1\leftarrow 2}B_f\left(\exp(\beta hf_{1\leftarrow 2})-1\right)
+b_{1\leftarrow 2} B_f \rho_M(f)\Delta E
=b_{2\leftarrow 1} \rho_M(f)\Delta E.
\end{equation}

Hence, \begin{equation}
a_{1\leftarrow 2}(1-B_f)+\left(b_{1\leftarrow 2} B_f -b_{2\leftarrow 1}\right)\rho_M(f)\Delta E=0.
\end{equation}

In the limit \(T\to\infty\), we have \(\beta=1/(k_BT)\to 0\) and so
\(B_f=\exp(-\beta h f_{1\leftarrow 2})\to 1\). This shows that
\(b_{1\leftarrow 2}=b_{2\leftarrow 1}\) must hold -- if there is any
absorption at all, then the corresponding ``induced emission'' process
must indeed occur, and the rate-coefficient must equal that for
absorption\footnote{This can also be reasoned out entirely quantum
  mechanically by calculating the actual matrix elements.}. With this
conclusion, we can next consider the limit \(T\to 0\) where we have
\(B_f=0\) (i.e.~no thermal excitation at zero temperature), and find
\(a_{1\leftarrow 2}=b_{2\leftarrow 1}\rho_M(f)\Delta E=b_{1\leftarrow
  2}\rho_M(f)\Delta E\).  Observing these relations between our three
rate-parameters, the detailed balance equation becomes:

\begin{equation}
c_1\underbrace{b_{2\leftarrow 1}\cdot(\rho_M(f)\Delta E)\cdot\frac{1-B_f}{1-B_f}}_{\text{Spontaneous\;Emission}}
+c_1\underbrace{b_{2\leftarrow 1}\cdot(\rho_M(f)\Delta E)\cdot\frac{B_f}{1-B_f}}_{\text{Induced\;Emission}}
-c_1\underbrace{b_{2\leftarrow 1}\cdot(\rho_M(f)\Delta E)\cdot\frac{1}{1-B_f}}_{\text{Absorption}}=0.
\end{equation}

From the perspective of a photon of frequency \(f_{1\leftarrow 2}\)
trying to use induced emission to self-replicate, we find that while
there is a ``self-replication'' term that increases the number of
copies, there is another process (absorption) that destroys copies
faster than they get created, and the rate-difference is made up by
spontaneous creation (spontaneous emission) of suitable photons. So,
the mere existence of a self-replication mechanism does not lead to a
transition to a replicator-dominated state. For the radiation field,
we rather find that ``light amplification by stimulated emission of
radiation'' (for which the acronym LASER was
coined~\cite{maiman1960stimulated}) requires specific circumstances
that are incompatible with thermodynamic equilibrium (in particular,
population inversion).

Also, if one takes extra measures to enable replication (by causing
population inversion and -- oftentimes, but not always -- also building an
optical resonator around the lasing medium), it is by no means true that
the most successful replicating mode would drive other possible laser
modes to extinction. This phenomenon is all too well known in quantum
optics as ``parasitic lasing''.

We end this exploration of the radiation field with a comment that puts
our derivation into physics context: while the key relations between the
relevant physical entities -- individual and joint thermodynamic
equilibrium for the gas and the radiation field -- are as in Einstein's
original derivation, we have taken Einstein's main conclusion (Planck's
blackbody radiation formula) as an additional input rather than an
output. Also, we did not elaborate on the precise form of the
mode-density factor that a physicist certainly would want to know the
specific form of. Finally, physics discussions of the Einstein
coefficients normally use spectral radiance (electromagnetic energy per
frequency interval per unit of time per area per solid angle) where we
instead reasoned with photon density (respectively the total number of
photons). As such, our coefficients
\(a_{1\leftarrow 2}, b_{1\leftarrow 2}, b_{2\leftarrow 1}\) are not
quite the Einstein coefficients, but closely related to them.

\hypertarget{conclusion-on-autocatalysis-mechanisms}{%
\paragraph{Conclusion on Autocatalysis
Mechanisms}\label{conclusion-on-autocatalysis-mechanisms}}

Overall, if a system contains a mechanism via which some entity can
create copies of itself with some positive rate \(a\),
i.e.~\(d/dt\,X = a\cdot X\), the key question is whether the \emph{net
  replication rate} that also takes into account \(X\)-destroying
processes with total rate \(b\), the total rate \(d/dt X=(a-b)X\) is
still positive. Clearly, being able to precisely work out total
reaction rates is important for understanding expected system
behavior. In some situations, this may require modeling that can
isolate a sub-aspect of the dynamics, such as the fate of a subset of
fragments for which one can come up with reasonably good
approximations for the impact of ``everything else'' on that
dynamics. It seems reasonable to think that being close to
thermodynamic equilibrium makes it hard for self-replication
mechanisms to work, and so a natural question is what specific aspect
of an inequilibrium situation gets exploited by any given working
self-replication mechanism. For systems such as the one discussed
in~\cite{alakuijala2024computational}, one may hypothesize that being
in an equilibrium situation would show in the distribution of the
relative proportions of different machine operations (or perhaps
operation sequences) that actually get executed by the (virtual) CPU,
and in nonequilibrium, there clearly is competition for adjusting
relative proportions.

\hypertarget{modeling-approach}{%
\section{Modeling Approach}\label{modeling-approach}}

Ultimately, one would want to have a complete ladder of models that
reach all the way from chemistry up to (collective) consciousness where
the relation between neighboring levels of frameworks is in spirit
equivalent to Dirac's claim that ``the underlying physical laws
necessary for the mathematical theory of \ldots{} the whole of chemistry
are thus completely known, and the difficulty is only that the exact
application of these laws leads to equations much too complicated to be
soluble''~\cite{dirac1929quantum}.

Relative to observations about evolution in computational toy models
as in~\cite{alakuijala2024computational}, the aspiration of this work
is to add a rung on this ladder between these computational models and
molecular chemistry, describing (computational) dynamics in terms of
chemical reaction kinetics. One likely would then want to add another
rung even further down that establishes contact with nonequilibrium
thermodynamics by resolving how interactions are implemented in terms
of reversible Hamiltonian dynamics. As is already well-understood for
even lower rungs of the extension of this ladder from chemistry to
quantum field theory, at each level we are making educated decisions
about modeling by starting from the more fundamental model, trying to
fully embed simple use cases of the higher level model into the more
fundamental one (such as trying to understand the nature of the
chemical bond in terms of quantum mechanics of the \(\text{H}_2^+\)
ion and then \(\text{H}_2\) molecule), identifying and separating
relevant detail from aspects that can be neglected with reasonable
justification (such as most relativistic corrections when going from
the Dirac to the Schr\"odinger equation via a Foldy-Wouthuysen
transformation), and then using the ``less precise'' model for an
enlarged set of use cases. Among these use cases, there is a blurry
boundary of ``small'' problems that are tractable by the higher level
model, but also more precisely (but with relevant effort) in the
more fundamental one.

In terms of theory ingredients, the construction presented here will
use basics of chemical reaction kinetics and thermodynamics at the
level of what is covered in the undergraduate chemistry core
curriculum (such as what is covered in Atkins's ``Physical Chemistry''
textbook ~\cite{atkins2023atkins}), plus basic theoretical concepts
about computing and programming languages as covered by core
curriculum computer science courses (such as in
particular~\cite{abelson1996structure}).

\hypertarget{general-principles}{%
\subsection{General Principles}\label{general-principles}}

Following this general approach, we want to base our modeling on the
principles listed below. A precise in-depth description of our modeling
approach can be found in appendix~\ref{appendix-b-modeling-details}.

\begin{enumerate}
\def\labelenumi{\arabic{enumi}.}
\tightlist
\item
  The basic entities that model chemical/computational processes are
  ``tapes'' (computational perspective) that equivalently can be
  interpreted as linear polymers (chemical perspective). These polymer
  strands are considered to be made of linking up monomer units from a
  finite (small) set of possible choices, the ``symbol alphabet''\footnote{
This is in close alignment with the role of RNA/DNA and somewhat
reasonably closely related with the role of oligopeptides in biological
information-processing systems. Given speculations that RNA's
autocatalytic properties might have played an important role in
abiogenesis, such focus appears justifiable.}.
\end{enumerate}

\begin{enumerate}
\def\labelenumi{\arabic{enumi}.}
\item
  Changes to the state of the system are modeled as arising from
  interactions between sub-sequences of tapes of finite length.
\item
  We are considering ``rarefied'' situations where polymer
  concentrations are sufficiently low that interactions happen when two
  polymer-subsequences come close to one another and finish over time
  scales that are very short in terms of the rate of such
  interaction-triggering encounters. This allows us to describe
  processes in terms of contributions to an effective rate-of-change of
  tape-subsequence occurrence probabilities, paralleling
  well-established (``textbook'') modeling approaches for chemical
  reaction kinetics.
\item
  The fundamental degrees of freedom described by the model are
  length-\(k\) subsequence occurrence probabilities. Tape composition is
  approximated by a Markov process that can accurately predict the
  probability to find any possible monomer unit given its prefix (or
  suffix) sequence of \(k-1\) monomer units.
\end{enumerate}

In this approach, modeling gets increasingly accurate in the limit
\(k\to\infty\). This approach admits an alternative interpretation of
subsequence probabilities in terms of chemical concentrations,
establishing full equivalence to reaction kinetics. Beyond this
probabilistic modeling of tape-content with limited \(k\), we strive for
not having any avoidable further inaccuracies as intrinsic properties of
the framework -- further approximations can be made, but this should be
the user's choice.

\begin{enumerate}
\def\labelenumi{\arabic{enumi}.}
\item
  The details of interactions between tape sub-sequences are fully
  user-definable, with possibilities ranging from interactions that
  closely model known (e.g.~measured) chemical reaction rates to ``total
  chemical fiction'' interactions that can describe e.g.~operation of a
  ``program'' stored on the first tape-sequence on ``data'' stored on the second
  tape-sequence according to some complex evaluation rules that have a stronger
  resemblance to Turing machine program execution than plausible
  chemical processes.
\item Given some initial subsequence probability-distribution (which
  provides a precise answer to some questions of the form ``if one
  were to probe the system at a random place, what is the probability
  to find \(\{\mbox{this particular sequence}\}\)''), plus
  user-definable transformations that cover all relevant cases (via
  nondeterministic evaluation) in a convenient and quantitatively
  precise way, we obtain the momentary rates-of-change of the system's
  degrees of freedom -- i.e.~subsequence probability
  rates-of-change. These we then integrate via numerical
  ODE-integration.
\item
  In alignment with fully user-specifiable interactions, no effort is
  made to model reaction constants, i.e.~how such reactions are powered
  specifically by entropy-generating processes. As with the material
  parameters we encounter in constitutive equations of much of
  engineering and condensed matter physics, such quantities are regarded
  as parameters that are provided by experiments or more fundamental
  modeling.
\end{enumerate}

Resolving the last point is the main aspect that will have to be handled
on the next rung down from this level of modeling towards molecular
chemistry. The flexibility to have arbitrary user-definable dynamics
will allow injecting chemically reasonable interactions at this level of
modeling, somewhat akin to how models of the mechanical properties of
simple gas molecules allow injecting reasonable estimates for the heat
capacities of different gases\footnote{Footnote: monoatomic gases have
  three translational degrees of freedom, hence we find \(c_V=3N_Ak_B/2\)
  to be a good approximation, while diatomic gases also have two rotational
  degrees of freedom, hence \(c_V\approx 5N_Ak_B/2\).}
into machine models that use engineering
thermodynamics. In this sense, this modeling approach can bridge the gap
between molecular chemistry and ``metaphor level'' computational models
of abiogenesis.

With respect to ODE-integration, we need to be careful about the detail
that generic user-provided interactions can readily give rise to very
small rates-of-change, if the conditions to encounter a particular
situation are unlikely. Especially in models where initial conditions
are such that some sequences do not occur at first, the kind of
numerical interpolation used by ODE integrators generally gives rise to
some amount of numerical noise, and that noise can make some
subsequence-probabilities take on negative values. Our modeling approach
is to regard such accidental numerically negative probabilities as
``zero probability within the bounds of the numerical accuracy of our
modeling''. From a purely numerical perspective, one possible
improvement might be to determine probability rates-of-change, but
represent the degrees of freedom of the model (on which ODE-integration
is performed) as logits \(\lambda_i=\log (1-p_i)/p_i\), and translate
the rates-of-change equations into logit space. One then would likely
also want to have a normalization constraint on the logits that
eliminates drift. Numerical noise from ODE-integration also can violate
the realizability constraint of the subsequence probability distribution
that we now want to discuss.

When using the framework, one has to pick a symbol-alphabet \(A\)
different types of monomer ``letters'' and then specify initial
conditions in the form of the initial probabilities to encounter any
of the \(k^A\) possible length-\(k\) (polymer) subsequences when
picking at random a cell on any tape and scanning forward until one
sees a complete sequence of \(k\) monomers. Here, one has to note that
not any vector of \(k^A\) probabilities that sum to \(1\) is
compatible with the idea of tape-composition being approximately
describable by a Markov process.  For example, if we have two types of
monomers, \(X, Y\), and model tape-composition in terms of length-2
subsequence probabilities, a situation such as
\(p_{XX}=p_{YY}=p_{XY}=p_{YX}=0.25\) would describe a completely
random chain, while \(p_{XX}=p_{YY}=0.48, p_{XY}=p_{YX}=0.02\) would
describe chains containing long sequences of \(X\)s, respectively
\(Y\)s, with rare switches of type along any such chain. In contrast,
\(p_{XX}=p_{YY}=0.48, p_{XY}=0.03, p_{YX}=0.01\) would describe an
\emph{impossible} situation: we can work out the probability for a
monomer at a random position to be an \(X\) as
\(p_X=p_{XX}+p_{XY}=0.51\) by summing over all possible suffixes, but
summing over possible prefixes instead gives us a conflicting answer,
\(p_X=p_{XX}+p_{YX}=0.49\).

An obvious criterion for realizability of a sequence probability
distribution is that the distribution aligns with the one obtained by
predicting the next token given a length-\((k-1)\) prefix, starting from
the current distribution. We can form a \(A^k\times A^k\) ``transfer
matrix'' \(T\) where the entry \((i,j)\) is the probability that the
\(j\)-th length-\(k\) sequence in lexicographic ordering is followed by
a symbol such that the length-\(k\) tail of the length-\(k+1\) extended
sequence is the \(i\)-th length-\(k\) sequence in lexicographic
ordering, and the prediction is made from the
next-symbol-given-the-prefix table obtained from the probability table.
Then, the probability distribution, written as a length-\(A^k\)-vector
of probabilities of length-\(k\)-sequences in lexicographic order, must
be an eigenvector of the matrix \(T\) for eigenvalue 1, and have entries
that all lie in the interval \([0, 1]\) and sum to 1.

In principle, a subsequence probability distribution can describe a
mixture of very long tapes. A simple example would be
\(p(XX)=0.5=p(YY)=0.5\), with all other probabilities being zero.
Generating tapes at random, one would either produce an all-\(X\) or an
all-\(Y\) tape, with equal probability, but never see any transition. We
will make use of such probability distributions in examples that model a
``tape'' as corresponding to a macromolecule in some ``solvent'' whose
composition also is relevant for the dynamics.

\hypertarget{interaction-rule-implementation}{%
\subsection{Interaction-rule
Implementation}\label{interaction-rule-implementation}}

In addition to specifying initial conditions, the user also has to
provide some definition of subsequence-interactions -- which in
degenerate cases may be completely independent of the content of one of
the two tape-sequences. Given that we want our modeling approach to
support rather complicated such interactions roughly at the level of
interpreting machine language operations, the only reasonable approach
is to allow specifying such an ``evaluator'' as a piece of code in some
programming language.

From the perspective of describing the time evolution of parameters that
describe a Markov process (here, length-\(k\) subsequence
probabilities), we are looking at an unusual type of ordinary
differential equation for which the \((d/dt)y(t)=f(y, t)\) right hand
side function \(f\) involves regarding the data content of \(y\) (here,
tape-state) as programs that are submitted to a program evaluator. It
might make sense to introduce the term ``algorithmic kinetics ODE''
(AK-ODE) for such a novel kind of ordinary differential equation.

An attractive choice for allowing users of the framework to specify
tape interactions is provided by the ``Scheme'' programming
language~\cite{steele1978rabbit, sperber2009revised6}, for three main
reasons:

\begin{itemize}
\item
  Scheme offers highly nontrivial features for implementing unusual
  evaluation semantics that we can put to good use for our intended use
  cases.
\item
  The language is rather simple and well-defined, and in particular is
  very well suited for formally specifying the behavior of evaluators in
  a purely mathematical (i.e.~functional, side effect free) way -- as is
  required for our approach to ``nondeterministic'' evaluation.
\item There are open source Scheme implementations which offer rather
  reasonable performance via compilation to machine code, and these
  are also reasonably easy to integrate into
  Python~\cite{van2007python}, for which there are convenient-to-use modules that
  can handle ODE-integration, plotting, and data processing.
\end{itemize}

More concretely, we would want the user to be able to write code that
directly aligns with some description such as: ``If the symbol at the
cursor is \(A\), we also check the symbol at the next cell; if this is
also \(A\), execution stops. Otherwise, if it is \(B\),
\ldots{}''. This is most straightforward if users can write code that
performs checks for which the framework then takes care of exploring
all possible relevant avenues. Given that we want tape-content to be
modeled by a Markov process, we cannot go for a naive approach that
merely runs the user's code on each of the possible tape-subsequences
of some particular length. The problem is readily understood by
considering a process where, whenever we encounter a subsequence
\texttt{101} on a binary tape that is modeled as a Markov process
parametrized by eight length-3 subsequence probabilities, this gets
re-written to \texttt{000} with some decay-rate \(\alpha\). Starting
from an initial condition in which symbols on the tape keep
alternating \texttt{...1010101010...}, if we were to only take the
\((d/dt)\,p_{101}=-\alpha p_{101}, (d/dt)\,p_{000}=+\alpha p_{101}\)
decay into account, we would start with
\(p_{010}(t=0)=0.5=p_{101}(t=0)\), and ODE-integration would get us to
an impossible situation of the general form
\(p_{000}(t)=\beta(t), p_{101}=0.5-\beta(t), p_{010}=0.5\) for which
we get conflicting probabilities \(p_0(t)=0.5+\beta(t)\) from summing
over length-2 suffixes and \(p_0(t)=0.5\) from summing over length-1
prefix and suffix. Rather, when \texttt{101} gets re-written to
\texttt{000}, mathematical consistency requires us to take into
account that \texttt{101} can be continued to the left to
\texttt{0101}, and from there to \texttt{10101} and then for \(t>0\)
also \texttt{00101}, so we also get context-induced contributions e.g. for
``re-writing \texttt{101} to \texttt{000} can turn \texttt{010} into
\texttt{000}''.

Generically, when we find that a particular combination of tape
sequences \((p, d)\) gets turned into \((p', d')\), we have to explore
the impact that the corresponding adjustments have on all Markov
process length-\(k\) subsequence probabilities where the subsequence
can overlap with tape-mutations. We in general want to do this in an
efficient way that does not explore the content of unexplored parts of
the tape unless this is unaviodable for determining rate-of-change
contributions. This is explained in detail in
appendix~\ref{appendix-b-modeling-details}.

With Scheme, we have the opportunity to conveniently implement a
function with unusual execution semantics such as \texttt{tape-get-sym}
in the framework presented here. The expression
\texttt{(tape-get-sym\ data-tape?\ i)} will evaluate to the symbol at
relative index \texttt{i} (which can be positive, zero, or negative) to
the starting index on the either the data tape (if \texttt{data-tape?}
is a Scheme boolean representing ``true'', \texttt{\#t}) or the program
tape (if \texttt{data-tape?} is the boolean \texttt{\#f} that represents
``false'') -- but in such a way that any such evaluation of a tape-lookup
for which the corresponding tape has not yet been ``unfolded'' to this
position according to the current tape-composition specified by the
Markov process's current probability-table will (effectively) split the
computation into a ``multiverse'' where each different computational
universe keeps track of its probabilistic weight, and while inside each
such universe, \texttt{tape-get-sym} evaluates to a single value,
looking across all universes, the expression will evaluate to
\emph{every possible value}. One would then in general also want to use
\texttt{tape-set-sym!} calls to adjust the content of the data- or
program-tape (possibly both) in one or multiple places. The framework
offers some additional scaffolding which, behind the scenes, will not
only keep track of the probability weights of each of the different
universes created by ``performing measurements of tape-content'', but
then also aggregate the observed adjustments into total rates-of-change
for the parameters of the Markov process. The first (and most basic)
example in section~\ref{example-1-radioactive-decay} should serve to
illustrate how this works.

The key idea behind implementing the \texttt{(tape-get-sym\
  data-tape?\ index)} function is to exploit the fact that Scheme
compilers, like many compilers for other functional programming
languages, as a first step translates code to continuation-passing
style (CPS)~\cite{steele1976lambda}. This general compiler technique
establishes symmetry between the notions of ``calling a function'' and
``returning a value from a function''. More concretely, code in
continuation-passing style, rather than having a function return a
value to its caller, passes on the value to a callable that performs
all the still-to-be-done data processing on this value. One benefit of
continuation passing style is that it trivializes some compiler
optimizations such as in particular \emph{tail call optimization}: if
the last action performed on the call frame of a function is to call
another function and then return its value to the current function's
caller, one can instead transform the call to the callee, which
nominally would receive a ``receive the return value and pass it on to
the caller's caller'' continuation, into a call that directly passes
the caller's return-continuation to the callee -- as the continuation
to use for returning a value. The Scheme programming language is
specified to contain a rather special function
\texttt{call-with-current-continuation}, via which code can get access
to its own continuation and then proceed to store it away and possibly
invoke that continuation multiple times. As is explained in detail
e.g.~in section 4.3 of the textbook~\cite{abelson1996structure}, this
approach not only trivializes implementing iterative processes via
recursive procedures, but in particular exposing the
\texttt{call-with-current-continuation} primitive then also allows
implementing unconventional execution semantics such as
``nondeterministic evaluation'' which nominally looks as if a function
could have multiple different return values, and all computational
pathways in which each of the different return values is used are
explored. For such an approach to work, one generally would want the
computations that get restarted at the point where some function
returns alternative values to be implemented in a purely functional
form, i.e.~\emph{as a pure mathematical data transformation without
  any side effects}. In the present framework, we \emph{do} have
functions that appear to affect state, since we can use
\texttt{(tape-set!\ data-tape?\ index\ value)} to adjust the content
of the two tape-sequences (``data tape'' and ``program tape'') in
focus, but these again use the behind-the-scenes machinery to keep
track of tape-adjustments in any one computational universe that gets
created by ``multiverse splitting'' whenever a measurement of
previously-unrevealed tape-state is performed. \emph{Aside from using
framework functions for mutating tape-state, no other side effects are
permissible in evaluator code in the sense that restarting
computations would interfere badly with adjustments to other mutable
state that gets shared across universes in a multiverse computation}.

Despite the usefulness for \texttt{call-with-current-continuation} to
implement unconventional execution semantics, it should be noted that
over time, insights have emerged that this approach is not without
problems. A general overview is given in~\cite{kiselyovcc}.

In the framework provided here, the Gambit-C Scheme
implementation~\cite{feeley1994gambit} was chosen as a a basis due to
its reasonably good performance that compares favorably with Python
and ease of integration into Python via its very simple foreign
function interface. Gambit-C supports Common Lisp style
(i.e.~non-Scheme-``hygienic'') macros, and using this feature, the
framework introduces a \texttt{tape-evaluator} macro for defining an
evaluator-function.

Within the body of a \texttt{(tape-evaluator\ \{alphabet\}\
  ...body...)} expression, one can use three bare-bones and three
higher-level functions for reading, respectively writing, tape-state,
plus performing tape-state independent nondeterministic choices with
adjustable per-case probabilistic weight that use the same underlying
multiverse-splitting machinery. In total, these functions are as
follows, with \texttt{data-tape?} being true (\texttt{\#t}) for
accessing the ``data-tape'' and
false (\texttt{\#f}) for accessing the ``program-tape''. In some
problems that do not have a clear ``program acting on data''
interpretation, it makes more sense to instead simply talk of the
``P-tape'' and ``D-tape''. Tape indices can be positive or negative,
but the further away they are from zero, the further the tape will
have to be unfolded from the starting point, leading to a
corresponding recursive splitting of the current universe. For
technical reasons, the \texttt{tape-evaluator} macro generally is not
used much directly, since we usually also want to embed a problem
specification in Scheme into Python (in particular to utilize
ODE-integration provided by widely known Python libraries), and
compiled Scheme programs will be selected from the Python side via
some name(-symbol). In this situation, it makes sense to introduce
another macro, \texttt{(register-problem\ \{problem-name\}\
  \{symbol-alphabet\}\ ...body...)}  that internally expands to a
\texttt{(tape-evaluator\ ...)} expression but also registers the
problem under a name via which the compiled code can be executed from
Python.

{\small
\begin{Shaded}
\begin{Highlighting}[]
\CommentTok{;; Getting the symbol the given tape index relative to the initial}
\CommentTok{;; randomly{-}placed tape{-}cursor:}
\NormalTok{(tape{-}get{-}sym data{-}tape? index)}

\CommentTok{;; Getting the number{-}in{-}the{-}alphabet of the symbol that is found}
\CommentTok{;; at tape the given index:}
\NormalTok{(tape{-}get data{-}tape? index)}

\CommentTok{;; Writing the symbol sym at the given index, relative to}
\CommentTok{;; the initial cursor position:}
\NormalTok{(tape{-}set{-}sym! data{-}tape? index sym)}

\CommentTok{;; Writing the k{-}th symbol of the alphabet at the given index:}
\NormalTok{(tape{-}set{-}sym! data{-}tape? index k)}

\CommentTok{;; Splitting the universe in a way that adjusts current{-}world probability}
\CommentTok{;; according to the given statistical weight of the result {-} the \textquotesingle{}options\textquotesingle{}}
\CommentTok{;; argument{-}list must be of the form:}
\CommentTok{;;   \textquotesingle{}((weight0 option0) (weight1 option1) ...)}
\CommentTok{;; The return value is one of \textasciigrave{}option0\textasciigrave{}, \textasciigrave{}option1\textasciigrave{}, ... {-} each being}
\CommentTok{;; returned in a different computational universe with the corresponding}
\CommentTok{;; relative probability{-}weight.}
\NormalTok{(choose options)}

\CommentTok{;; Lower{-}level variant of choose with arguments v{-}probabilities and v{-}choices,}
\CommentTok{;; where both arguments need to be same{-}length vector objects with the}
\CommentTok{;; first one\textquotesingle{}s entries being probabilities that must sum to 1.}
\CommentTok{;; The i{-}th entry from v{-}options gets returned in an universe with relative}
\CommentTok{;; probability adjusted by the i{-}th entry of v{-}probabilities.}
\NormalTok{(vector{-}choose v{-}probabilities v{-}choices)}
\end{Highlighting}
\end{Shaded}
}

As the examples presented in the next section will show in detail,
having these functions with unconventional evaluation semantics
available allows us, as promised, to write \emph{straight code} that
implements tape-transformations which would be far more complicated to
express if one instead had to first work out the impact of some
particular evaluator-definition on tape-composition parameters.

\hypertarget{ode-perspective}{%
\section{ODE Perspective}\label{ode-perspective}}

In order to simplify proving theorems about our construction, we give a
mathematical description of the approach.

To do so, we need to first introduce some auxiliary definitions. Given
an alphabet \(A\) of \(\#A\) symbols
\(A=(\alpha_0, \alpha_1, \ldots, \alpha_{\#A-1})\), a tape-sequence
\(\bar t\) is a triplet \(\bar t=(m, n, a_t)\) with
\(\mathbb{Z}\ni m\le 0\le n\in\mathbb{Z}\) where \(a_t\) is a function
\(\{k\in\mathbb{Z}|m\le k\le n\}\mapsto A\). Informally, a tape-sequence
maps each index in an index-range to an alphabet symbol, and the
index-range must contain the index zero. An Evaluator-function \(E\) is
a mapping of a pair \((\bar p, \bar d)\) of tape-sequences (the P-tape
(or ``program-tape'') and the D-tape (or ``data-tape'')) to a finite set
of triplets \((p_k, \bar p'_k, \bar d'_k)\) with \(\sum_k p_k=1\) where
we want to interpret \(p_k\) as the probability for evaluation (which
can be nondeterministic) when operating on tape-sequences
\((\bar p, \bar d)\) to result in a pair of tape-sequences
\(\bar p'_k\), \(\bar d'_k\) whose index-ranges match those of
\(\bar p\) respectively \(\bar d\). Intuitively, these will be the
sequences that will replace \(\bar p\) and \(\bar d\) on the P-tape,
respectively D-tape. We further require that for any evaluator-function
\(E\), we can find integers
\((p_\text{min}, p_\text{max}, d_\text{min}, d_\text{max})\) such that
for no \((\bar p, \bar d)\), any \(\bar p'\) from the result-set tuples
differs from \(\bar p\) at an index smaller than \(p_\text{min}\) or
larger than \(p_\text{max}\), and likewise for
\(\bar d', \bar d, d_\text{min}, d_\text{max}\).

Then, we can define \(\mathcal{E}(E)\) to be the set of all possible
pairs
\((\bar p=(p_\text{min}, p_\text{max}, f_p), \bar d=(d_\text{min}, d_\text{max}, f_d))\).
If we specify a Markov process in terms of probabilities of length-\(N\)
words \(\vec w\in A^N\) such that \(\sigma_{\vec w}\) is the probability
to observe the sequence of symbols \(\vec w\) when probing the tape at a
random position, and define
\(S(\bar t)=S((m_t, n_t, f_t)):=\{i|m_t\le i\le \text{min}(m_t+N-1, n_t)\}\)
as the set of possible length-\(N\) sub-word starting indices in
\(\bar t\), as well as
\(R(\bar p, k)=R((p_m, p_n, f_p), k)=(k, k+N-1, f_p|_{k\ldots k+N-1})\)
the length-\(N\) sub-word of \(\bar p\) starting at index \(k\), we can
formalize the ordinary differential equation describing tape-content
evolution via the following ``master equation'':

\begin{equation}\boxed{\boxed{\label{eq:master}\begin{array}{lcl}
\frac{d}{dt}\sigma_{\vec w}(t)&=&\sum_{(\bar p, \bar d)\in\mathcal{E}(E)}\sum_{(p,\bar p', \bar d')\in E(\bar p, \bar d)} p\cdot p_{\sigma_{\vec w}}(\bar p) \cdot p_{\sigma_{\vec w}}(\bar d)\times\Biggl(\\
&&\sum_{m\in S(\bar p), n\in S(\bar d)}\Bigl(\phantom+\delta(\vec w, R(\bar p', m)) + \delta(\vec w, R(\bar d', n))\\
&&\phantom{\sum_{m\in S(\bar p), n\in S(\bar d)}\Bigl(}-\delta(\vec w, R(\bar p, m)) - \delta(\vec w, R(\bar d, n))\Bigr)
\Biggr).\end{array}}}\end{equation}

Here, \(p_{\sigma_{\vec w}}(\bar p)\) is the probability for the Markov
process described by parameters \(\sigma_{\vec w}\) to produce the
tape-sequence \(\bar p\), so
\(p\cdot p_{\sigma_{\vec w}}(\bar p) \cdot p_{\sigma_{\vec w}}(\bar d)\)
is the probability when starting at a random position on both the D-tape
and P-tape to find tape-sequences \(\bar p\), \(\bar d\) for which the
evaluator produces the replacement-proposals \(\bar p', \bar d'\).
Finally, \(\delta(\vec w, \bar v)\) is 1 if the tape-sequence \(\bar v\)
has length \(N\) and for all \(k\), its \(k\)-th symbol aligns with the
\(k\)-th symbol of \(\vec w\) -- otherwise, \(\delta(\cdots)=0\).

\textbf{Uniqueness of the Solution}: A sufficient
criterion for Eq.~(\ref{eq:master}) to have a unique solution is
Lipschitz continuity\footnote{For some of the example systems
presented in section~\ref{example-systems}, solution uniqueness is
immediately evident via other structural properties of the
problem. For the first example, this directly follows from a
factorization argument.}: If we can show that the magnitude of
\(\frac{\partial}{\partial\sigma_{\vec
    u}}\left(\frac{d}{dt}\sigma_{\vec w}\right)\) is bounded by a
constant, then the Picard-Lindel\"of Theorem allows us to conclude
that our for given initial conditions, the solution to our ODE is
unique. We only sketch the key idea for a possible proof here -- it is
advantageous to not approach this from spelling out multiverse
world-probabilities in terms of conditional-probabilities factors but
instead focus on what subsequences can turn into other subsequences
and derive a contradiction from ``partial derivatives can get larger
than any threshold \(M\)'', using guaranteed termination while looking
at no more than two bounded-length stretches of tape (this is a
problem-specific property that may be violated for some
problem-definitions) plus the fact that the tape-rewriting code cannot
inspect the current world-probability in the multiverse.

If every program execution can be guaranteed to finish while only
looking at tape-sites that are contained in two intervals of length no
more than $K$ each, i.e. touching at most \(2(1+K-N)\) length-$N$
words, the magnitude of the probability rate-of-change
\(\frac{d}{dt}\sigma_{\vec w}\) for any symbol-vector $\vec w$ is
bounded by \(2(1+K-N)\) (saturating if any of the touched sequences in
one time-step either turn into \(\vec w\), or cease to be \(\vec w\),
and the probability to encounter this situation is 1).  Every
tape-stretch that turns into \(\vec w\) (or alternatively ceases to be
\(\vec w\)) has been (respectively, turns into) some other sequence
\(\vec u\), and as such, the maximal impact that \(\vec u\)-sequences
can have on \(\vec w\)-sequences arises if every \(\vec u\)-sequence
gets turned into a \(\vec w\)-sequence (or vice-versa) -- but for just
about every problem definition, some of these possible transmutations
(for any choice of \(\vec w\)) will fail to execute. Hence,
\(\left|\frac{\partial}{\partial\sigma_{\vec
      u}}\left(\frac{d}{dt}\sigma_{\vec w}\right)\right|\le
2(1+K-N)\): if this were violated, we could (since
world-probabilities are hidden from the the user-provided evaluator)
identify a fraction of sequences \(\delta \sigma_{\vec u}\) that turn
into a ``too large'' fraction of $\sigma_{\vec w}$-sequences (respectively
vice-versa).

\hypertarget{example-systems}{%
\section{Example Systems}\label{example-systems}}

In general, one can classify systems that can be studied with the
framework presented here as belonging into one of four classes:

\begin{enumerate}
\def\labelenumi{\arabic{enumi}.}
\item
  Systems where the update rules are simple enough that one can
  analytically derive closed-form expressions for time evolution with
  little effort.
\item
  Systems where the dynamics is too complicated to allow closed-form
  analytical treatment, but we can still solve the underlying ODE
  numerically to any desired accuracy.
\item
  Systems that are tractable-in-principle with the framework presented
  here, but their behavior is too complicated to allow analysis that
  does not resort to further approximations on any computer we could
  ever hope to build.
\item
  Systems with fundamentally too complicated dynamics to be analyzed
  with this framework without substantial simplifying approximations.
\end{enumerate}

An obvious problem that can clearly put a system into class 4 is that,
with this framework, while there is a requirement for every
computational path to terminate, showing that this always is the case
can amount to solving an arbitrarily hard mathematical problem.

From a purely chemical perspective, one would argue that in such
situations, i.e.~when there are cases for which program execution takes
a very large number of steps, the ``rarefication'' approximation breaks
down that program interaction is always fast on the relevant time scales
for tape-segments to come close to one another and (by some additional
mechanism not modeled in this framework) get ``energized'' by some
entropy-generation opportunity needed to power program execution. In
this sense, if the overall focus is on chemically plausible systems,
this computational classification of different types of systems has
little relevance.

We proceed to discuss some concrete examples -- as well as their dynamics
as determined with the framework provided here.

\hypertarget{example-1-radioactive-decay}{%
\subsection{Example: ``Radioactive Decay''}\label{example-1-radioactive-decay}}

The purpose of our first example is to anchor explanations for how the
framework operates, using a system that is simple enough to allow exact
analytic treatment. Being able to independently compute, or at least
estimate, probabilities, is generally useful when reasoning about the
behavior of (e.g.~when debugging) more complicated settings.

We are considering tapes made of two symbols only, \texttt{A} and
\texttt{B}, where ``program execution'' degenerates to depending only on
the content of one of the two tapes (which we take to be the
``D''/``data''-tape). Also, we consider simple ``radioactive decay'' of
\texttt{B}-symbols to \texttt{A}-symbols with a constant decay rate that
does not depend on any other circumstances, so not on the structure of
the vicinity of the \texttt{B} symbol on the tape. Despite this, we
decide to model dynamics in terms of a Markov process with a context
window length of three symbols. A fully random initial state is
described by each of the eight possible 3-symbol sequences \texttt{AAA}
to \texttt{BBB} having probability \(1/8\). The definition of the
dynamics then can be given as follows, using the
\texttt{register-problem} Scheme macro provided by the framework:

{\small
\begin{Shaded}
\begin{Highlighting}[]
\CommentTok{;; Example: Radioactive Decay}
\NormalTok{(register{-}problem}
 \StringTok{"ex1{-}radioactive{-}decay"}
\NormalTok{ \#(A B)}
\NormalTok{ (}\KeywordTok{if}\NormalTok{ (}\KeywordTok{eq?}\NormalTok{ (tape{-}get{-}sym }\DecValTok{\#t} \DecValTok{0}\NormalTok{) \textquotesingle{}B)}
\NormalTok{     (tape{-}set{-}sym! }\DecValTok{\#t} \DecValTok{0}\NormalTok{ \textquotesingle{}A)))}
\end{Highlighting}
\end{Shaded}
}

This registers a problem under the \texttt{"ex1-radioactive-decay"} name
string (with which it is identified from the Python side of the
framework), for the alphabet \texttt{\#(A\ B)}, where we inspect the
data-tape at index \texttt{\#0}, and if (and only if) we find a
\texttt{B}-symbol there, we replace this with an \texttt{A}-symbol.

Inside the body of this expression, we can use the (universe-splitting)
functions \texttt{tape-get-sym} and \texttt{tape-set-sym!} to probe and
set symbols at any (negative or non-negative) index of both the data and
program tape, but using far-out indices will lead to behind-the-scenes
splitting of the universe into very many sub-universes, which can
substantially slow down program execution. This program only unfolds and
affects one tape, at a single position. Once program-execution finishes,
the framework compares the original and final state of the visible
window on the tape, and handles extracting parameter updates for all
Markov process context windows that overlap with changes induced by code
execution. The magnitude of each adjustment is given by the probability
to (after all universe-splitting operations) be in a world where all
observations are as found.

In our case, we only affect the data tape, and only at index
\texttt{\#0}. As is explained in more detail in
appendix~\ref{appendix-b-modeling-details}, this will lead to an
unfolding of the relevant visible tape-window as-needed to cover all
relevant sites -- in a way that is not length-constrained by the Markov
model's prefix-length.

If the symbol at data type index \texttt{\#0} was an \texttt{A},
program execution did not change any state, so in this universe, which
has a statistical weight of \(p=0.5\), nothing happens. The universe
in which element \texttt{\#0} is a \texttt{B} gets split further in
order to work out the impact on all length-3 sequences. For
performance reasons, the algorithm that implements this is more
sophisticated than the basic approach (for which there also is a
reference implementation in the provided code), but ultimately, the
end result is the same as if one further split the universe by
exploring tape-content to the left and the right of the farthest-out
revealed change until every window that is possibly affected by the
change has its contents determined. Here, this would mean looking up
two further cells both to the left and to the right, further splitting
the computational universe into \(2^4=16\) ``baby universes'' each
with a statistical weight of \(p=(1/2)\cdot(1/2)^4=1/32\).

Collecting rates-of-change across all possible universes, then at the
initial ``fully random binary sequence with equal probabilities for both
symbols'' starting point, the rates of change come out as:

\begin{equation}\begin{array}{lclclcl}
(d/dt)p_{AAA}&=&+0.375,&\qquad&(d/dt)p_{AAB}&=&+0.125\\
(d/dt)p_{ABA}&=&+0.125,&\qquad&(d/dt)p_{ABB}&=&-0.125\\
(d/dt)p_{BAA}&=&+0.125,&\qquad&(d/dt)p_{BAB}&=&-0.125\\
(d/dt)p_{BBA}&=&-0.125,&\qquad&(d/dt)p_{BBB}&=&-0.375
\end{array}\end{equation}

This is in full alignment with expectations: the 3-symbol sequence
\texttt{BBB} has initial probability \(p_{BBB}(t=0)=1/8\), and there
are three different pathways for decay that are equally likely,
depending on which symbol \texttt{B} is affected. Since our
``interaction rarefication'' approach (see
appendix~\ref{appendix-b-modeling-details}) introduces the time scale
in such a way that one unit of time, \(\Delta t=1\) corresponds to
``expected one program start per program tape symbol'', we get a decay
rate of \((d/dt)\,p_{BBB}=-3/8\). If, for example, the middle
\texttt{B} decays to an \texttt{A}, this also contributes \(+1/8\) to
\((d/dt)\,p_{BAB}\), but since this sequence also has two pathways
along its \texttt{B}-symbols decay to \texttt{A}-symbols, the net rate
of change for \(p\_\{BAB\}\) is
\((+1/8)-2\cdot(1/8)=-1/8\), and likewise for the
other two sequences with exactly two
\texttt{B}-symbols. Symmetrically, the rate-of-change for symbols with
exactly two \texttt{A}-symbols is \(+1/8\), and the momentary
rate-of-change for \texttt{AAA} at time \(t=0\), which can be produced
via three channels, is \(+3/8\).

\begin{figure}
\centering
\includegraphics[width=\textwidth,height=0.25\textheight]{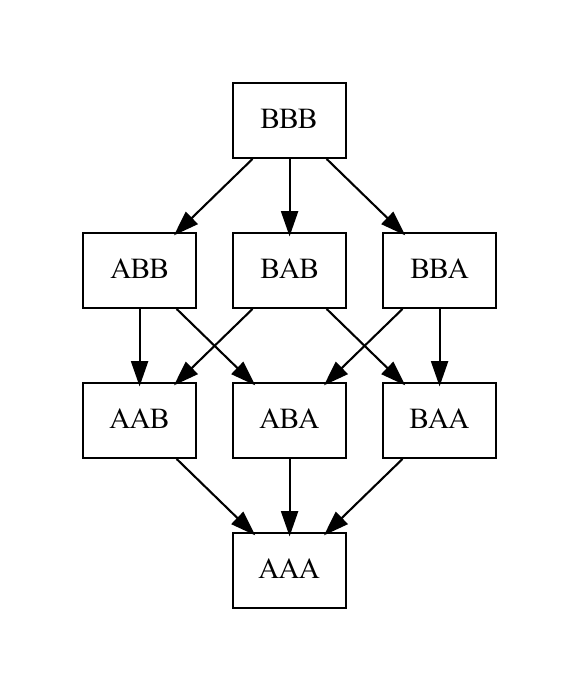}
\caption{Radioactive Decay Example: Parameter Changes}
\label{fig:radioactivegraph}
\end{figure}

In total, the rate equations on length-3 subsequences defined by this
simple update rule can be described diagrammatically as shown in
figure~\ref{fig:radioactivegraph}, where every arrow indicates a
probability rate-of-change contribution that moves one unit of
probability from the arrow's end to its tip.

We not only can fully understand this system, we also can readily write
down the full time evolution by doing away with the unnecessarily wide
context window, describing single-symbol dynamics as
\(p_B(t)=0.5\exp(-t), p_A(t)=1-p_B(t)\), and using
\(p_{XYZ}(t)=p_X(t)p_Y(t)p_Z(t)\). As explained earlier, the purpose of
this example is to illustrate basic aspects of how the framework works.

\textbf{Variant}: Adjusted decay rate.

As later examples will show, when studying computational models with
this framework, one in general would want to have symbols have an
interpretation of machine language opcodes with non-probabilistic
behavior. When studying chemical systems on the other hand, it can make
eminent sense to not have two structurally equivalent reactions have the
same reaction rate, but allow a dependency of the rate on the chemical
neighborhood. The framework provides this via a function \texttt{choose}
which, like \texttt{tape-get-sym}, performs behind-the-scenes
universe-splitting. More precisely, the expression
\texttt{(choose\ \textquotesingle{}((w0\ val0)\ (w1\ val1)\ (w2\ val2))}
evaluates to the value \texttt{val0} with statistical weight
\texttt{w0}, to the value \texttt{val1} with statistical weight
\texttt{w1}, etc. -- returning each possible value in a different
sub-universe. We can use this to amend the previous example to one where
the overall decay-rate is further reduced to \(1/8\) of its original
value as follows:

{\small
\begin{Shaded}
\begin{Highlighting}[]
\CommentTok{;; Example: Radioactive Decay (reduced{-}rate variant)}
\NormalTok{(register{-}problem}
 \StringTok{"ex1var1{-}radioactive{-}decay"}
\NormalTok{ \#(A B)}
\NormalTok{ (}\KeywordTok{if}\NormalTok{ (choose \textquotesingle{}((}\FloatTok{1.0} \DecValTok{\#t}\NormalTok{) (}\FloatTok{7.0} \DecValTok{\#f}\NormalTok{)))}
\NormalTok{     (}\KeywordTok{if}\NormalTok{ (}\KeywordTok{eq?}\NormalTok{ (tape{-}get{-}sym }\DecValTok{\#t} \DecValTok{0}\NormalTok{) \textquotesingle{}B)}
\NormalTok{         (tape{-}set{-}sym! }\DecValTok{\#t} \DecValTok{0}\NormalTok{ \textquotesingle{}A))))}
\end{Highlighting}
\end{Shaded}
}

This ``rejection sampling'' style approach (but without actual sampling,
since the framework nowhere uses random numbers, instead keeping track
of universe-probabilities) will be used in subsequent examples to
implement specific reaction constant ratios for forward- and
backward-reactions in alignment with thermodynamic stability and the law
of mass action.

Here, the framework makes it perfectly valid to make the
multiverse-splitting evaluation part of a conditional -- the behavior of
this code is equivalent to the above:

{\small
\begin{Shaded}
\begin{Highlighting}[]
\CommentTok{;; Example: Radioactive Decay (reduced{-}rate variant)}
\NormalTok{(register{-}problem}
 \StringTok{"ex1var1{-}radioactive{-}decay"}
\NormalTok{ \#(A B)}
\NormalTok{ (}\KeywordTok{if}\NormalTok{ (}\KeywordTok{and}\NormalTok{ (}\KeywordTok{eq?}\NormalTok{ (tape{-}get{-}sym }\DecValTok{\#t} \DecValTok{0}\NormalTok{) \textquotesingle{}B)}
\NormalTok{          (choose \textquotesingle{}((}\FloatTok{1.0} \DecValTok{\#t}\NormalTok{) (}\FloatTok{7.0} \DecValTok{\#f}\NormalTok{))))}
\NormalTok{     (tape{-}set{-}sym! }\DecValTok{\#t} \DecValTok{0}\NormalTok{ \textquotesingle{}A)))}
\end{Highlighting}
\end{Shaded}
}

\hypertarget{example-2-classical-ferromagnetic-spin-chain}{%
\subsection{Example: ``Classical Ferromagnetic Spin Chain''}\label{example-2-classical-ferromagnetic-spin-chain}}

The purpose of this example is to quantitatively clarify the relation
between Monte Carlo simulations, analytically tractable approximations,
and the Markov Process Dynamics model introduced here. We will also
explore the role of finite subsequence window size.

We want to consider a long one-dimensional chain of magnetic elements,
each of which can be in ``up'' or ``down'' configuration. The state of
the system is represented by a vector
\(\vec \sigma,\;\sigma_i\in\{-1,+1\}\). The system's total energy comes
from two contributions: Nearest-neighbor coupling, parametrized by
uniform nearest-neighbor coupling strength \(J\), favors adjacent
magnetizations to be in alignment. Coupling to an external magnetic
field, whose strength is parametrized by \(h\), energetically favors
alignment with the field, i.e.~\(\sigma_i=+1\) if \(h>0\):

\begin{equation}E_{\text{total}}=\sum_{i,j\in\{\text{Sites}\},\;j=i+1}(-J)\sigma_i\sigma_{j} +\sum_i (-h)\sigma_i.\end{equation}

Sign conventions are as usual: extracting work from the system lowers
its energy.

In ferromagnetic model systems of this general structure, the
nearest-neighbor interaction does \emph{not} correspond to the familiar
attractive or repulsive force between magnetic objects from everyday
experience with macroscopic bodies: The force between two
centimeter-scale magnets is a long range interaction (i.e.~decays
according to a power law) that tries to lower total magnetic field
energy in the system
\(E_{\text{magnetic}}=\int\,d^3x\,\vec B\cdot\vec B/{(2\mu)}\) by aligning
one magnet's north pole with the other magnet's south pole. If only this
force governed the physics of magnetism, macroscopic bodies could not
exhibit noticeable intrinsic magnetization.

Here, we are instead modeling the coupling between magnetic elements at
nanometer scale which is governed by the interplay of the Coulomb
(i.e.~electromagnetic) force with quantum mechanical exchange. Between
closeby elements, this effective force is far stronger than the
macroscopically experienceable coupling of magnetic moments, typically
equivalent to some \(100-300\,\text{T}\) of magnetic field strength
whereas a ferromagnet's surface field strength typically is in the
ballpark of \(0.1-1\,\text{T}\) (and hence in comparison negligible). As
this force decays exponentially with distance with characteristic length
scales in the nanometer range, it makes sense to try to approximate this
by a nearest-neighbor-only interaction.

Both classically and also quantum mechanically, magnetization is not
forced to assume only one of two possible directions. If an electron can
be ``spin-up'' \(|\uparrow\rangle\) or ``spin-down''
\(|\downarrow\rangle\) relative to an external magnetic field, the
superposition
\((1/\sqrt2)\cdot\left(|\uparrow\rangle+|\downarrow\rangle\right)\)
would correspond to one of the two independent ``spin perpendicular to
the magnetic field'' directions. One could however imagine engineering
(such as via nanolithography) a chain of non-spherical magnetic dots
which due to their shape have preferred magnetization direction but are
nevertheless small enough to have nearest-neighbor interaction dominated
by the exchange interaction. This would allow for an experimental
realization of a system for which this model is a good approximation.

We want to consider such a classical (i.e.~no quantum superpositions, no
entanglement, each site has a binary ``magnetization-up or
magnetization-down'' degree of freedom) system in a heat bath where
thermal excitations induce magnetization flips at a rate \(R\) that is
in alignment with Arrhenius-N\'eel theory,
\(R_{B\leftarrow A}=\tau_0\exp\left(-\Delta E_{B\leftarrow A}/(k_BT)\right)\), with
\(\tau_0\) being the ``attempt time'', i.e.~characteristic time scale
over which sufficient redistribution of energy among thermal degrees of
freedom has happened to regard the new situation as an independent
attempt at performing the transition.

For the specific example studied in this section, we take
\(\beta=1/(k_BT)=1.0, J=1.0, h=-0.25\). Also, we want to prepare the
system to start out in a state where almost all elements have
magnetization pointing ``down'', with the exception of some rare
neighboring pairs with magnetization up. The main objects of interest
are the probabilities \(p(L=\lambda)\): For each island size
\(\lambda\), this is the probability for a randomly selected site to be
the leftmost magnetization-``up'' site of an island extending of total
size \(\lambda\) extending to the right. Due to \(h<0\), the applied
field tries to force elements into a ``magnetization down'' state. Our
initial configuration shall be described by \(p(L=2)=1/250\) and
\(p(L=k)=0\) for \(k\neq 2\). The idea behind this specific choice is
that we can here easily obtain an analytically tractable approximation
of the dynamics that we can compare with other methods.

In the Monte Carlo approach, this is realized by starting from a ``every
magnetization down'' state, generating one uniformly distributed random
number between 0 and 1 per site and setting the magnetization of every
site for which this number is smaller than the \(p(L=2)\) threshold -
and also its right neighbor -- to ``up''. While this produces some
initial probability to encounter islands larger than size-2, their
occurrence is suppressed by extra powers of \(p(L=2)\), which we here
can take to be negligible.

In this model, the energetic situation makes different relevant
processes happen at different effective rates: For any given element,
its own magnetization state as well as the relation to its two
neighbors' magnetization states determines both the transition-type as
well as its effective rate. The fastest flip-rates are observed for
isolated magnetization-up elements:
\((\cdots\downarrow\uparrow\downarrow\cdots)\to(\cdots\downarrow\downarrow\downarrow\cdots)\).
In a basic Monte Carlo simulation, it makes sense to take the
corresponding flip rate as the reference rate \(R_0\) that is directly
tied to the time scale, and handle all other relevant processes via
rejection sampling: We choose the unit of time such that over a time
interval \(\Delta T=0.001\), we would expect (close to) \(1/1000\) of
all sites whose immediate neighbors are observed to form the
configuration \(\downarrow\uparrow\downarrow\) to transition to
\(\downarrow\downarrow\downarrow\).

With our example parameters, the to the external magnetic field is
smaller than the nearest-neighbor-coupling. Here, the second-fastest
process is the ``melting'' of a magnetization-up island at an endpoint:
\((\cdots\downarrow\uparrow\uparrow\cdots)\to(\cdots\downarrow\downarrow\uparrow\cdots)\)
and also the symmetric case
\((\cdots\uparrow\uparrow\downarrow\cdots)\to(\cdots\uparrow\downarrow\downarrow\cdots)\).
If a site is in this configuration, the associated rate for this process
is \(R_1=R_0\cdot\exp(-4\beta J)\approx0.0183R_0\) (since we are no
longer gaining \(4J\) energy-units from removing the interface at both
sides of an island). The third-fastest process then is the expansion of
a magnetization-up island via the reverse of the previous process, at a
rate of \(R_2=R_1\exp(+2\beta h)\approx0.0111R_0\). Due to the rarity of
islands, we can ignore processes associated with
\((\cdots\uparrow\downarrow\uparrow\cdots)\to(\cdots\uparrow\uparrow\uparrow\cdots)\)
island-fusion (rate \(R_0\exp(2\beta h)\approx0.6065R_0\)) and its
reverse, island-splitting (rate \(R_0\exp(-8\beta J)\approx0.0003R_0\)).
Conversely, since there are many more magnetization-down sites in the
initial configuration than magnetization-up sites, we cannot neglect
spontaneous creation of size-1 islands despite their low formation-rate
\(R_3=R_0\exp(-8\beta J+2\beta h)\approx0.0002R_0\) due to the
corresponding configurations being more than two orders of magnitude
more frequent than islands in our example.

As our main objective is to compare Markov process parameter dynamics
not only to Monte Carlo simulations but also to analytic results based
on simplifications that make the dynamics more tractable, we can afford
to keep our model simple at the expense of making it unphysical: We want
to regard only the single-site-flip processes described above that are
associated with rates \(R_0, R_1, R_2, R_3\) as relevant for the
dynamics. For an actual physical system, the situation would be more
complicated -- and adding that aspect to our model would be doable, but
require some extra ingenuity. A major problem is that, for external
field \(h=0\), the energy change from flipping a long magnetization-up
island all at once (as well as the reverse process) does not depend on
the length of the island, and thermodynamics would ask us to also take
such collective processes into account. For \(h<0\), we would even find
that that the energetic advantage for flipping an entire
magnetization-up island in one step grows linearly with the length of
the island, and hence expect long islands to disappear rapidly.

For a nanoscale magnetic realization as sketched above, the discrepancy
between the model and the physical system would arise from magnetic
elements being made of many individual magnetic moments, which we only
aggregate into per-element magnetizations without taking into account
that the magnetic substructure allows for collective ``bending'' of
magnetization that is not visible in the ``one magnetization per site''
approximation. As hard disk engineers know, the energy barrier for
flipping a magnetic particle that is made of very many individual atomic
magnetic moments depends on the lowest-energy continuous deformation of
the position-dependent mangetization in the particle that manages to
reach the saddle point in the energy landscape separating the two
minima.

Correspondingly, for a quantum system, we would have to take entangled
superpositions in the tensor product Hilbert space made of the
individual sites' Hilbert spaces into account -- which computationally
would usually be an even bigger challenge. For the classical magnetic
system at hand, this deviation is resolvable with some work that would
lengthen our explanations, but one may wonder whether these
considerations hold a deeper caveat: for any modeling approach (such
as in~\cite{alakuijala2024computational}) where we take change to come
from localized discrete changes, there is a need to clarify to what
extent processes that the model cannot capture (such as by involving
low-energy pathways that involve quantum superpositions stretching
multiple individual sites) indeed can be reasoned to only play a
negligible role. It may well be conceivable that in particular
abiogenesis were to involve quantum chemistry in a sufficiently subtle
way to render all approaches to capture the dynamics in terms of
spatially isolated step-by-step adjustments as seriously off. But even
if so -- it certainly would be interesting to have a better
understanding of the power and especially the limitations of the
corresponding class of simplistic models.

\hypertarget{approximate-analytic-approach-aa}{%
\paragraph{Approximate Analytic Approach
(AA)}\label{approximate-analytic-approach-aa}}

For the parameters described above, we can make this problem tractable
via simple analytic means.

For low initial size-2-island concentration, we can consider all the
size-\(k\) magnetization-up islands to generally be sufficiently far
apart to ignore contributions to the dynamics that come from effects
where they are close to one another: We can treat the probabilities
\(p_k\) of a random site to be at the start of a size-\(k\) island as
the fundamental degrees of freedom of a simple (affine-)linear and
time invariant model. As we ignore island-splitting processes as
sufficiently suppressed to be negligible, any site-flip will either
enlarge, or shrink the size of such an island by one site, or create a
new length-1 island. Formally, we would be dealing with an infinite
linear system of transition rates between size-\(k\) island densities,
where size-\(1\) islands melt away and get created from all-down
configurations at given transition rates, and all other transitions
increase or decrease island-size by 1. This gives us a band-diagonal
transition-rate matrix plus an inhomogeneous contribution from
spontaneous creation.

Clearly, if the initial probability for a random site to be a length-2
chain start is \(p_2(t_0)\), the typical distance between such islands
is \(D\sim 1/p_2(t_0)\). As such, these probabilities only make sense
for \(k\ll D\). We naturally expect for islands that undergo a random
walk in their size where shrinking is more probable than growing to take
on a size-distribution that makes islands increasingly rare the larger
they are at an exponential rate: thanks to the external field, the
energy of an island has a contribution that is proportional to its size,
and as such, in thermodynamic equilibrium, the size distribution would
follow a barometric law, \(p_L\propto\exp(-\lambda L)\). Here, we evolve
a non-equilibrium distribution, but we still would naturally expect
\(p_L(t)\) to decay about-exponentially with length, at least beyond the
first few \(L\). As such, truncating the ODE to only a finite set of
island-lengths seems justifiable, and one can indeed confirm that moving
the cutoff \(L\) will not noticeably affect the numerical values. In
principle, one could try to diagonalize the infinitely large transition
matrix via a Fourier transform, but to keep things simple, we will
instead simply prune at maximal length \(L=50\).

Our ODE for the probabilities-vector
\(\vec p(t):=(p_1(t), p_2(t), \ldots)\) hence looks as follows:

\begin{equation}(d/dt)\,\vec p(t)=R_0\left(\begin{array}{cccccc}
-1&A&0&0&0&\\
B&-A-B&A&0&0&\\
0&B&-A-B&A&0&\\
0&0&B&-A-B&A&\ldots\\
0&0&0&B&-A-B&\\
&&\ldots&&&\\
\end{array}\right)\vec p+\vec b\end{equation} where \(A=2(R_1/R_0), B=2(R_2/R_0)\),
and \(\vec b\) corresponds to the spontaneous size-1 island creation
rate, \(\vec b=(R_0\exp(-8\beta J-2\beta h), 0, 0, 0)\). Here, we use
that the total fraction of sites occupied by islands is small --
so we take this background creation rate to be constant.

\hypertarget{monte-carlo-simulation-mc}{%
\paragraph{Monte Carlo Simulation
(MC)}\label{monte-carlo-simulation-mc}}

A basic Monte Carlo based approach to studying system dynamics focuses
on a fraction \(\phi\ll 1\) of sites (selected with replacement) at
every time step, determines the energy-adjustment associated with
flipping this site, and from that works out the rate-suppression
factor relative to the fastest process modeled, dissolution of a
length-1 island. Using rejection sampling, transitions are then
performed if a random number is below the threshold for accepting the
transition. A good introduction to Monte Carlo approaches is available
in~\cite{krauth2007introduction}.

As always with sampling based approaches, halving the size of error bars
will require four times as much data-collection effort: such approaches
are good to get reasonable first ballpark estimates and validations, but
painful to use for obtaining highly accurate estimates.

\begin{figure}
\centering
\includegraphics[width=\textwidth,height=8cm]{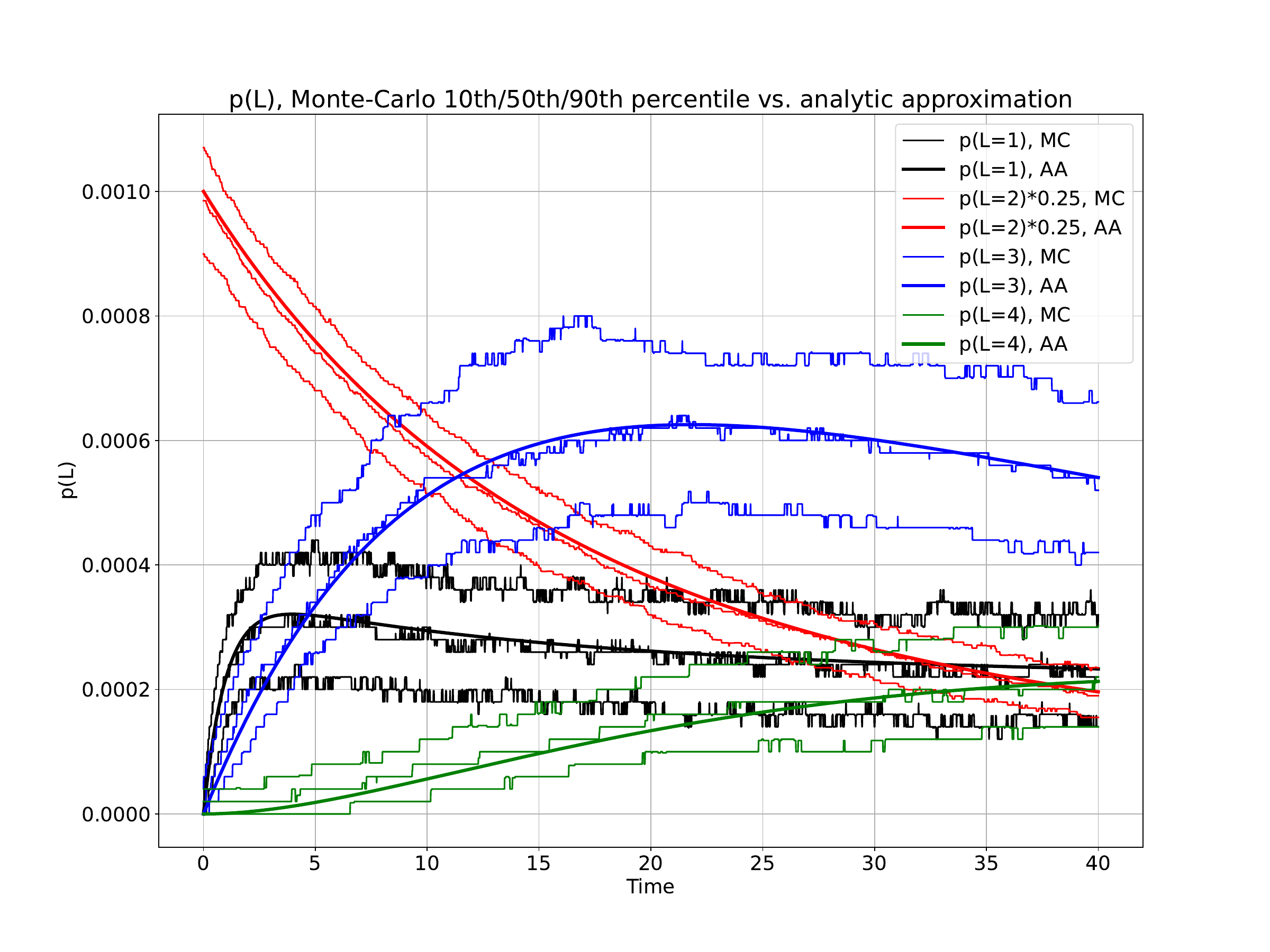}
\caption{Analytic Approximation vs.~Monte Carlo Simulation}
\label{fig:aamc}
\end{figure}

As figure~\ref{fig:aamc} shows, statistics gathered on 100
length-$50\,000$ chains (actually loops) are in good agreement with
the approximative analytic model, but illustrate how a modest number
of samples leads to a large spread between 10th and 90th percentile
counts. For better diagram space utilization, the probability for a
random site to be the start of a length-2 island has been scaled down
to \(1/4\). The apparent systematic deviation between the length-2
50th percentile curve and the analytic approximation can be attributed
to this specific approach having a fraction of (to leading order)
\(4p(L=2)=0.016\) length-2 chains being generated at not-isolated
positions.

\hypertarget{markov-process-parameter-dynamics-mpd}{%
\paragraph{Markov Process Parameter Dynamics
(MPD)}\label{markov-process-parameter-dynamics-mpd}}

The framework Scheme code implementing the Markov process dynamics for
this problem can be found in appendix~\ref{c.2-classical-ferromagnetic-spin-chain}.

Diligence is required in setting up the initial state. If we imagine
rare length-2 magnetization-up islands that are generally spaced
well-apart and (due to their rarity) have negligible probability to show
up as close neighbors, and we parameterize the system in terms of
length-4 subsequence probabilities, we would expect the probability
distribution to be the same as is observed by randomly probing a
sufficiently long loop with equidistant islands. For \(p(L=2)=1/250\),
we would conclude to take
\(p_0(\downarrow\uparrow\uparrow\downarrow)=p(L=2)=1/250=p_0(\downarrow\downarrow\uparrow\uparrow)=p_0(\downarrow\downarrow\downarrow\uparrow)=p_0(\uparrow\uparrow\downarrow\downarrow)=p_0(\uparrow\downarrow\downarrow\downarrow)\),
\(p_0(\downarrow\downarrow\downarrow\downarrow)=1-5p_0\),
\(p_0(\text{anything else})=0\).

Clearly, this specific probability-distribution would leave some Markov
process probabilities undetermined: Since we nowhere encounter a
\(\uparrow\downarrow\uparrow\)-prefix in the initial state, the relative
probabilities for such a prefix to be followed by \(\uparrow\),
respectively \(\downarrow\) encounter a numerical \(0/0\) division. We
get well-defined probabilities by splitting the total probability-weight
1 evenly across all possible next symbols for each such impossible
prefix. Subtly, with \(p(L=2)=1/250\), the probability for a
\(\downarrow\downarrow\downarrow\) prefix to be followed by a
\(\uparrow\) state is \(1/246\), since on a loop with 250 sites and one
length-2 island, four out of the 250 length-3 prefixes contain at least
one \(\uparrow\)-state, and of the remaining 246 such prefixes, only one
is followed by an \(\uparrow\)-state. All other next-symbol
probabilities are either complementary to this one, or 1, or \(1/2\) due
to a prefix that has initial probability zero. The corresponding (as
usual, non-symmetric) \(2^3\times 2^3\) length-3 sequence transfer
matrix that is obtained with the Python code below does indeed have a
single eigenvalue 1, which (after rescaling) is identical to our initial
probability-distribution.

{\small
\begin{Shaded}
\begin{Highlighting}[]
\KeywordTok{def}\NormalTok{ ctm\_from\_mpp(num\_alphabet, num\_context, mpp):}
  \CommentTok{"""Computes a Context Transfer Matrix from Markov Process Parameters.}

\CommentTok{  Args:}
\CommentTok{    num\_alphabet: Number of symbols in the alphabet.}
\CommentTok{    num\_context: length of the context{-}prefix subsequence from which}
\CommentTok{      the next symbol\textquotesingle{}s probability is determined.}
\CommentTok{    mpp: [num\_alphabet]*(num\_context+1) numpy.ndarray such that}
\CommentTok{      \textasciigrave{}mpp[*prefix, i]\textasciigrave{} is the probability for the symbol{-}index}
\CommentTok{      sequence \textasciigrave{}prefix\textasciigrave{} to be followed by the symbol with index \textasciigrave{}i\textasciigrave{}.}

\CommentTok{  Returns:}
\CommentTok{    [num\_alphabet**num\_context, num\_alphabet**num\_context]{-}ndarray}
\CommentTok{    with prefix{-}index{-}sequence transition probabilities.}
\CommentTok{  """}
\NormalTok{  result }\OperatorTok{=}\NormalTok{ numpy.zeros([num\_alphabet }\OperatorTok{**}\NormalTok{ num\_context] }\OperatorTok{*} \DecValTok{2}\NormalTok{)}
\NormalTok{  result\_stepwise }\OperatorTok{=}\NormalTok{ result.reshape([num\_alphabet] }\OperatorTok{*}\NormalTok{ (}\DecValTok{2} \OperatorTok{*}\NormalTok{ num\_context))}
\NormalTok{  mp\_stepwise }\OperatorTok{=}\NormalTok{ mpp.reshape([num\_alphabet] }\OperatorTok{*}\NormalTok{ (}\DecValTok{1} \OperatorTok{+}\NormalTok{ num\_context))}
  \CommentTok{\# There may be more elegant ways to express this multiindex operation,}
  \CommentTok{\# but this is likely clearest:}
  \ControlFlowTok{for}\NormalTok{ indices }\KeywordTok{in}\NormalTok{ itertools.product(}\BuiltInTok{range}\NormalTok{(num\_alphabet),}
\NormalTok{                                   repeat}\OperatorTok{=}\NormalTok{num\_context }\OperatorTok{+} \DecValTok{1}\NormalTok{):}
\NormalTok{    prob }\OperatorTok{=}\NormalTok{ mp\_stepwise[indices]}
\NormalTok{    result\_stepwise[indices[}\DecValTok{1}\NormalTok{:] }\OperatorTok{+}\NormalTok{ indices[:}\OperatorTok{{-}}\DecValTok{1}\NormalTok{]] }\OperatorTok{+=}\NormalTok{ prob}
  \ControlFlowTok{return}\NormalTok{ result}
\end{Highlighting}
\end{Shaded}
}

\begin{figure}
\centering
\includegraphics[width=\textwidth,height=8cm]{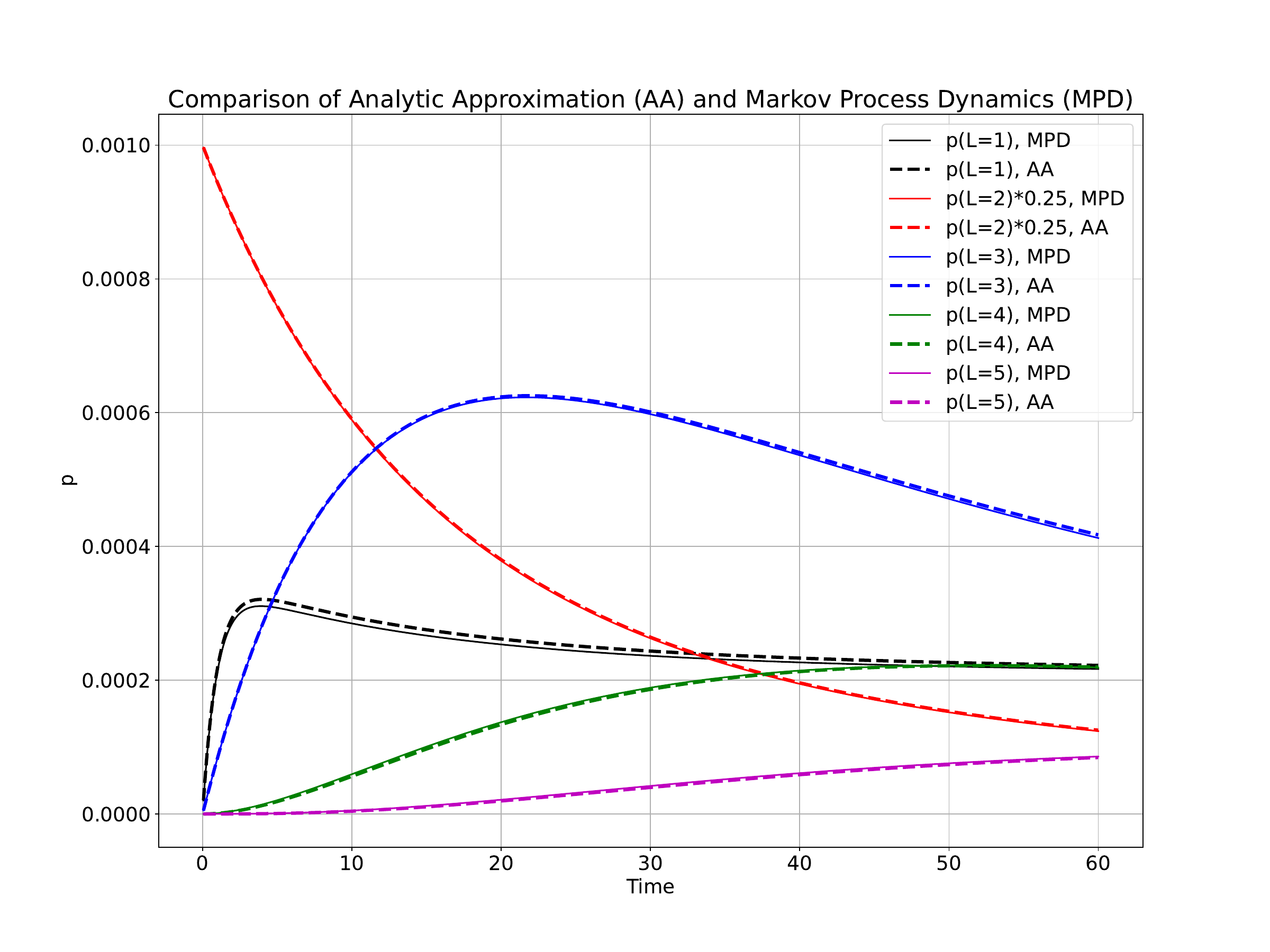}
\caption{Comparison between approximate analytic calculation and Markov
  process dynamics (for subsequence length 7).}
\label{fig:aampd}
\end{figure}

As is shown in figure~\ref{fig:aampd}, the evolution of the toy model
as computed via Markov process dynamics agrees well with the analytic
approximation. The analytic approximation predicts a slightly higher
length-1 island concentration. This which can be attributed to the
approximation over-estimating the spontaneous creation rate by
ignoring that the presence of other islands will reduce the fraction
of sites at which a length-1 island can get created. As such, one
would expect a corresponding refinement of the approximative analytic
model to come even closer to the the Markov process dynamics result.

\begin{figure}
\centering
\includegraphics[width=\textwidth,height=8cm]{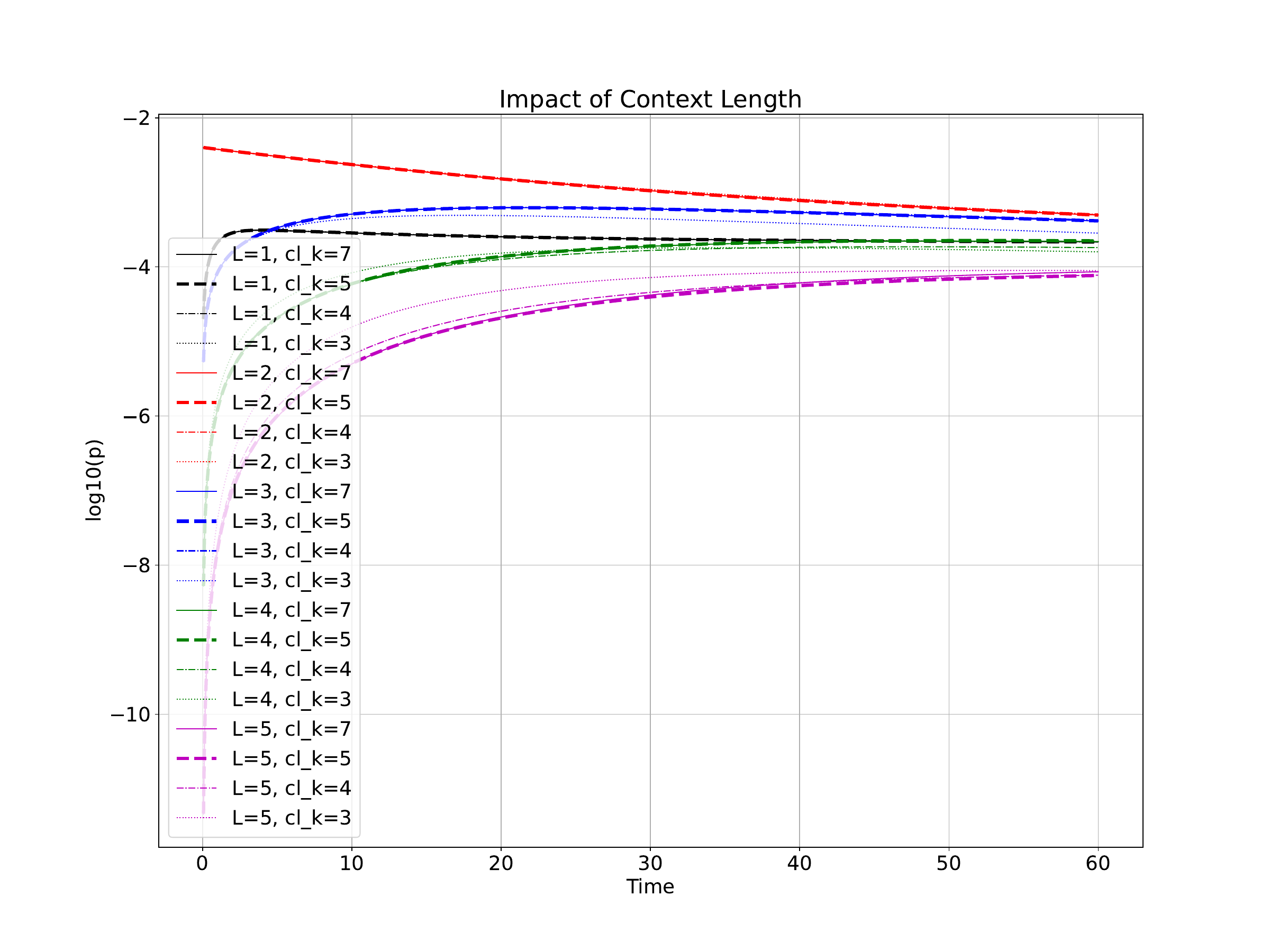}
\caption{Dynamics for different subsequence-lengths}
\label{fig:seqlen}
\end{figure}

The analytic approximation uses a cutoff for island size that is being
tracked. The Markov process approximation also uses a cutoff, but
implemented in a very different way: If we keep track of length-\(k\)
subsequence probabilities, then the probability for a random site to
be the start of a length-\(m\ge k\)-island is related to the
probability for such a site to be the start of a length-\(m+n\) island
by \(p(L=m+n)=p(L=m)\cdot \alpha^n\), where \(\alpha\) is the
probability for a length-\((k-1)\) \(\uparrow\)-sequence to be
followed by another \(\uparrow\)-site. Since we expect island-lengths
to follow an exponential size-distribution in thermodynamic
equilibrium, the static situation can for this particular example be
modeled precisely even with finite subsequence length. If we are
interested in dynamics, we expect to see deviations. Including longer
sequences and plotting log-probabilities to handle the very low
long-island occurrence probabilities, we observe in
figure~\ref{fig:seqlen} that for the specific parameters chosen for
this example, predicted sequence-probabilities converge quickly as the
subsequence length increases. One in particular notes that for
subsequence length \(\ge 4\), the Markov process approximation yields
good quantitative predictions for the concentration of islands that
are too large to fit into that window -- for example, the purple
dash-dotted line describes the evolution of the concentration of
size-5 islands (i.e.~involving seven sites,
\(\downarrow\uparrow\uparrow\uparrow\uparrow\uparrow\downarrow\))
using only length-4 subsequence probabilities reasonably well.

\hypertarget{example-3-co-polymerization}{%
  \subsection{Example: ``Co-Polymerization''}\label{example-3-co-polymerization}}

The previous examples only considered processes involving a single
stretch of tape. The purpose of this example is to illustrate how to
quantitatively study chemical transformations that occur due to some
stretch of tape encountering a different stretch of tape -- a key design
goal of the framework. It is reasonable to think that, at least in the
rarefied (i.e.~low-concentration) limit, such two-piece processes
dominate the dynamics and multi-strand processes can be neglected.

We again first consider a very simple toy system, and then explore
slightly more complicated variations. The general setting is
co-polymerization of two types of monomers. A chemical model that
provides inspiration for the design of this example is polycondensation
of a dicarboxylic acid (such as sebacic acid) with a diamine (such as
hexamethylene diamine) to a synthetic polyamide, i.e.~some nylon, in
which monomers form an \texttt{...ABABAB...} chain.

Given that the modeling framework describes processes on ``tapes'', a
bit of inventiveness is needed to use this approach for modeling
dissociated monomers. While this may be addressed differently in
future refinements of the framework, our strategy for modeling this
with the present implementation that only knows about interacting
tape-segments shall be as follows: At every point in time, we will
have some leftover monomers, some short chains, and some long
chains. Taking the effective monomer-density (units per volume) found
in any chain, we imagine space to be filled with a mix of actual
monomers and ``phantom monomers'' which merely take up space the way
monomers would, but only for bookkeeping, in the sense that the space
attributed to ``phantom monomers'' is actually filled by solvent
(e.g.~ethanol) only. We also consider all true and phantom monomers in
the system to be enumerated, the only constraint on the enumeration
being that for any polymer chain in the system, starting at the end
with the lower index, if we proceed along the chain, the index
increases by 1 at every step. Overall, our alphabet shall consist of
four types of monomers, a ``phantom monomer'' \texttt{O}, an ``di-acid
monomer'' \texttt{A}, and a ``di-amine monomer'' for which we take the
mixture to provide two different variants, \texttt{N} and
\texttt{M}. For the first instance of this example problem, these are
taken to behave equivalently. We want to use a context window of
length 4, so in total need to describe the evolution of \(4^4=256\)
Markov process parameters. If we take the initial concentration of
\texttt{A}-monomers to equal the total concentration of \texttt{M}
plus \texttt{N}-monomers, which we take to be present in 1:1 ratio,
and have the \texttt{A}-monomers take up 1/50 of the available space,
a reasonable approximation of the initial state in terms of a Markov
process model is to have the sequence \texttt{OAOO} have probability
\(1/50\), just like the sequences \texttt{AOOO}, \texttt{OOAO},
\texttt{OOOA}. Correspondingly, the sequences \texttt{NOOO},
\texttt{ONOO}, \texttt{OONO}, \texttt{OOON} and also \texttt{MOOO},
\texttt{OMOO}, \texttt{OOMO}, \texttt{OOOM} shall each have initial
probability \(1/100\). This leaves an initial probability of
\(1-16/100=84/100\) for the sequence \texttt{OOOOO}, and zero
probability for all other sequences. In principle, we would consider
to also take into account a nonzero initial probability for sequences
such as \texttt{OAOA}, at order-of-magnitude \((1/50)^2\), but both
numerically and chemically, we do not expect dynamics to be
substantially different if we make the extra assumption that at these
low concentrations, monomers initially are at statistically slightly
unusually high distance from one another. Taking the underlying
chemical system literally, we would find that monomers form
\texttt{-CO-NH-} bonds not the way they are found in natural
oligopeptide polymers, i.e.~with the sequence along the chain being
\texttt{CO-NH\textbar{}CO-NH\textbar{}CO-NH} but
\texttt{CO-CO\textbar{}NH-NH\textbar{}CO-CO}, like in a synthetic
polyamide. As such, a chain such as \texttt{AMAMO} could connect to a
chain such as \texttt{NANAO} and form a chain such as (note reversal
of the second chain) \texttt{AMAMANAN}. While we could indeed model
such chain-to-chain addition with possible reversal if we are willing
to perform an approximation that stops unfolding a Markov chain at
some fixed maximal length, we will here for the sake of keeping our
second example still relatively simple assume that dissociated
monomers come in an ``activated'' form, that only ``activated''
monomers can connect to chains or other monomers, and monomers at the
end of a chain never are considered ``activated'', effectively
eliminating chain-to-chain addition reactions. As with the
``ferromagnetic chain'' example, this deviation between reality and
experiment can be fixed with some extra work.

In computational settings, one would in general call one of the two
interacting tape-fragments as the ``program tape'' and the other one as
the ``data tape'', but in settings like this that are somewhat remote
from computation, it appears to make sense to talk in a slightly more
abstract fashion of the ``P-tape'' and ``D-tape''.

Our ``program execution'' rule shall then codify the following idea:
if the P-tape has an isolated monomer at index zero (i.e.~we have
\texttt{O} at indices \texttt{-1} and \texttt{+1}, but not at index
\texttt{0}), and the D-tape has a complementary monomer at index zero
with at least one \texttt{O}-neighbor, then the monomer unit gets
removed from the \texttt{P}-tape and added to the \texttt{D}-tape,
where each pathway that fills one \texttt{O}-neighbor has the same
probability. One way to implement this system is shown in
appendix~\ref{c.3-co-polymerization}, which also shows code for the
variants discussed in this section.

Given that initial monomer concentrations are of the order of
\(\sim10^{-2}\), and roughly every monomer-monomer encounter is
effective, the probability for a monomer to hit another monomer and
dimerize over the first time step is \(\sim10^{-2}\), and the time scale
over which much of the dynamics plays itself out is \(\sim10^{2}\). We
show the evolving base-10 logarithm of the probabilities of encountering
various subsequences of interest when probing the solution at random
tape-position (which includes ``phantom monomers''). Specifically,
\(p_{ANAN}\) is the probability that probing the solution at a random
spot (which may be in the location of a ``phantom monomer'') finds an
\texttt{A}-unit that is followed by a \texttt{N} unit followed by
another \texttt{A} and then a \texttt{N}-unit. In this setting, we
observe that monomers get used up at an asymptotically approximately
exponential rate -- showing as asymptotically straight lines in a
log-plot. With no chemical difference in the behavior of \texttt{M} and
\texttt{N} monomers, the probability to encounter, when starting at a
random point, the sequence \texttt{ANAM} is always the same as the
probability to encounter \texttt{ANAN}, and so the corresponding two
plotted curves sit exactly on top of one another.

\begin{figure}
\centering
\includegraphics[width=\textwidth,height=8cm]{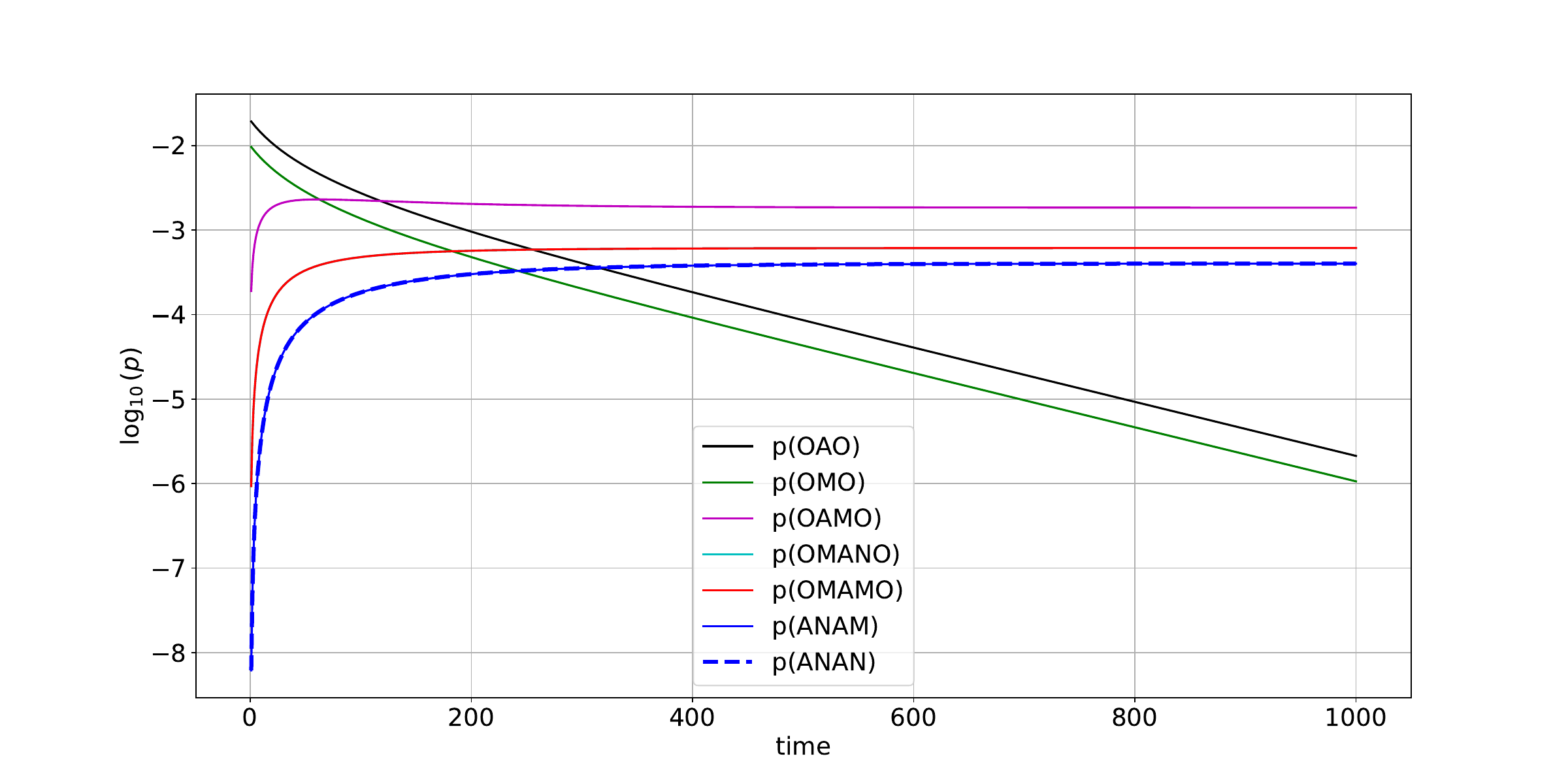}
\caption{Copolymerization: log-concentration evolution}
\end{figure}

Looking closely, one also finds that the concentration of
\texttt{AM}-dimers passes through a maximum, since at some point, the
concentration of monomers has decreased so much that dimer formation
becomes negligible, but some of these monomers still convert dimers to
trimers, hence in the long run, dimer concentration falls very slightly.
This happens at about the time when \texttt{A}-monomer concentration
equals dimer concentration, since then a given monomer is as likely to
encounter another monomer to create a dimer as it is to encounter a
dimer to form a trimer.

\begin{figure}
\centering
\includegraphics[width=\textwidth,height=8cm]{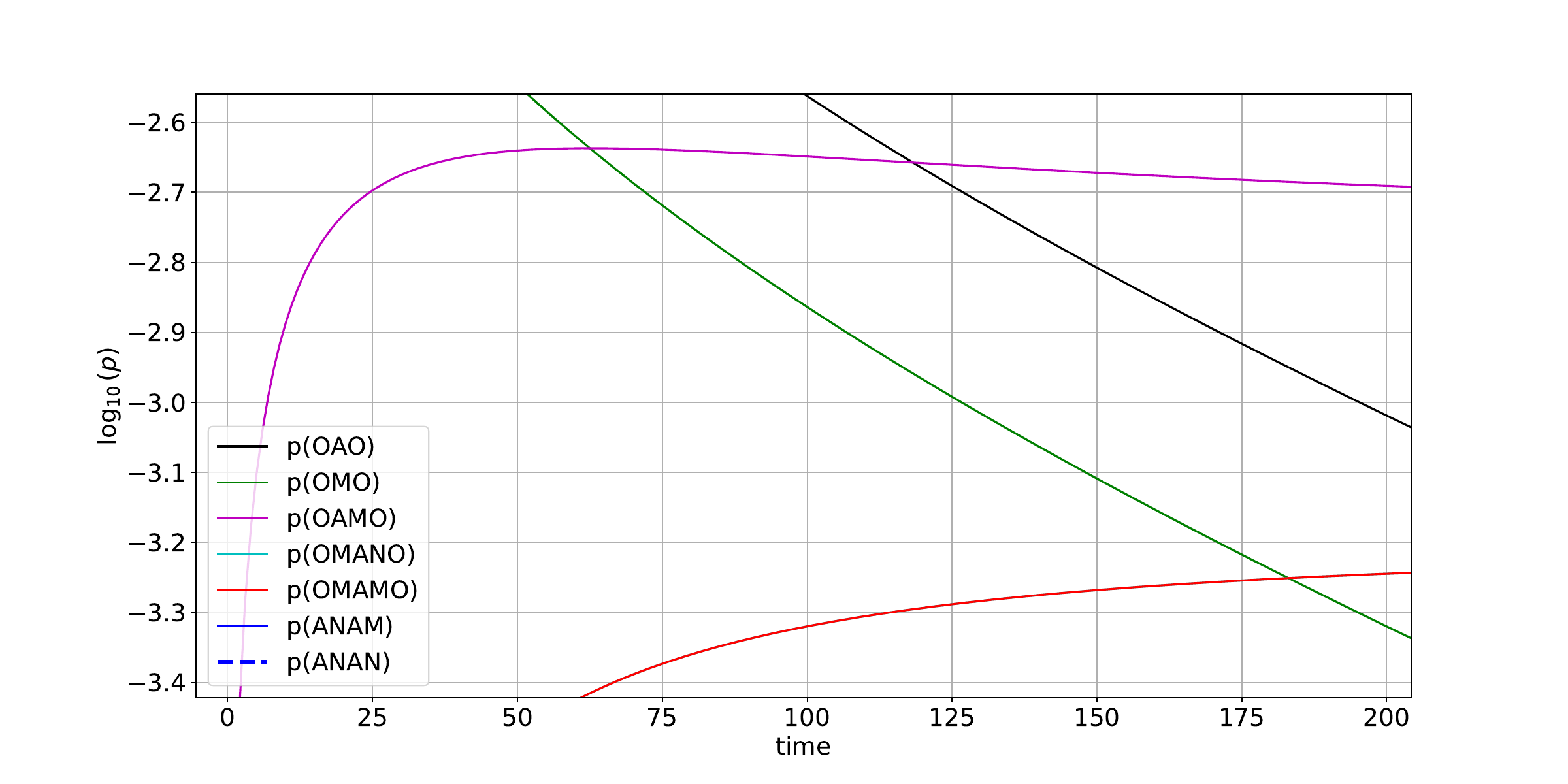}
\caption{Copolymerization: log-concentration evolution detail}
\end{figure}

\textbf{Variant 1}: Slight preference for \texttt{AMANAMAN} alternation.

It is conceivable that in such a system, minor chemical differences
between the \texttt{M}- and \texttt{N}-monomers might lead to an
\texttt{AMAOO}-chain having some preference to bind a \texttt{N}-monomer
over a \texttt{M}-monomer, and vice-versa. There are different plausible
mechanisms for such preferences, such as different shapes or space
requirements of \texttt{M}- and \texttt{N}-units, or perhaps differences
in inductive effect, i.e.~if a \texttt{M}-unit has a higher
electron-pull, such as due to presence of a somewhat electronegative
component such as a chlorine atom in the compound, a component with some
relative electron-pull would, in terms of binding strength, prefer to be
followed by a component with some relative electron-push, giving
preference to an alternating \texttt{AMANAMAN}-pattern.

\begin{figure}
\centering
\includegraphics[width=\textwidth,height=8cm]{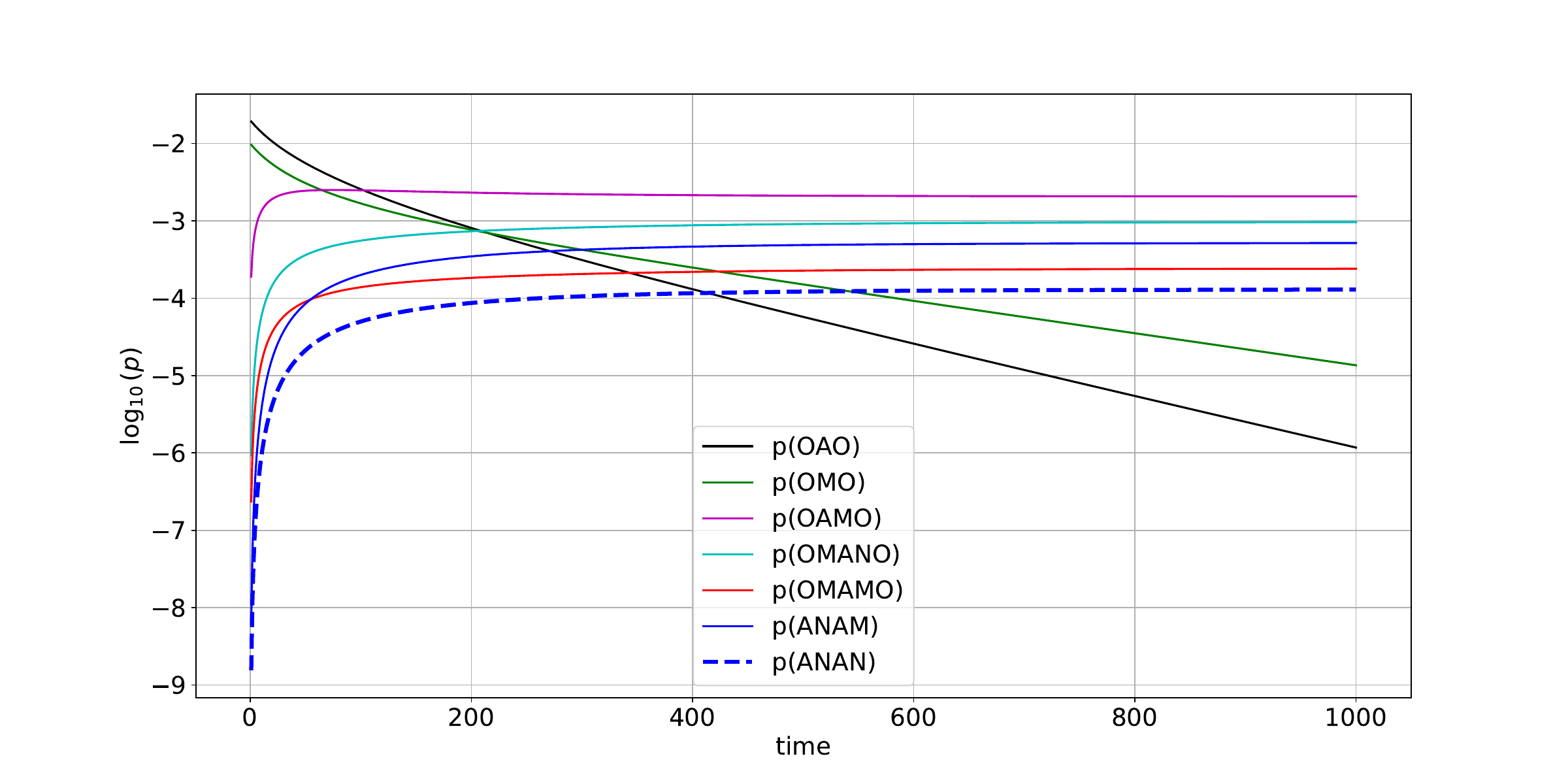}
\caption{Copolymerization: Preference for alternation}
\label{fig:aman}
\end{figure}

For this variant, figure~\ref{fig:aman} shows the dynamics as obtained by
integrating the rates-of-change with a numerical ODE integrator. One
finds that modeling this problem in terms of length-4 subsequence
probabilities gives results that are in good alignment with length-5 and
length-6, apart from reporting visibly lower trimer concentrations. We
show plots for length-6.

Qualitatively and semiquantitatively, observed behavior fully aligns
with expectations due to the definition of the dynamics. At time \(t=1\)
(one unit with respect to the graph plotting resolution used), we expect
every monomer to have encountered another monomer it can connect to with
a probability of about \(1/50\), so the concentration of dimers is
expected to rise to \(\sim(1/50)^2\) over about-unit time (where the
graph starts). Following the same reasoning, the concentration of
trimers (which are at early times expected to make up a large fraction
of e.g.~the \texttt{MANO} sequences, i.e.~the by far most likely prefix
makes this extend as \texttt{OMANO}) is expected to rise, over the same
time scale, to \(\sim10^{-5}\) and climb from there. In the longer run,
since with the given dynamical rules, dimers cannot connect to dimers,
polymerization will make the concentration of still available monomers
fall so much that some residual dimers fail to get any opportunity to
grow to trimers -- the long-term end state will not only have long
chains, but also have some fraction of dimers, and then also (more
rarely) trimers etc., in the final state. Since with the given rules, a
polymerization reaction that connects an \texttt{A}-unit is always
effective whereas we modeled reactions that connect a \texttt{M}-unit or
\texttt{N}-unit to sometimes fail, given on the distance-two neighbor,
\texttt{A}-units get consumed at initially the same rate as \texttt{M}-
and \texttt{N}-units combined, i.e.~the monomer-probability curves keep
constant distance of \(\log_{10}2\). Once there is an appreciable amount
of dimers in the solution, this changes, since \texttt{AMA} trimers form
whenever an \texttt{A} connects to the left side of a \texttt{MA} dimer,
but a \texttt{M} dimer has some chance of being rejected by an
\texttt{AM} dimer. In the long run, we expect to end up with a
probability-ratio of \texttt{AMAN}-to-\texttt{AMAM}-subsequences that
reflects the preference for alternation built into the rules.

\textbf{Variant 2} : Reversible reactions.

At the microscopic level, chemical reactions are based on reversible
microscopic dynamics, and so should be seen as all being (at least in
principle) reversible. In some situations, the circumstances necessary
for a given reaction (such as a chlorine radical reacting with a
hydrogen molecule by forming a \(\text{HCl}\) Molecule and a hydrogen
radical in a highly exothermic reaction) to happen in reverse would
require random thermal fluctuations to concentrate so much energy in
one place that the reverse reaction-rate becomes exceedingly small,
but in general, in thermodynamic equilibrium, the ratio of the
per-encounter forward- and backward-reaction rates will match the
ratio of product-concentrations to reactants-concentrations, making
the net product creation rate zero in equilibrium. If we want to
establish contact between data-processing and molecular chemistry, we
hence will need to be able to model reversibility.

One finds that, using a reaction-constant ratio of \(50:1\) as in the
code shown in the appendix, the system very rapidly reaches equilibrium.
Monomers no longer get used up at an asymptotically exponential rate,
and correspondingly, the concentrations of sequences longer than dimers
is reduced. This is in alignment with the entropic benefit of a monomer
to remain dissociated in this low(ish)-concentration example.

\begin{figure}
\centering
\includegraphics[width=\textwidth,height=8cm]{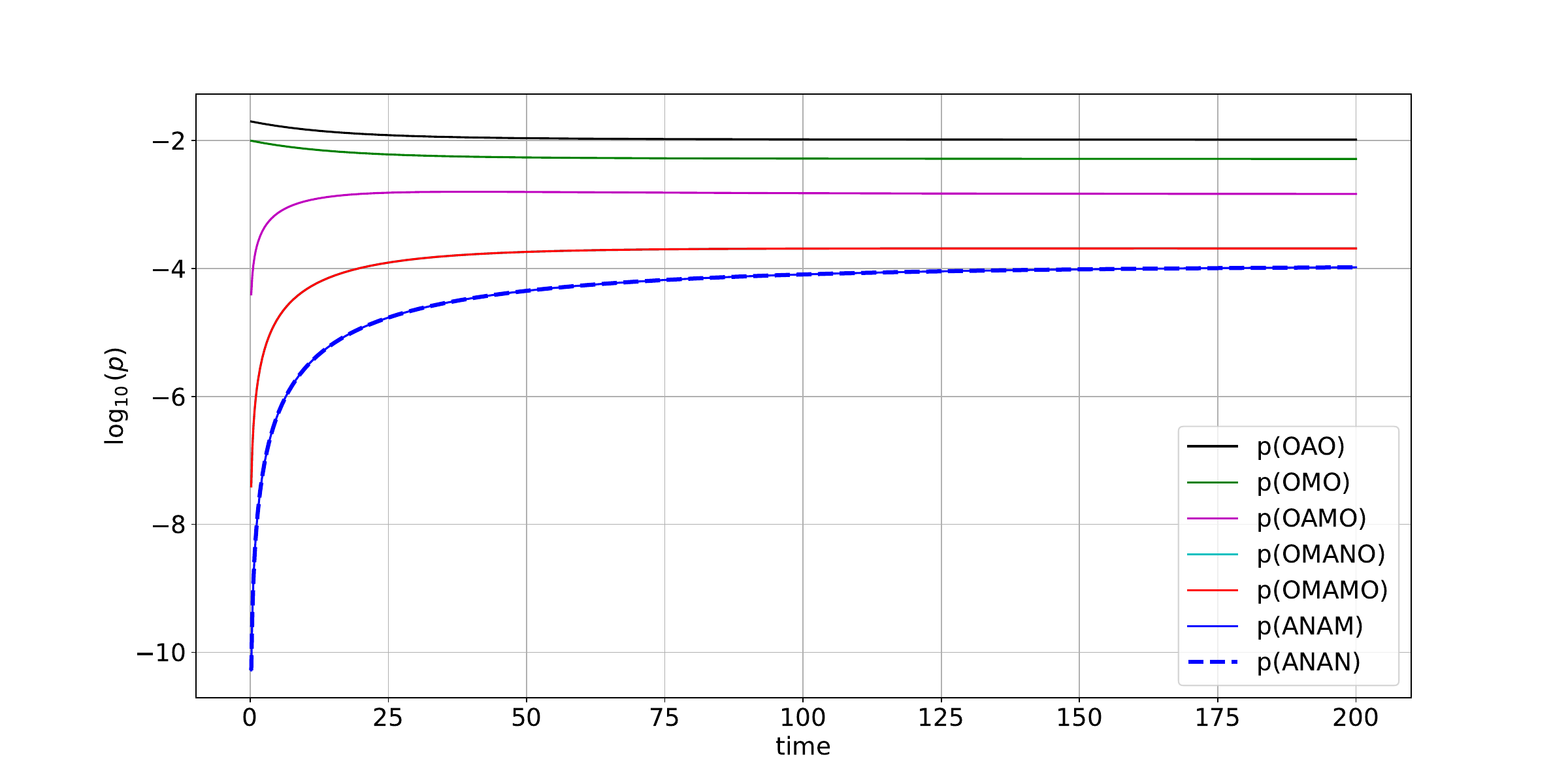}
\caption{Copolymerization: Reversible reactions}
\end{figure}

While for this example, we picked reaction rates in a way that is
cognizant of thermodynamics, we do have the freedom to choose them
arbitrarily. So, if for example our ``phantom monomer'' approach for
modeling solution as a tape were to turn out statistically questionable,
the problem could be absorbed into a redefinition of reaction rejection
rates. As the example in the next section will show, care must be taken
when trying to read off reaction-constant ratios from code and mapping
these to differences in Gibbs free energy of formation.

While we here resorted to simplifications that only consider monomer
addition/dissociation mostly for pedagogical reasons (i.e.~to keep the
code simple), extending the code to also allow oligomer
addition/dissociation would be mostly straightforward. One here has to
pick a cutoff length for sequences that can be added/dissociated, since
termination of the rate-function needs to be guaranteed,. The
expectation is that predictions converge reasonably quickly to the
infinite-length limit as one increases that cutoff.

\hypertarget{example-4-chemically-reversible-turing-machine}{%
\subsection{Example: ``Chemically Reversible Turing Machines''}\label{example-4-chemically-reversible-turing-machine}}

Having seen how the framework handles interactions between tapes, also
in situations where we have to model monomer concentrations via
fictitious effective tape-sequences, we can proceed from models that
somewhat closely represent well-understood chemical processes to models
that have a stronger focus on implementing computational procedures in a
thermodynamically plausible setting. Our focus will be on providing a
somewhat nontrivial but otherwise as-simple-as-possible example.

The underlying computational model is a slight variant of a Turing
machine -- the main difference being that staying in alignment with
thermodynamics requires us to retain notion of reverse-reactions. One
important caveat about this approach is that Turing machines are an
abstraction that was introduced primarily as a reasoning tool for
proving theorems about computations by employing a minimalistic
machine model that comes without avoidable extras. Biochemical systems
shaped by evolution will in general have no such direct requirement
for conceptual minimality and instead can utilize steps and mechanisms
that are not in straightforward one-to-one correspondence with
convenient data-processing primitives. This clearly is the case for
getting biochemical machines -- in particular enzymes -- to perform
certain tasks by having their spatial structure match the needs of the
process, such as initially docking some specific molecule. Also, in
biochemical context, one useful property for a biological machine to
have is that errors in its construction (such as a mutation changing a
DNA base pair in the blueprint) often produce a likely still somewhat
useful machine.  The genetic code is known to have this
property~\cite{freeland1998genetic}. This aspect is out of scope for
idealized-as-perfect abstractions of computational procedures such as
Turing machines.

Out of multiple different options for implementing a Turing machine in
the given framework, we here want to explore an approach in which every
state-transition is modeled by a (in principle, reversible) chemical
transformation of the data-tape, and we also want to include in our
modeling the need to power data-transformation operations. As such, the
transition rules of the Turing machine (i.e.~its programming) are here
taken to be not part of the tape-state, but considered to belong to the
definition of the system under study. At the conceptual level, this is
not a restriction, as there are well-known compact constructions of
``universal'' Turing machines which -- loosely speaking -- perform
data-transformations where the program itself also is encoded on the
tape~\cite{rogers1987theory, rogozhin1996small}.

Aligning with the notion of a Turing machine, we want to model a
binary tape. Our alphabet will have to contain two symbols that
represent zero and one. As these, we want to take the letters
\texttt{O} and \texttt{I}. As usual for a Turing machine, we want to
start from a tape with all-zeroes. There are different options for how
to represent the cursor on the tape. Here, we want to use letters
\texttt{A}, \texttt{B}, \texttt{C} that are inserted on the tape right
before the symbol the cursor is on, where the type of letter
represents the Turing machine's state, and \(A\) is the initial
state. The Turing machine's transition rules then can be cast into the
form of a string rewriting system (semi-Thue
system~\cite{post1947recursive}). Modeling tape-content as a Markov
process has two implications: (a) our on-tape context window
effectively shrinks in length by 1 if we have a symbol indicating
cursor-position in the window, and (b) we are then automatically
considering a system where there are multiple active cursors on a long
tape. As with earlier examples, we have some control over initial
spacing: if we use subsequence length 5, and give all sequences that
contain either of the sequences \texttt{AA} and \texttt{AOA} initial
probability zero, this guarantees that cursors initially will be
spaced at least two fields apart. It is in general more useful to
control the initial occurrence rate of nearby cursors by making the
initial-state symbol rare: if only~\(1\) in~\(1\,000\) symbols is
\(A\), this would make finding two A's spaced less than 10 symbols
apart rare at the percent level -- evolution of the system will mostly
describe independently-acting identical Turing machines.

If we want to model the thermodynamics of such a system, we have reason
to introduce further symbols. Chemically, one might imagine \texttt{O}
and \texttt{I} as representing two different configurations of a
substituent that is connected to the tape backbone chain, and we might
have an evaluator-molecule that can effect state-changes, whose point of
attachment to the tape is represented by a cursor-symbol. Since we want
to model data-processing as coming with an entropic cost, we need some
way to represent that. The simplest possible idea would be to embed
entropy-production in the definition of the evaluator rules that might
e.g.~correspond to energetic activation from ultraviolet light. We here
want to instead model a chemical basis for powering the machinery.

As with the previous ``nylon'' example, we want to represent solvent as
``phantom monomers'', for which we will use the symbol \texttt{S}, and
in that solvent, we want to have two kinds of small molecules dissolved
that represent an ``energized'', respectively ``de-energized'' form of
an energy-carrier. These we want to represent with the symbols
\texttt{P} (``powered'') and \texttt{X} (``de-powered''). In a biological
system, these might e.g.~correspond to ATP and ADP. (In cytoplasm, we
also have an assumed-constant concentration of phosphate species, mostly
\(\text{HPO}_4^{2-}\) and \(\text{H}_2\text{PO}_4^{-}\) at pH
\(\sim 7\), that we do not model). The structure of our interaction
rules then is as follows:

When the ``data-tape'' with an Turing-machine-evaluator-molecule
attached (indicated by some cursor symbol \(A,B,C,\ldots\)) meets a
\texttt{P} from the ``program-tape'' (which we here abuse to represent
the solvent around the evaluator-molecule), a reaction occurs that
depletes the \texttt{P} to \texttt{X}-form, and performs one
data-transforming step. The idea is that this transformation is
``powered'' by the underlying chemical reaction releasing energy to the
thermal environment, hence increasing the statistical weight of
microstates that belong to the ``product'' side: If \(P\to X\) releases
heat \(-\Delta H\), thermalization will distribute that heat to the
system's environment at temperature \(T\), increasing the total entropy
of system plus environment by \(\Delta S=\Delta q/T=-\Delta H/T\).

With this as the key fundamental reaction, we also have to take the
reverse reaction into account. This corresponds to an
\texttt{X}-molecule connecting to the evaluator-molecule and rare
thermal fluctuations running the machinery in reverse, turning
\texttt{X} into \texttt{P} and making the Turing machine perform an
operation that is the reverse of the operation it executed. If the
corresponding forward-operation was ``write an \texttt{I}, move cursor
one to the right, and transition from state \texttt{B} to state
\texttt{C}'', then, in case there are no other transitions into state
\texttt{C}, there may be \emph{two} options for the reverse operation:
``Transition to state \texttt{B}, move cursor one to the left, and (a)
write an \texttt{I} or (b) write an \texttt{O}''.

Here, care has to be taken when going back and forth between reaction
constants (implemented via step-rejection probabilities) and differences
in Gibbs free energy of formation \(\exp(-\Delta G^\circ/(k_BT))\). If,
for example, we have a process where the forward-reaction is of the form
\(\mbox{\tt A}\{\mbox{\tt O or I}\}\to \mbox{\tt IB}\) and overwrites
one bit on the tape, the backward reaction superficially is
\(\mbox{IB}\to\mbox{\tt}\{\mbox{O or I}\}\), but if we want to attribute
the same \(\Delta G^\circ\) to all \texttt{A?} and \texttt{?B}, we have
to regard this as two pairs of equilibria,
\(\mbox{\tt AI}\leftrightharpoons \mbox{\tt IB}\) and
\(\mbox{\tt AO}\leftrightharpoons \mbox{\tt IB}\). At the Scheme code
level, this means that if the reverse-reaction update for
\(\mbox{\tt A?}\leftarrow\mbox{\tt IB}\) is implemented along these
lines:

{\small
\begin{Shaded}
\begin{Highlighting}[]
\NormalTok{(}\KeywordTok{let}\NormalTok{ ((sym (choose \textquotesingle{}((}\FloatTok{1.0}\NormalTok{ I) (}\FloatTok{1.0}\NormalTok{ O)))))}
\NormalTok{  (tape{-}set{-}sym! }\DecValTok{\#t} \DecValTok{0}\NormalTok{ \textquotesingle{}A)}
\NormalTok{  (tape{-}set{-}sym! }\DecValTok{\#t} \DecValTok{1}\NormalTok{ sym))}
\end{Highlighting}
\end{Shaded}
}

then the corresponding forward-reaction must be implemented as follows:

{\small
\begin{Shaded}
\begin{Highlighting}[]
\NormalTok{(}\KeywordTok{if}\NormalTok{ (choose \textquotesingle{}((}\FloatTok{1.0} \DecValTok{\#t}\NormalTok{) (}\FloatTok{1.0} \DecValTok{\#f}\NormalTok{)))}
\NormalTok{  (}\KeywordTok{begin}
\NormalTok{    (tape{-}set{-}sym! }\DecValTok{\#t} \DecValTok{0}\NormalTok{ \textquotesingle{}I)}
\NormalTok{    (tape{-}set{-}sym! }\DecValTok{\#t} \DecValTok{1}\NormalTok{ \textquotesingle{}B)))}
\end{Highlighting}
\end{Shaded}
}

Without the
\texttt{(choose\ \textquotesingle{}((1.0\ \#t)\ (1.0\ \#f)))}
rate-halving, the forward-reaction to backward-reaction rate-ratio at
the level of individually reversible chemical processes would be \(2:1\),
corresponding to the product being thermodynamically more stable
relative to the reactants by \(\Delta G=-k_B\ln 2\) (per particle).

A sketch for a chemically plausible forward process that has such a
reverse might be that \texttt{I} and \texttt{O} differ in whether the
four groups around a steric center are in R- or S-configuration, and
\texttt{P} binds to the evaluator-molecule in such a way that it enables
it to remove some anionic group (such as perhaps \(\text{OH}^-\)) from the
steric center, which gets transported away, intermediately stabilizes
the carbocation, and then induces later re-binding of an equivalent
group. As the intermediate carbocation is flat, this loses steric
information, and the reverse process may produce either form. It
definitely would be interesting to see whether a precise chemical
realization of the Turing machine described here can be given that uses
some such simple process.

As one immediate consequence of this form of reversibility, the sheer
presence of the evaluator plus energized/de-energized molecules
activating its machinery gives us a background rate of some
forward-reaction followed by a backward-reaction that erases the bit at
the cursor, causing ``irreversible aging'' of the tape-content.

In the interest of staying with a simple example that may have a hope of
actually being confirmed as fully chemically realizable, we consider the
following simple 4-state Turing machine that writes the sequence
\texttt{IOI} to the (initially empty, i.e.~\texttt{OO...OOAO...OO})
tape:

\begin{longtable}[]{@{}cccccc@{}}
\toprule\noalign{}
Tape Symbol & State & Writing & Moving & Transitioning & Halt? \\
\midrule\noalign{}
\endhead
\bottomrule\noalign{}
\endlastfoot
O & A & I & +1 & B & No \\
I & A & I & +1 & B & No \\
O & B & O & +1 & C & No \\
I & B & O & +1 & C & No \\
O & C & I & +1 & D & Yes \\
I & C & I & +1 & D & Yes \\
\end{longtable}

While this problem would appear to ask for a 9-letter alphabet
\texttt{(A,\ B,\ C,\ D,\ I,\ O,\ P,\ X,\ S)}, we observe that some
subsequences such as \texttt{SSOSS} or \texttt{OOXO} are not considered
possible. In some situations, one might want to exploit this in order to
reduce the size of the alphabet. Here, we could for example use that no
two cursors can be next to one another, and so we can use
\texttt{...AAA...} rather than \texttt{SSS} to represent solvent, and
use \texttt{...ABA...} to represent an ``energized'' molecule in the
solvent, which we otherwise would have written \texttt{...SPS...}, as
well as \texttt{...ACA...} to represent a de-energized molecule. This
way, we could use the 6-symbol alphabet
\texttt{(A,\ B,\ C,\ D,\ I,\ O)}. For this example, we do not follow
this approach for two reasons: First, using such a complicated encoding
creates edge cases that complicate the Scheme code. This easily gives
rise to incorrect implementations. Second, we here also want to
demonstrate how the framework is able to even handle problems which from
the ODE perspective have high-dimensional state-vectors.

An implementation of this problem is given in
appendix~\ref{c.4-chemical-turing-machine} -- again alongside code for
the variants discussed in this section. Here, we set the success-rate
of backwards-reactions to 5\% of the success-rate of the
forward-reactions. If we consider a model system where the different
tape-states are equivalent in terms of thermodynamic stability, and
likewise for the attached executor-molecule's states that correspond
to Turing machine states, this would, due to
\(0.05=\exp(-\Delta G_\text{1 molecule}/(k_BT)=\exp(-\mu/(k_BT))\),
correspond to de-powering a single energy-carrier providing an exergy
(i.e.~thermodynamically extractable work) of
\(-k_BT\cdot\log0.05\approx 3\,k_BT\).

To illustrate the dynamics of this system, we explore a setting where
25\% of all sites are tape-related sites and 75\% represent ``solvent''.
Given that completing a single computation requires de-powering three
energy-carrying molecules, the tape-to-solvent ratio of 1:3 means that a
1:1 concentration ratio of cursors on the tape to energy-carriers in the
solvent would provide just enough power to run every possible
computation to completion.

Setting the cursor-fraction to 1\% of all sites that are attributed to
the tape carrying an \texttt{A}-symbol (Turing machine in its starting
state), we consider two scenarios: In the first, there is a 4:1
over-abundance of energy-carriers. In the second, there is a 1:1 ratio.

In the first setting, we quickly reach an equilibrium where the most
abundant Turing machine state is the end-state, indicating completion of
the calculation. There still is some residual equilibrium concentration
of Turing machines that did not fully complete the computation, with
concentrations roughly declining by a constant factor for every
computational step that was not completed. As a numerical validation
check, the total concentration of tape-cursors in all computational
states (red dashed line) remains (effectively) constant. Having some
residual ``unfinished computation'' states is expected since for any
individual evaluator, the entropy-contribution for being in state
\(A, B, C, D\) is \(\propto-\log p_{A,B,C,D}\). Hence, populating a
previously exceedingly rare state by transitioning from a less rare
state will increase entropy.

In the first setting, we get close to an equilibrium where (as expected)
just a little bit less than the amount of powered molecules needed to
complete every computation got de-powered. With an initial 3:1
over-supply, the ratio of powered to de-powered energy-carrier at the
ODE-integration endpoint here is \(0.02258 / 0.007415\approx 3.045\).
Given that for every successful backward-reaction, there are 20
successful forward-reactions in equilibrium, we would hence expect
equilibrium concentrations of state-D Turing machines to state-C Turing
machines to be about \(60.915\). We observe
\(p(\mbox{\tt OIOID})/p(\mbox{\tt OIOCO})\approx0.002417515/3.968644\cdot10^{-5}\approx60.915\).

\begin{figure}
\centering
\includegraphics[width=\textwidth,height=8cm]{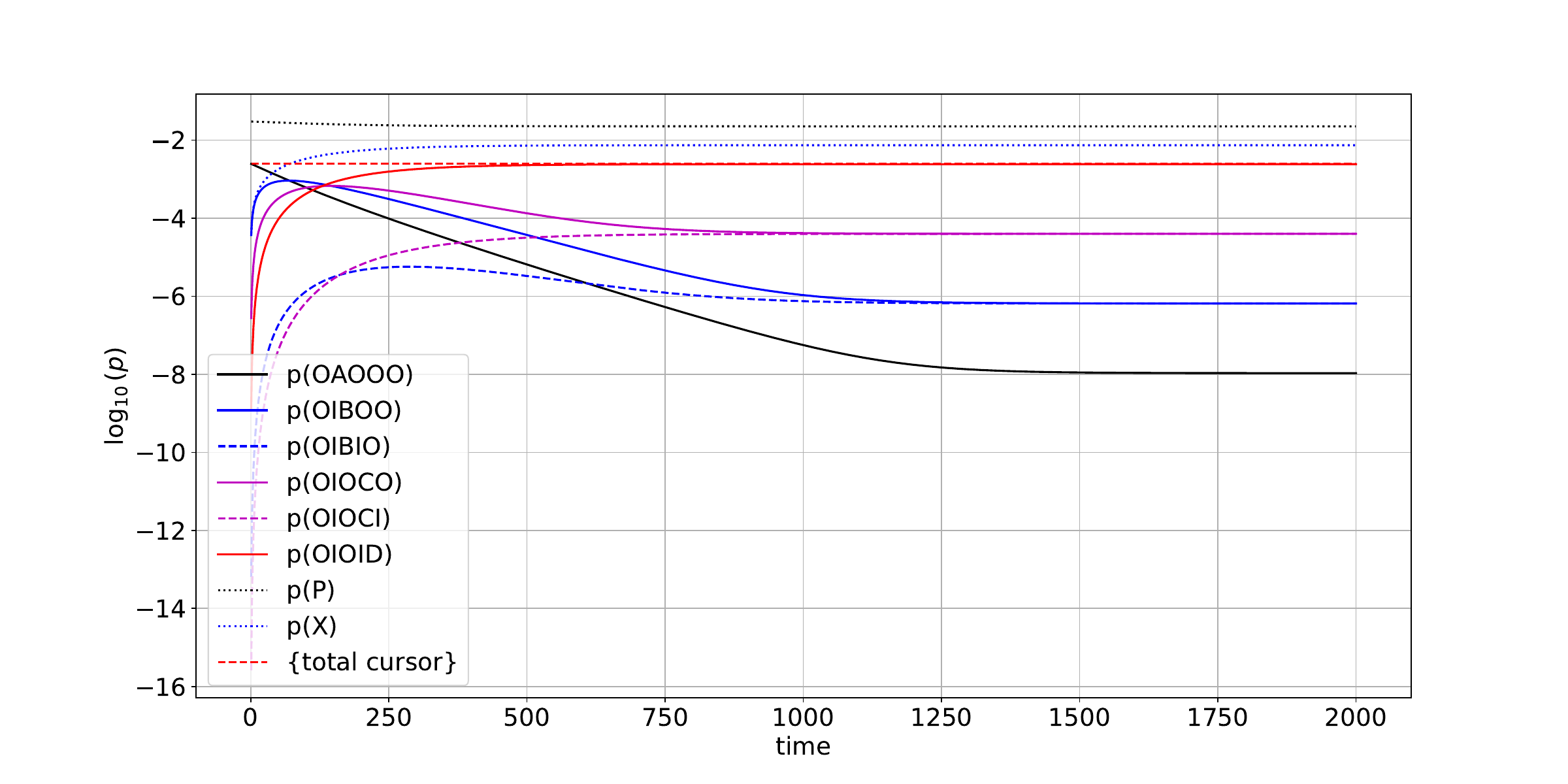}
\caption{Chemical Turing Machine Dynamics -- excess energy-carrier}
\end{figure}

If the availability of energy-carrier molecules is constrained,
relations between the relevant species follow the same principles, but
in this example with a final de-powered:powered ratio of
\(\approx2.94\). Concentration-differences between individual
evaluator-states are less pronounced.

\begin{figure}
\centering
\includegraphics[width=\textwidth,height=8cm]{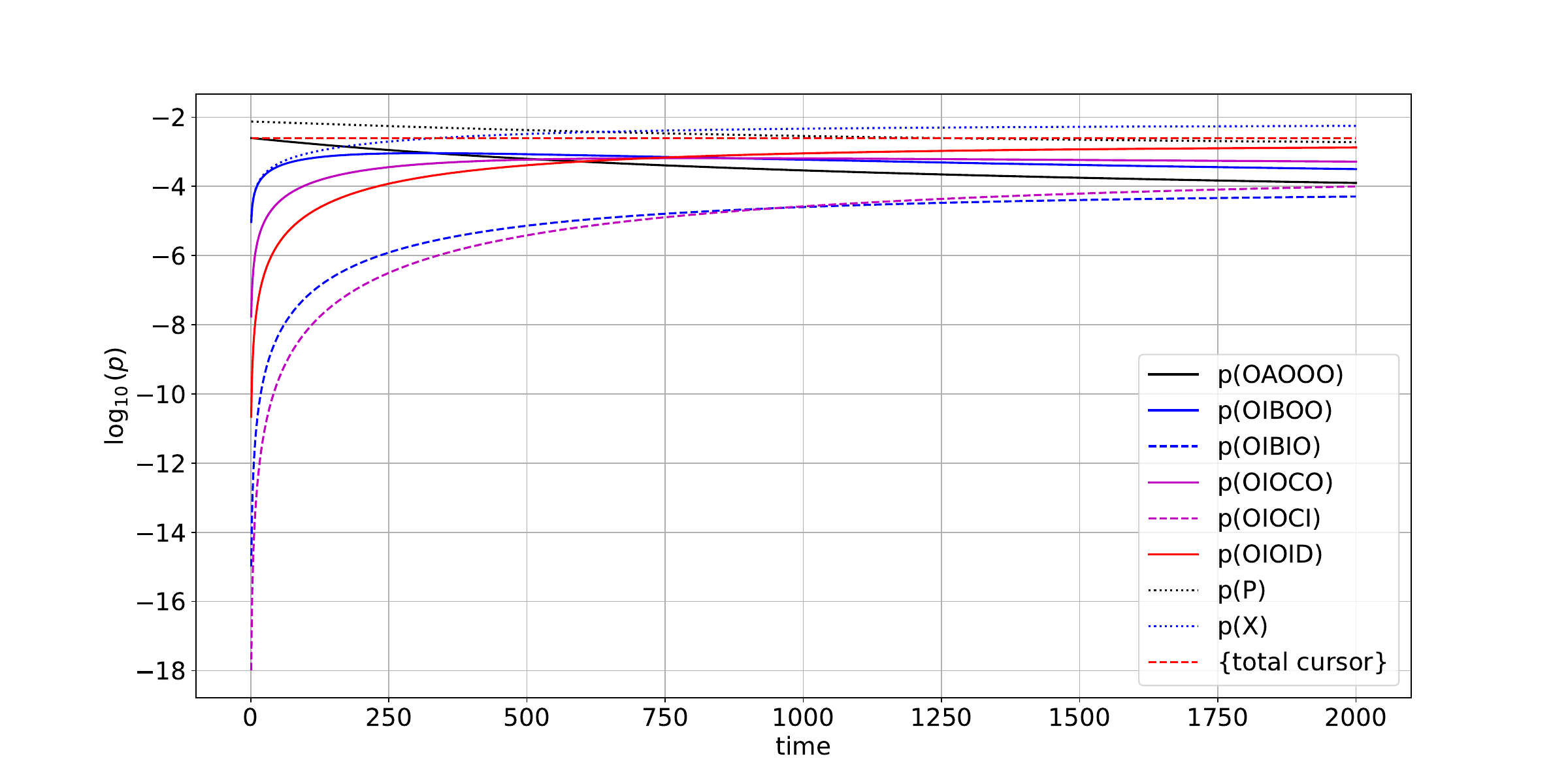}
\caption{Chemical Turing Machine Dynamics -- energy-limited}
\end{figure}

\textbf{Variant 1}: Equal thermodynamic stability

If one were to adjust this example to raise the success-rate for
reverse-reactions from 5\% to 100\%, this would correspond to lowering
the difference in Gibbs free energy of formation between the ``powered''
and ``de-powered'' form of the energy carrier molecules to zero, and
the only driving force in the system then would be the entropy-gain
equilibriating the concentrations of the ``powered'' and
``de-powered'' form -- where the former initially needs to be
higher. (To a lesser extent, the entropy gain from populating multiple
cursor-states would also drive the reaction.) A biological example for
such a (partially) entropy-powered process would be chemiosmotic ATP
synthesis~\cite{mitchell1966chemiosmotic, mitchell2011chemiosmotic}.
Here, the ``concentration chain'' difference in proton concentrations
across a cell membrane provides Gibbs free energy
\(\Delta G=\Delta H-T\Delta S\).

For our example Turing machine, initial tape-state is all-zeros but also
irrelevant for the concrete program under consideration. As we want to
explore the entropic perspective, we want to start with a tape where
each cell has equal probability to be in \texttt{O}- or
\texttt{I}-state, with no correlations.

\begin{figure}
\centering
\includegraphics[width=\textwidth,height=8cm]{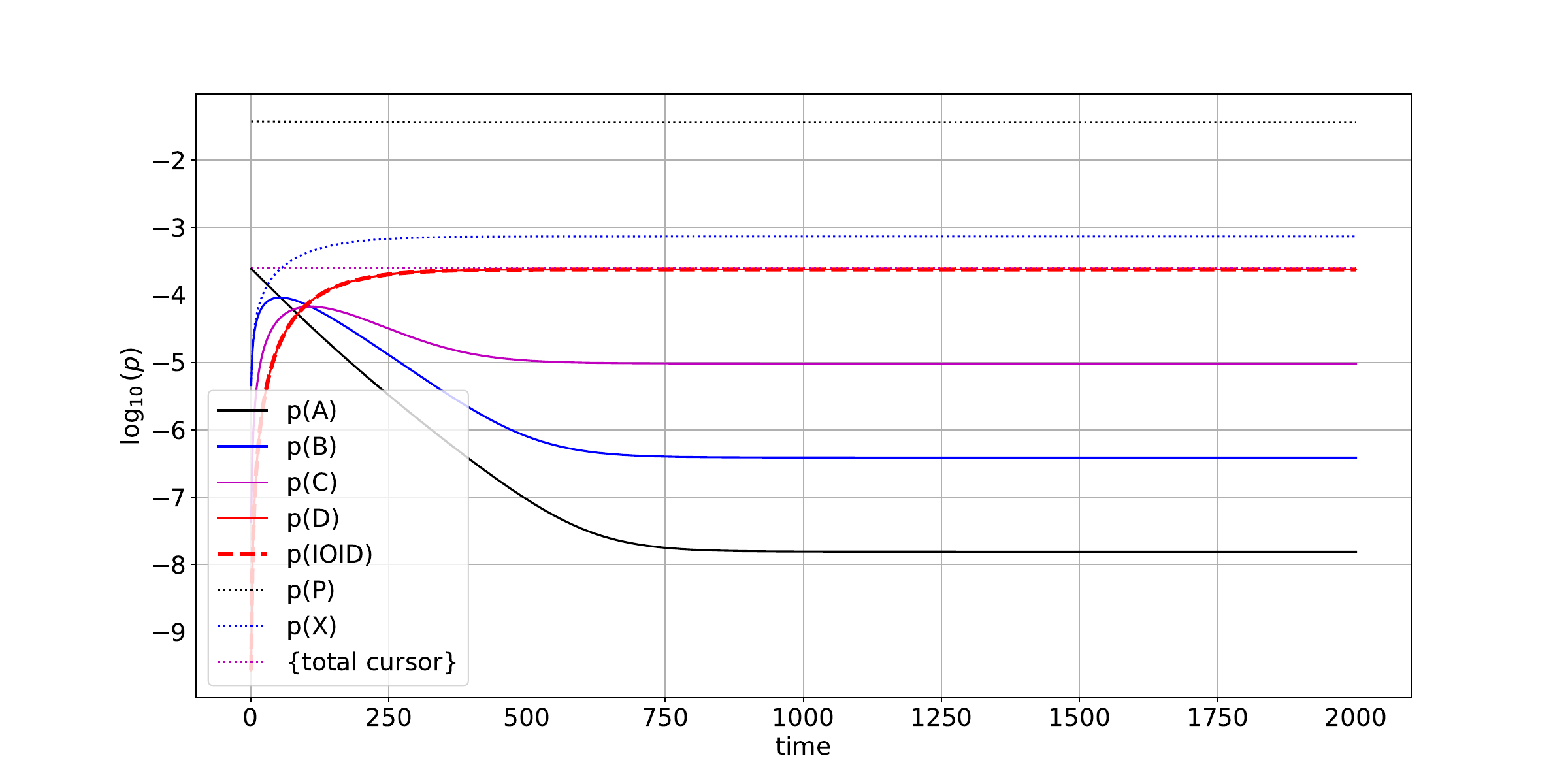}
\caption{Chemical Turing Machine Dynamics -- Equal Gibbs Enthalpies,
Randomly Initialized Tape}
\end{figure}

\hypertarget{chemical-turing-machines-and-the-landauer-limit}{%
\paragraph{Chemical Turing Machines and the Landauer
Limit}\label{chemical-turing-machines-and-the-landauer-limit}}

It is interesting to study this system from the entropic perspective. As
we have no changes in internal entropy or free energy from chemical
reactions, total entropy in the system equals the Markov process
entropy:

\begin{equation}S_\text{Markov} = \sum_{\text{prefix}\;w}p_{w} \sum_{\text{Next\;symbol}\;s} -p(s|w)\log p(s|w).\end{equation}

We again want to explore a situation where the tape-to-solvent ratio
is 1:3, with 0.1\% of all on-tape sites starting out as an
\texttt{A}-symbol and 5\% of all solvent-sites carrying a
\texttt{P}-symbol. At \(t_\text{final}=2000\), we observe
\(p_\text{final}({\mbox{\tt IOID}})\approx2.3981\cdot10^{-4}\),
corresponding to a 95.9\% ``yield'', with only \(<0.06\%\)
of all \texttt{D}-sites not being prefixed by
\texttt{IOI}. Comparing entropy at \(t=0\) with entropy at
\(t_\text{final}\), we find that Markov chain entropy increased by
\(\approx 3.168\cdot10^{-4}\), or \(\approx 13.2\) nepit (\(19.1\)
bit) per unit of result. As every successful instance of running the
computation loses information about 3 bit of tape-state (overwriting
the three bit to the left of the \texttt{D}-marker), hence introduces
a factor-\(1/2^3\) shrinkage of the number of microstates representing
tape-state, a thermodynamically spontaneous process (where we lose
information due to the end-macrostate containing more microstates than
the start-macrostate) will have to produce at least 3 bits of entropy
elsewhere, such as from partial equilibriation of \texttt{P} and
\texttt{X}-concentrations. This is the famous ``Landauer
limit''~\cite{landauer1961irreversibility}\footnote{See
  e.g.~\cite{feynman2018feynman} for a pedagogical introduction.}.
With a large over-supply
of \texttt{P} with respect to the concentration of cursor-sites, the
driving force from \texttt{P}/\texttt{X} equilibriation (and to a
lesser extent \texttt{A}/\texttt{B}/\texttt{C}/\texttt{D}
equilibriation) here manages to give better-than-\(20:1\) yield with
entropy production being a mere factor \(\sim 6.4\) above the Landauer
limit -- rather than being many orders of magnitude larger than the
theoretical limit as for a typical semiconductor circuit based
implementation of a computational process.

\begin{figure}
\centering
\includegraphics[width=\textwidth,height=8cm]{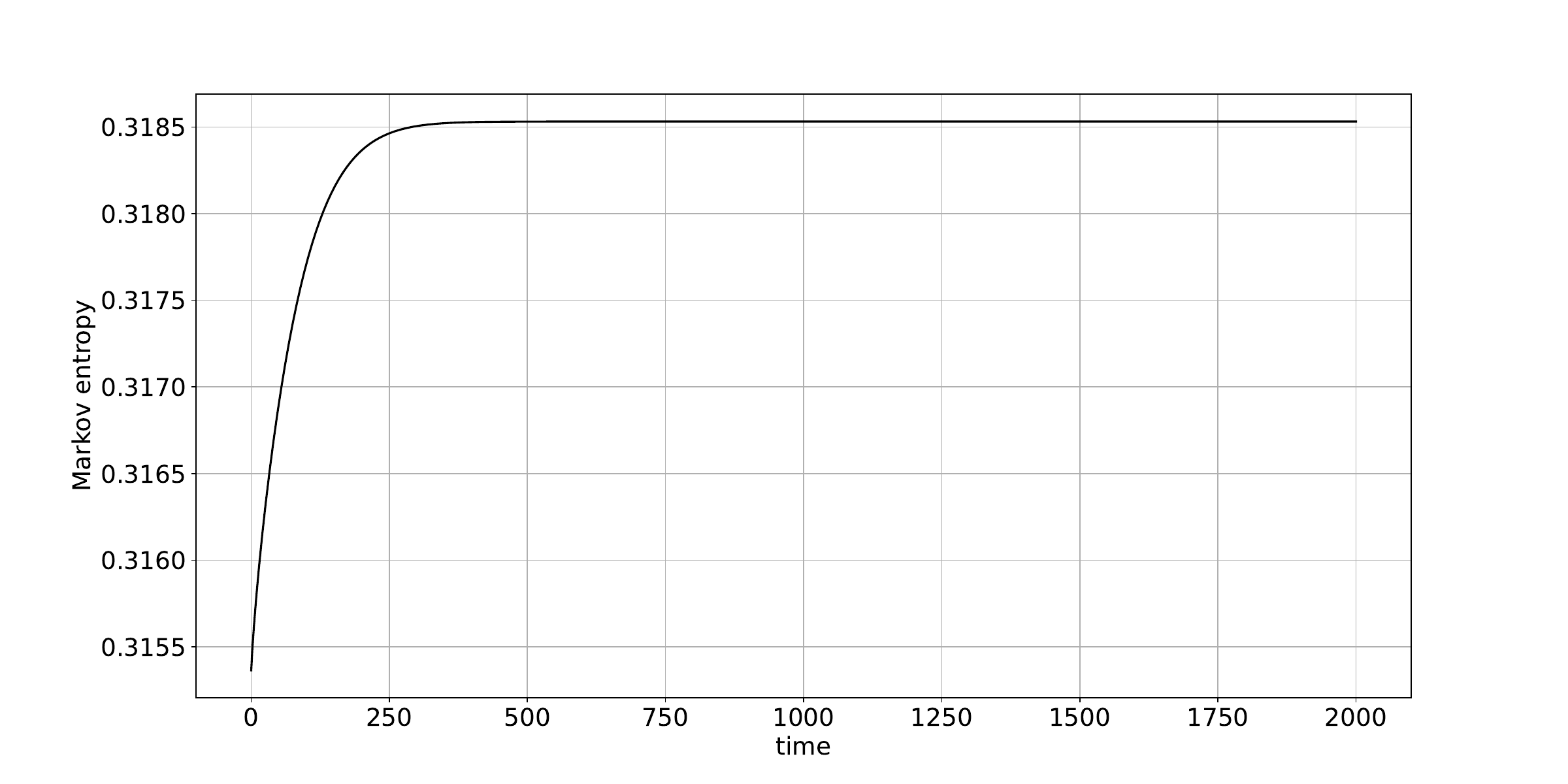}
\caption{Chemical Turing Machine Dynamics -- Entropy evolution}
\end{figure}

The Landauer bound is only attainable in a reversible calculation
(i.e.~no growth in the number of microstates), but since in this
example, irreversibility powers the computation and moves it forward
quickly, we expect to be somewhat above that bound. Still, the
observed ratio we get from reasonable parameter choices for an
extremely simple computation gives a glimpse on how much evolutionary
refinement must have happened for some bacteria to come within
\(<10\times\) the theoretical entropy production minimum for
replication, as claimed in~\cite{england2013statistical} ! It appears
plausible that an understanding of the entropic perspective will be
instrumental to understanding self-replication and evolution in
biological systems. Even if in such a chemical system, the entropy
budget is mostly used for \texttt{CO-NH} peptide bonds, this would
have had to arise as an evolved design choice.

\textbf{Variant 2}: Detachable Evaluator-Molecule

One feature of this system one might argue to be unnatural that also has
a profound consequence for its behavior is that we consider the
tape-transforming evaluator-molecule to be always attached to the tape,
and unable to detach once the computation ends. Even without such
detachment, the presence of data-erasing reverse-steps has observable
consequences, such as causing the occurrence of \texttt{IBI}-sequences
on a tape that initially only had \texttt{O}- and \texttt{A}-symbols.
While this situation is unreachable with only forward-transitions, it
can be reached via
\(\mbox{\tt AOO}\to\mbox{\tt IBO}\to\mbox{\tt IOC}\to_\text{(rev)}\mbox{\tt IBI}\).

Still, if thermodynamic equilibriation mostly drives the computation to
completion, the new species that can only arise via processes involving
reverse-reactions will only show up with low concentration.

With our present framework, tape-sequences get rewritten into
tape-sequences of equal length, and processes that change tape length
require some ingenuity to be modeled, as was illustrated with the
earlier copolymerization example. Since in our approach, an A/B/C/D
symbol marks both the position of the cursor and the state of the Turing
machine, it would be incorrect to include a tape-detaching transition
such as \(\mbox{\tt DO}\to\mbox{\tt OO}\): the resulting tape would not
be a cell-by-cell re-writing of the initial tape, and as such, since the
cells on the initial and re-written tape are not in a one-to-one
correspondence, the differential equation for evolving subsequence
probabilities would fail to be applicable.

In some situations, one might want to model tapes as having ends, and
model tape-transforming molecules to run along the tape and detach when
they reach an end. In other situations, one would want to model the
tape-attaching and -detaching process differently. A generic recipe is
to have tape-cell-state be a product-state of the form
\(\{\text{value in the cell}\}\times\{\text{state of the machine
attached at this cell, or `not attached'}\}\).

For our toy model, we can employ a trick that keeps the number of states
small: If the evaluator-cursor sits at a given tape-cell, the state of
that cell is well-defined in the forward-reaction and random in the
backward-reaction. In neither case do we really have to model the state
of that tape-cell, since for both transitions, we know the outcome. we
hence can, \emph{for this particular toy system that writes data without
paying attention to tape-state}, introduce a variant where the state of
the tape-cell under the cursor is a third one, ``undefined (but with an
evaluator-molecule attached)''. Chemically, this may make sense in some
situations -- as the evaluator attaches, the data-carrying group on the
polymer strand gets transformed into an ``attached and not carrying
data'' state. One obvious consequence then is that ``attaching the
evaluator in A-state to a tape cell in \texttt{O}-state'' has a reverse
reaction that leaves the cell in random state. This introduces a
low-rate pathway for scrambling the tape, via such
attachment-and-subsequent-detachment processes. For this example, we
expand the alphabet with another symbol \texttt{E} that represents a
dissociated evaluator-molecule in the solvent.

An evaluator can attach and detach to the tape freely if in
\texttt{A}-state or \texttt{D}-state, but Gibbs free energies of formation
are chosen such that an evaluator in solvent will more readily attach in
\texttt{A}-state than in \texttt{D}-state (correspondingly, will more
readily detach when in \texttt{D}-state).

The Scheme code that implements the described system is more involved
than earlier examples, mostly since we need to make sure that reaction
rates are compatible with thermodynamics, which we here accomplish by
directly computing them from differences in Gibbs free energy of formation.
As we have seen in the previous example, we also need to be careful
about rate-factors between forward and backward reactions if information
gets erased.

The specific example we want to study has
\((\Delta G^\circ(P), \Delta G^\circ(X), \Delta G^\circ(E))=(6,0,1)\),
as well as
\((\Delta G^\circ(A), \Delta G^\circ(B), \Delta G^\circ(C), \Delta G^\circ(D))=(-1, -1, -1, 1.5)\),
and \(\beta=1/(k_BT)=1\). ODE-integration starts with a tape-composition
where \(25\%\) of all sites are tape-sites initialized to \texttt{O},
and the rest is solvent, with \(4\%\) of the solvent-sites initially
carrying a dissolved evaluator-molecule \texttt{E} and \(10\%\) of the
solvent-sites carrying an energy-providing \texttt{P}-molecule.

This system exhibits complex dynamics; re-running simulations with
different ODE integrator tolerance settings indicates that the computed
dynamics is likely reliable up to at least \(T=10^4\). At the start,
dissolved evaulators first have to attach to the tape, and in alignment
with relative thermodynamical stability, this can happen either in
\texttt{A}-state or \texttt{D}-state, with \texttt{A}-state being more
likely. The \texttt{B}- and \texttt{C}-Turing machine states, which are
only reachable via program-processing, are rare to encounter initially,
but build up their concentration. At the end of the simulation interval,
the most frequent attached Turing machine state is the \texttt{C}-state,
in alignment with the \texttt{D}-state being easy to detach. The
available energy-carrier mostly gets used up, ultimately reaching a low
residual concentration, and the tape-attached \texttt{D}-state
evaluators preceded by the expected program output ultimately also
decline as \texttt{D}-state evaluators detach. Over this time frame, the
concentration of ``final result with no evaluator attached'' sequences
on the tape keeps increasing, accidentally reaching about the same
concentration as \texttt{B}-state evaluators in this example. Due to
tape-scrambling such as in particular from catalytic processes where an
evaluator attaches and immediately detaches again, leaving the
underlying tape-cell in indeterminate state, the concentration of
sequences that only can be produced by such tape-corruption -- like
\texttt{IIII} -- does increase in the long run, as expected.

\begin{figure}
\centering
\includegraphics[width=\textwidth,height=8cm]{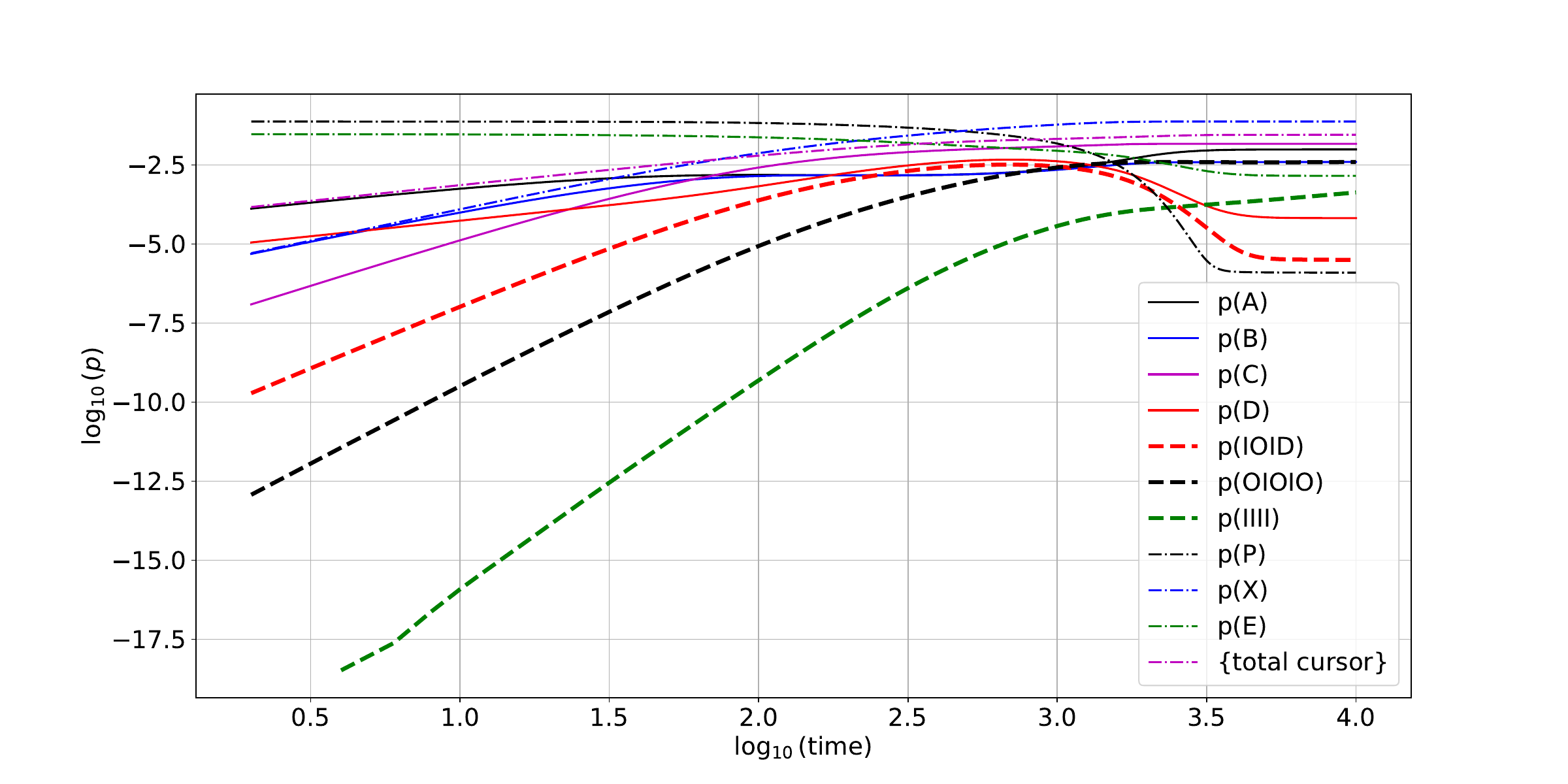}
\caption{Chemical Turing Machine Dynamics -- Detachable Evaluator}
\end{figure}

Clearly, while long term behavior will be dominated by scrambling, this
would look different if attachment/detachment processes were
non-scrambling, so for example always attached/detached to a zero (or
stretch of zeros). This particular modeling choice exemplifies how
data-aging processes such as due to spontaneous hydrolysis can be
included in such models.

\textbf{A Hypothesis}

While there are open questions around the extent to which such a
simplistic implementation of chemistry-based data-processing might
actually provide insights about biology, these observations are
suggestive and may hold an interesting lesson. Biologists generally
struggle to give a stringent definition of ``what is life''. The
definition one normally encounters at the start of the secondary
education curriculum on biology includes a somewhat fuzzy and list of
about six to twelve (depending on where one went to school) criteria
that feels a bit ad-hoc, but in particular includes being made of cells,
having a metabolism, having reproduction, and having repair mechanisms.

Approaching the problem from a data-processing perspective, and starting
from the observation that reversibility of individual chemical steps may
well mean that a data-processing chemical system is highly constrained
in its ability to communicate data into its long-term future, constantly
having to fight data-corruption, one may feel compelled to come up with
the following proposition:

\begin{center}
\emph{``Life'' is any data processing that has the potential to affect
how data processing happens in its own far (such as: geological)
future.}
\end{center}

While there is some residual conceptual fuzziness around questions such
as in particular what qualifies as ``data processing'', and this
approach leaves the question open whether our universe may admit
solutions to the problem posed here that are very different from ``life
as chemistry with reproduction'' (including e.g.~``robotic life''), it
is possible that having some form of reproductive mechanism may be the
only viable solution to the implied problem. Pragmatically, one might
want to consider any mechanism which allows adjusting some form of
``input program'' to make it produce any out of a large number (say,
\(>10^{100}\)) of possible different target sequences as being a likely
candidate for a data-processing system. For biological systems, it is
now widely known that mRNA technology can be used to make ribosomes
produce almost arbitrary oligopeptide sequences.

From this perspective, ``having a metabolism'' and ``having repair
mechanisms'' may then well be unavoidable necessities to make
data-processing reach into the far future, rather than separate aspects.
This unification of properties generally regarded as separate is not so
dissimilar from how Kepler's three laws of planetary motion follow from
Newton's law of gravitational attraction (plus the framework that is
needed to even meaningfully talk about e.g.~forces).

The idea to define phenomena in terms of their relation to their
far-into-the-future / far-into-the-past effect is not novel and has been
useful -- for example -- both for discussing scattering processes in
quantum field theory and also in defining event horizons in general
relativity.

At a more general level, one might argue that such a definition might
potentially regard societies as ``being alive'' in a way that goes
beyond their member organisms fitting this definition -- in the sense
that their prolonged existence also depends on their ability to
transport information into the future, beyond the lifetimes of
individuals. While the characteristic time scale over which
civilizations fail appears to be short in comparison to geological time
scales, it might be that at least some elements of societal behavior
that involve passing on of information (such as perhaps pastoralism per
se) may involve long time scales. One thought that naturally emerges
from this perspective is that societal stability may well be strongly
impacted by a society's ability to keep important information alive -
such as via teachings and traditions -- and therefore, any changes that
directly affect inter-generational information propagation likely need
to be assessed very carefully for their risk potential.

\hypertarget{example-5-a-simple-machine-language}{%
\subsection{Example: ``A Simple Machine Language''}\label{example-5-a-simple-machine-language}}

This example serves a dual purpose: while it explains how to use the
framework to study the behavior of systems with more complicated
definitions that more closely resemble ``evolving programs'' (albeit for
a somewhat nonsensical machine language), it also shows how life-like
behavior (in the previous example's sense -- by having
replicator-patterns carry information into the future) can require
precursor steps that first create the building blocks necessary for
data-copying. Whereas in earlier examples that involved interaction
between different stretches of tape, an interpretation of the P-tape
content in terms of ``providing computer code instructions'' would have
been somewhat unnatural, we here explicitly have such an interpretation.

In the interest of keeping the construction as conceptually simple as
possible, the definitions of machine instructions have been designed in
such a way that there in total only is a (small) finite number of
different possible programs that are of interest. The overall
construction might still be sufficiently close to chemistry that one
might be able to find a plausible chemical realization.

Whereas previous examples (except the very first one) generally included
the perspective of microscopic reversibility, this construction more
closely aligns with the extant literature on ``artificial life'' like
constructions that largely do not include this perspective. Here, the
hope is that illustrating in detail how to handle microscopic
reversibility but also how to align with the literature on evolving
problems will enable the reader to explore more deeply the subtleties
that arise when including reversibility in the modeling of computation --
as will inevitably be necessary when trying to greatly reduce the
entropic cost of computation.

We will again study two variants of this example: the first one shows
emergence of a replicator after a necessary precursor step. The second
one illustrates how a small change to the overall rules that does not
affect the replicator per se can render replicator-patterns
near-ineffective. This shows that the mere possibility for the existence
of an entity that can create copies of itself -- or even the existence of
such entities -- does not imply that the system will naturally transition
into replicator-dominated behavior, in alignment with earlier comments
on autocatalysis made in section~\ref{on-the-role-of-autocatalysis}.

We want to consider an alphabet of five symbols, \((M, S, R, T, F)\)
which we take to be encoded, in that order, by the numbers \(0-4\). The
tape, whose contents are modeled in terms of probabilities of
length-(at-least)-4 subsequences, starts out as as a perfectly random
sequence of only three of these symbols, \(S, T, R\), with equal
probability.

Each symbol has an interpretation as a machine instruction (to be
executed sequentially) that can be roughly described as follows:

\begin{verbatim}
S: [START] Start execution. "Charges" the instruction-counter
   to allow execution of at most 4 operations in total.
T: [TRANSFER] Transfer data from P(rogram)-tape to D(ata)-tape,
   "if activated".
R: [ROTATE] Cyclically increment the content of the current
   D(ata)-tape cell, M->S->R->T->F->M.
F: [FORWARD] Move the Data-cursor forward.
M: [MULTIPLE] If the previous operation was T or R, repeat
   that operation 3x, then stop.
\end{verbatim}

The implementation of these machine operations in purely functional code
needs to forward state between instruction-invocations, for which it
uses the following ``register'' variables:

\begin{verbatim}
Q: "Generalized instruction counter". Starts at 4.
   Decremented by one at every instruction-execution.
   M-operation makes Q jump to -1.
   Execution will stop at either $Q=0$ or $Q=-4$.
Is: P(rogram)-tape start-of-execution index.
Ip: P(rogram)-tape index.
Id: D(ata)-tape index.
Op: Operation executed in the previous step.
NT: 1 if a T-operation has been invoked since program start.
NR: 1 if a R-operation has been invoked since program start.
NF: 1 if a F-operation has been invoked since program start.
\end{verbatim}

Our rules shall be such that a \texttt{T} operation only will copy data
(and advance both the \texttt{Is} and \texttt{Id} index) if, since
program start, both a \texttt{F} operation and a \texttt{T} operation
already have executed. The precise semantics of these operations is
defined by the implementation given in appendix~\ref{c.5-simple-machine-language}.

By inspection, one observes that, according to these rules, the sequence
\texttt{SFTM} can create a copy of itself (not at the original random
position of the data-cursor, but starting one cell after that, which
still is an equivalent randomly-chosen position), while no other
sequence has this ability. Clearly, starting from a tape with only
\texttt{M}, \texttt{S}, \texttt{R} symbols, the initial probability to
find such a replicator is \emph{exactly zero}, but there are nontrivial
programs that affect tape-composition by executing
\texttt{R}-operations, which then over time give rise to other
sequences, such as also \texttt{SFTM}. The system has been designed in
such a way that replicators can only arise as a consequence of earlier
interactions between elements (in particular, the \texttt{R}-operation)
that play no role in the replicator itself.

One finds that the dynamics is rather rich and nontrivial, and the
\texttt{STFM} sequence in the long run manages to establish a
probability/concentration that is substantially above the random average
of \((1/5)^4=0.0016\). Still, with the given rules, its probability
first overshoots and is affected by oscillations that in the longer run
die down. Overall, it does not succeed at completely taking over
tape-composition.

\begin{figure}
\centering
\includegraphics[width=\textwidth,height=8cm]{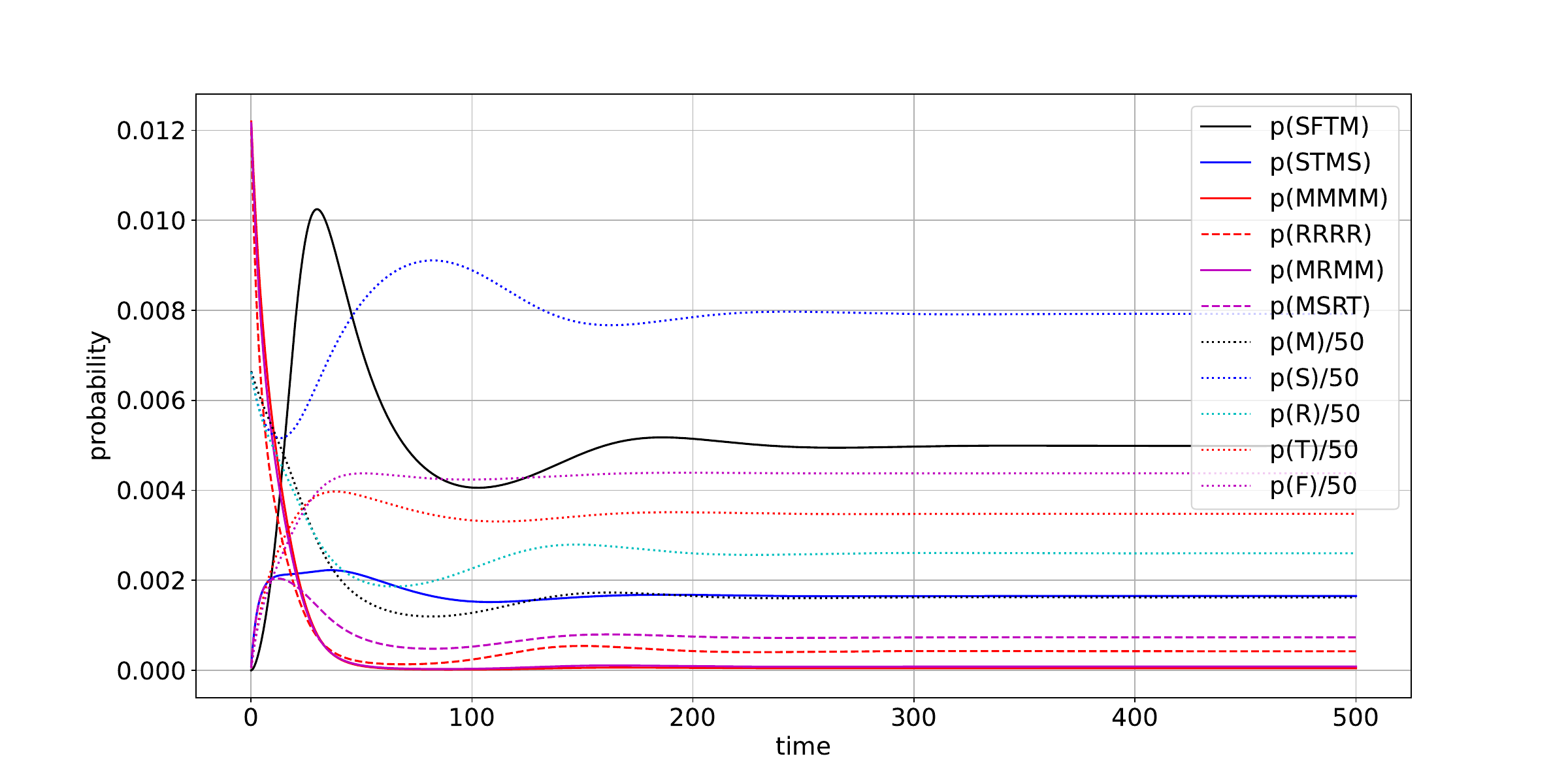}
\caption{Dynamics for some key sequences}
\end{figure}

According to the given rules, an attempt to start program invocation at
a random tape-position is unsuccessful unless the tape contains a
\texttt{S}-instruction at that position. If one were to consider a
slight variation of this system where an attempt to invocate a program
can also be at least partially successful in some other situation, such
as by adding the rule ``if program-invocation starts at a \texttt{R}-
rather than \texttt{S}-operation, a single \texttt{R}-operation is
executed (and then the program stops), this adjustment does not touch
any aspect of the mechanism as-seen-in-isolation via which the
\texttt{SFTM}-replicator creates copies of itself.

Still, with these modified rules, which are obtained by switching the
\texttt{(let\ ((single-R-can-execute\ \#f))} assignment to \texttt{\#t}
in the first line of this example, dynamics changes rather dramatically:
This replicator here only is able to establish an equilibrium
concentration that is a little bit above the average for a perfectly
random sequence under perfectly randomizing dynamics.

\begin{figure}
\centering
\includegraphics[width=\textwidth,height=8cm]{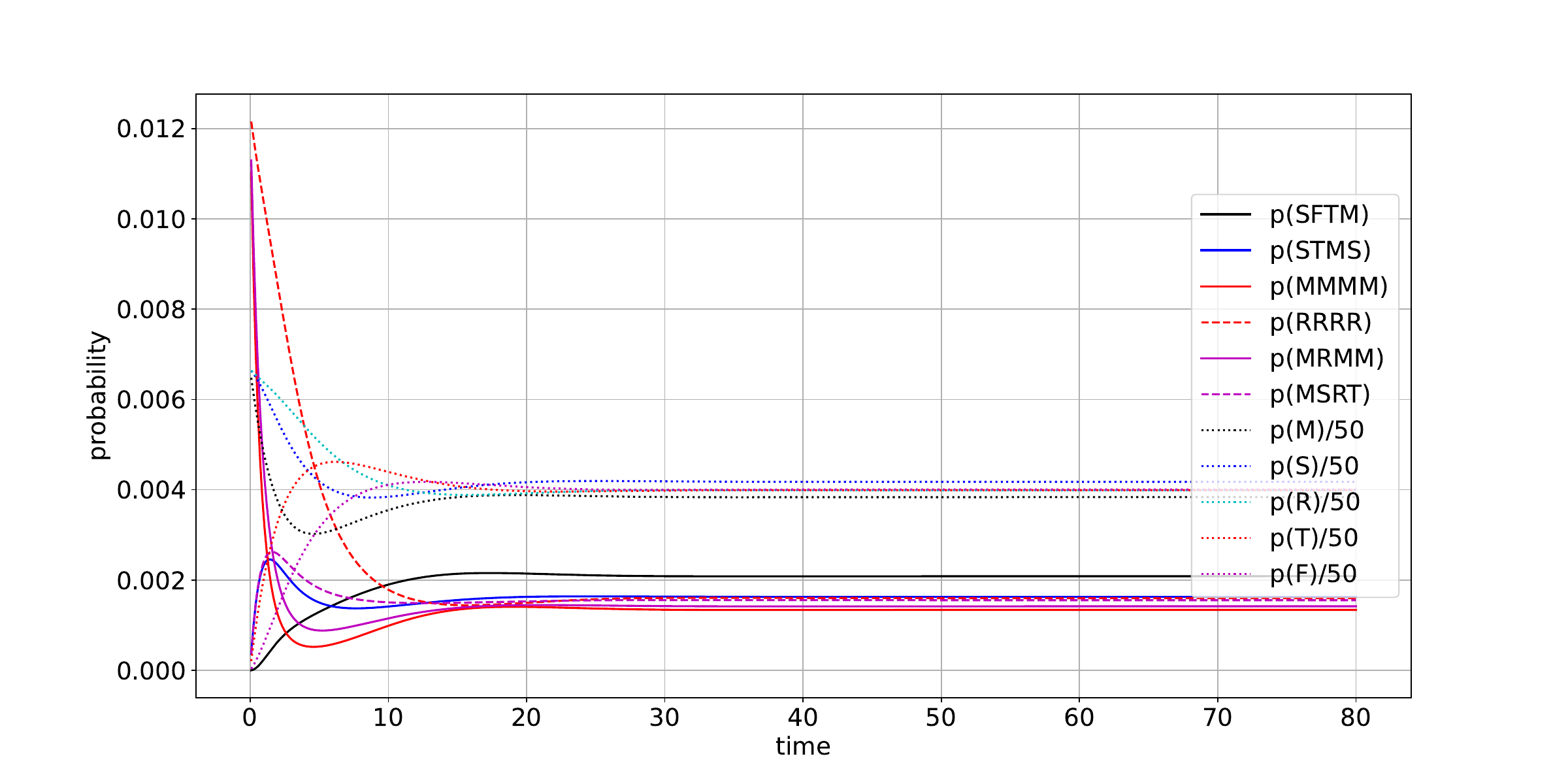}
\caption{Modified dynamics for some key sequences}
\end{figure}

Here, the most frequently encountered sequence at \(t=100\) is
\texttt{SFTS} at \(p=0.00229\), and the least frequently encountered
sequence is \texttt{MMMM} at \(p=0.00134\), both not far from
\(p=(1/5)^4=0.0016\).

This example demonstrates that in systems which allow self-replication,
even if such self-replication-capable patterns keep emerging
spontaneously, it is by no means true that this would automatically lead
to them taking over the composition of the substrate.

Overall, if one were to do an in-depth exploration of some aspects of
the dynamics of a system like this, such as trying to confirm the
presence of the oscillations observed in the basic model, perhaps even
trying to derive a simplifying but reasonably accurate approximative
analytical model along the lines of the ``ferromagnetic spin chain''
example, the value of having access to a quantitatively accurate
numerical probability model that does not have sampling noise should be
self-evident.

\hypertarget{example-6-the-bff-type-system-of-1}{%
\subsection{Example: The ``BFF-type system'' of~\cite{alakuijala2024computational}}\label{example-6-the-bff-type-system-of-1}}

The purpose of our final example is to illustrate that the ``language
of problems expressible in this framework'' is, if we ignore
constraints due to the inability to build arbitrarily powerful
computers, expressive enough to contain systems such as the one
described in~\cite{alakuijala2024computational}. Oftentimes, there
are interesting insights to be gained from discussing the structural
constraints on a particular formalization of a problem even in
situations where the actual calculation is out of reach -- for example,
while we are unable to quantitatively describe a quantum system that
contains an intelligent observer (the ``Wigner's Friend'' thought
experiment), it still makes sense to ponder whether one would expect a
breakdown of unitary time evolution, i.e.~Schr\"odinger's equation, in
such a setting.

The BFF language uses 10 nontrivial machine instructions out of a symbol
alphabet of 256, all other symbols codifying no-operation (no-op)
instructions. Despite this, a truncation to a smaller alphabet -- such as
16 symbols encoded by 4 bits each -- would not be fully faithful due to
differing fractions of no-op vs.~other instructions. Even if one were to
adjust symbol probabilities to address this, this would not faithfully
represent the distances between machine operations in terms of the
number of increment-operations required to transform one such operation
into another. Given the overall nature of the construction, it
nevertheless seems unlikely that many of the details, such as using a
256-symbol alphabet, are essential for key observations about the
system's behavior.

While implementation of opcode-evaluation is straightforward and can
closely follow the structure of the previous example, other aspects of
the documented evaluation strategy for this system, in particular
fusing and splitting pairs of tapes, is less straightforward to
implement. One relevant discrepancy between the approach chosen by
this framework and the BFF-construction is that in the latter, the
program- and data-cursor sit on the same tape. From a chemical
perspective, both the modeling approach chosen here -- having random
segments of polymer strands interacting with one another, as well as
having ``attachment points'' that play a special role appear
justifiable and interesting. While it would be possible to encode this
other approach via extra symbols that indicate a tape-end, we instead
will here focus on the ``long tape'' variant discussed in section 2.4
of~\cite{alakuijala2024computational}, where P-tape and D-tape cursors
are positioned effectively at random.

While a full simulation of the BFF-like system using this approach would
be computationally prohibitive, the construction presented here may be a
useful basis for constructing simplified models that allow exploration
of specific aspects such as key steps that happen prior to the emergence
of a replicator. One natural simplification would be to shrink the
alphabet by fusing machine instructions into groups and have a random
instruction from the given group get picked and executed. Such
techniques may be used to shed some light on the evolution of
tape-composition before the first successful replicators arise.

The construction shown in appendix~\ref{c.6-mini-bff}, which uses a
12-symbol alphabet, can be used as a starting point for constructions
that more closely resemble BFF than the previous example.

\hypertarget{conclusion-and-outlook}{%
\section{Conclusion and Outlook}\label{conclusion-and-outlook}}

The obvious benefit we gain from the intermediate-level modeling
approach presented here is that it allows precise quantitative
investigations into the behavior of many systems that involve polymer
strand interactions and may or may not have a computational
interpretation. In some situations, this new tool in the toolbox
simplifies establishing conclusions that would be harder to obtain with
only Monte Carlo based simulation approaches. For many systems for which
the number of parameters of the underlying Markov model is lower than
about \(200\,000\), detailed quantitative numerical explorations should
be within reach.

Via utilization of nondeterministic (``multiverse'') evaluation,
transformations can be specified in a simple way with straight code, and
(as has been shown in the ``Nylon polymerization with preferences''
example) we readily can insert reaction constants as they may be
provided by a more fundamental theory into our models, thus allowing us
to establish contact, at the quantitative level, between microscopically
reversible dynamical theory and computation. This enables us to
quantitatively study achievable entropic costs of data processing in
chemical models.

Plausibly, this framework may be of interest to research in
statistical mechanics, molecular biology, (perhaps) polymer chemistry,
abiogenesis, and ``artificial life'', especially where these fields
overlap. Despite its conceptual convenience, a price to pay for users
is that they may have to invest some effort into gaining basic
proficiency with functional programming in Scheme (should they not
have this already) -- or find a collaborator with such expertise. On
the positive side, the Scheme programming language has been designed
as a minimalistic, straightforward language with a very compact
standard\footnote{In this work, we only use $\text{R}^5\text{RS}$
  scheme features, plus some extensions provided by Gambit-C such as
  Common Lisp DEFMACRO style macros.}~\cite{sperber2009revised6}, and
the (even smaller) subset of the language that is relevant for these
explorations is comparatively easy to master. Still, the underlying
idea to utilize continuation support as proposed is not intrinsically
linked to the Scheme programming language, and alternative approaches
to implementing an equivalent framework are conceivable, which may
even be computationally somewhat more efficient. Given the exponential
nature of tape-unfolding, such implementation-level efficiency
improvements would nevertheless only lead to small shifts of the
boundary between what can be done with reasonable computational effort
without further simplifying approximations and what can only be done
with such simplifications.

Focusing on low level languages, the C programming language offers deep
call stack unwinding via \texttt{setjmp(3)/longjmp(3)} which potentially
might be used as a basis for multiverse evaluation that somewhat
resembles the present construction. A more appealing alternative
approach might be to translate the continuation-based construction to a
``continuation monad'' in Haskell (see e.g.~\cite{liucps}). For the
current article, basing the construction on Scheme appeared especially
appealing, since the powerful C interface of the Gambit Scheme
implementation allows one to keep the amount of code small that is
required to efficiently integrate Scheme with Python -- which here is
used for ODE integration and data analysis, including plotting.

As the ``nylon polymerization'' example discussed in
section~\ref{example-3-co-polymerization} illustrates, some
applications - especially if they are more on the side of chemistry --
will require some ingenuity in order to cast them into a form where
they are tractable with the framework described here that only models
tape/tape interactions. This seems to ask for further
generalizations. The current construction starts from the idea that
``dynamical processes happen when two sequences meet at a random
position'' and the ``rarefication'' assumption that these events are
sufficiently rare that we can model them as happening independently of
one another and in the ``large (i.e.~\(\gg 10^{20}\)) number of
sequences'' chemical limit contributing to total tape-composition
changing chemical reaction rates.

In this limit, where our natural unit of time is such that over an
interval of \(\Delta t=1\), every sequence site experiences sparking an
expected number of \(N=1\) interactions. If one were to extend the model
to have processes other than tape/tape interactions, such as
tape/monomer interactions in addition to tape/tape interactions, this
would require both modeling non-tape constituents of the chemical setup,
alongside a more diligent modeling of the relative reaction rates of
processes involving different combinations of types of chemical
constituents.

With the current approach, system state is modeled in terms of Markov
process parameters that govern tape-composition, where there is only one
form of tape. Biology knows many processes for which it would be natural
to consider different kinds of ``tapes'' -- RNA sequences, polypeptides,
DNA sequences, but perhaps also e.g.~polysaccharides or even fatty
acids. While one could model such a situation in terms of just one kind
of tape and a ``total sum'' alphabet of elements with Markov chain
probabilities being such that an amino acid never connects to a DNA
sequence, it would be far more natural (and efficient, since this would
do away with a need to keep track of the probabilities of many
``impossible'' sequences) to approach this with ``different
probabilistic models for different kinds of tapes''.

Looking in another direction, the general mechanism of making a
``measurement'' of so-far-unrevealed tape-content split the universe in
a ``multiverse that ultimately keeps track of statistical weights of
universes'' approach could in principle be generalized from sequences to
more complex graphs, but especially if these have cycles, extra
diligence is required to ensure that statistical weights are correct.

While the current modeling approach is viable for small and simple
systems, combinatorial explosion prevents us from modeling complex
computational settings, unless we have specific questions in mind that
are simpler to answer, such as ``what sort of processes change the
composition of the soup before an observed transition to
replicator-dominance?'' As has been shown experimentally
in~\cite{alakuijala2024computational}, replicators mostly do not
arise as a consequence of being assembled by random mutations, and
re-starting a simulation with freshly sampled random tapes after a
number of steps long enough for a replicator to take over but short in
comparison to observed time-to-replicator-dominance time scales will
strongly suppress the emergence of replicators. While there always is
the option to add actual random sampling back into the current
modeling framework (which however needs to be done carefully, in order
to not upset the ODE integrator's smoothness assumptions), a more
promising approach may be to see if problem structure can be exploited
in a way that replaces the subsequence probabilities table with some
reasonably-accurate but for the actual problem at hand more powerful
model -- akin to how an approximate analytic model was constructed for
the ``ferromagnetic chain'' example.

In any case, the exponential increase in complexity as a function of the
number of relevant components (i.e.~tape cells) parallels the situation
for quantum mechanics. Here as well, the dimensionality of the Hilbert
space of a composite system is the product of the dimensionalities of
the Hilbert spaces of the subsystems, making multiparticle systems (in
principle) computationally challenging to handle. Despite this, quantum
mechanics is clearly indispensible for understanding the properties of
molecules, and advances that happened decades after the initial
formulation of quantum mechanics made even rather complex systems
tractable to useful accuracy.

In a sense, this fallback to sampling parallels the aspect that in the
real world, the notion of dynamics of a ``chemical concentration''
breaks down for concentrations so low that these would have to be
interpreted as corresponding to at most a few discrete molecules. One
still would be able to capture much of the dynamics at a quantitative
accuracy that would be very difficult to obtain with a purely
sampling-based approach, but accepting some noise for very low
probabilities. A relevant difference to purely sampling-based modeling
approaches remains in the way tape content is modeled via parameters of
a Markov process. On the computational side, care has to be then taken
that allowing some low-amplitude sampling based noise in the computed
rates of change does not upset the numerical ODE integrator.

Finally, applications: Given that the quantitative modeling approach
presented here is expected to establish contact with chemical
kinetics, one should be able to extract experimentally verifiable
quantitative predictions such as the evolution over time of the
viscosity of a solution in which polymerization processes take
place. For purely computational model systems, one would expect that
the predicted reaction rates closely align with the behavior of
sampling-based simulations in the ``infinite number of samples''
limit, and as such should allow giving good quantitative answers to
questions for which sampling-based simulations leave it unclear
whether failure to observe some specific process might have been
merely due to bad luck. It is hoped that this framework -- very likely
in combination with simplifying approximations -- may turn out to be
useful to shed some light on deep and potentially highly practically
relevant questions such as how even a biochemically simple system such
as \emph{Escherichia coli} can apparently (as is speculated
in~\cite{england2013statistical}) utilize its entropic budget for
self-replication -- which includes in particular the associated
data-processing required for copying its DNA -- at an efficiency that
is far from anything present-day engineered data-processing systems
are able to achieve.

\begin{appendices}
\renewcommand{\thesection}{\Alph{section}} 

\hypertarget{appendix-a-competing-autocatalytic-species-in-a-flow-equilibrium}{%
\section{Competing autocatalytic species in a flow equilibrium}\label{appendix-a-competing-autocatalytic-species-in-a-flow-equilibrium}}

This appendix illustrates the generic behavior of the autocatalytic
A-Piece/B-Piece/Monomer system described in section~\ref{on-the-role-of-autocatalysis}:
\(2M\leftrightharpoons A\), \(2M\leftrightharpoons B\),
\(2M+A\leftrightharpoons 2A\), \(2M+B\leftrightharpoons 2B\). The system
is parametrized in terms of a vector
\((\sigma_A, \alpha_A, s_A, \sigma_B, \alpha_B, s_B, a, w)\), where
\(\sigma_{A,B}\) are the spontaneous formation reaction constants,
\(\alpha_{A, B}\) are the autocatalytic formation reaction constants,
\(s_{A,B}\) are the species-stability factors (ratios of forward- to
back-reaction rates), \(a\) is the monomer-addition rate, and \(w\) is
the mixture withdrawal-rate. Intuitively, \(M=10^{-6}\) would correspond
to removing one millionth of the amount of every species present per
unit time. With concentrations being \(A(t), B(t), C(t)\), the rate
equations are:

\begin{equation}\begin{array}{lcl}
(d/dt)A(t) &=& \sigma_A M(t)^2 + \alpha_A A(t) M(t)^2\\
&&-(\sigma_A/s_A) A(t) - (\alpha_A/s_A) A(t)^2 - wA(t)\\[2ex]
(d/dt)B(t) &=& \sigma_B M(t)^2 + \alpha_B B(t) M(t)^2\\
&&- (\sigma_B/s_B) B(t) - (\alpha_B/s_B) B(t)^2 - wB(t)\\[2ex]
(d/dt)M(t) &=& a+2(A(t)\cdot \sigma_A/s_A + B(t)\cdot\sigma_B/s_B)\\
&&+ 2(A(t)^2\cdot \alpha_A/s_A + B(t)^2\cdot\alpha_B/s_B)\\
&&-2(\sigma_A(t)+\sigma_B(t))M(t)^2\\
&&-2(\alpha_A A(t)+\alpha_B B(t))M(t)^2-wM(t)
\end{array}\end{equation}

Comparing with Eqs. (4) and (5) from appendix~C of
\cite{lifson1997crucial}, these rate equations do include
contributions that model catalytic acceleration of the
reverse-reaction that is a consequence of catalysts not being able to
shift thermodynamic equilibrium -- not having these terms violates
thermodynamics.

The ``default'' example that is used as a reference baseline in
subsequent numerical explorations is specified by
\((A(0), B(0), M(0)) = (0, 0, 1)\) as well as
\((\sigma_A, \alpha_A, s_A, \sigma_B, \alpha_B, s_B, a, w)=(0.001, 20,
10, 0.001, 50, 20, 0, 0)\).  For each example dynamics shown, we list
the adjustments made from these default settings.

While A-piece and B-piece dimer spontaneously are formed at the same
rate, the B-piece is thermodynamically more stable in this default
setting, and also a more effective autocatalyst. Numerically integrating
the time evolution, we find the behavior shown in diagrams 1-3.

Diagram 1 illustrates that, without addition or withdrawal, as long as
the relative stabilities of the A- and B-species are as given, the
system always reaches the same equilibrium -- the thermodynamic state of
maximal entropy: Solid (-) = default configuration, dashed (--) =
\((A(0), B(0), M(0))=(0.2, 0.1, 1-2\cdot(0.2+0.1))\), dash-dotted (-.) =
\(\alpha_B=80\), dotted (.) = \((\alpha_A, \alpha_B)=(50, 20)\). Even
swapping the autocatalysis rates (dotted) gets us to the same
equilibrium, even if in this case, the A-species starts out rapidly
building up a large concentration.

\begin{figure}
\centering
\includegraphics[width=\textwidth,height=8cm]{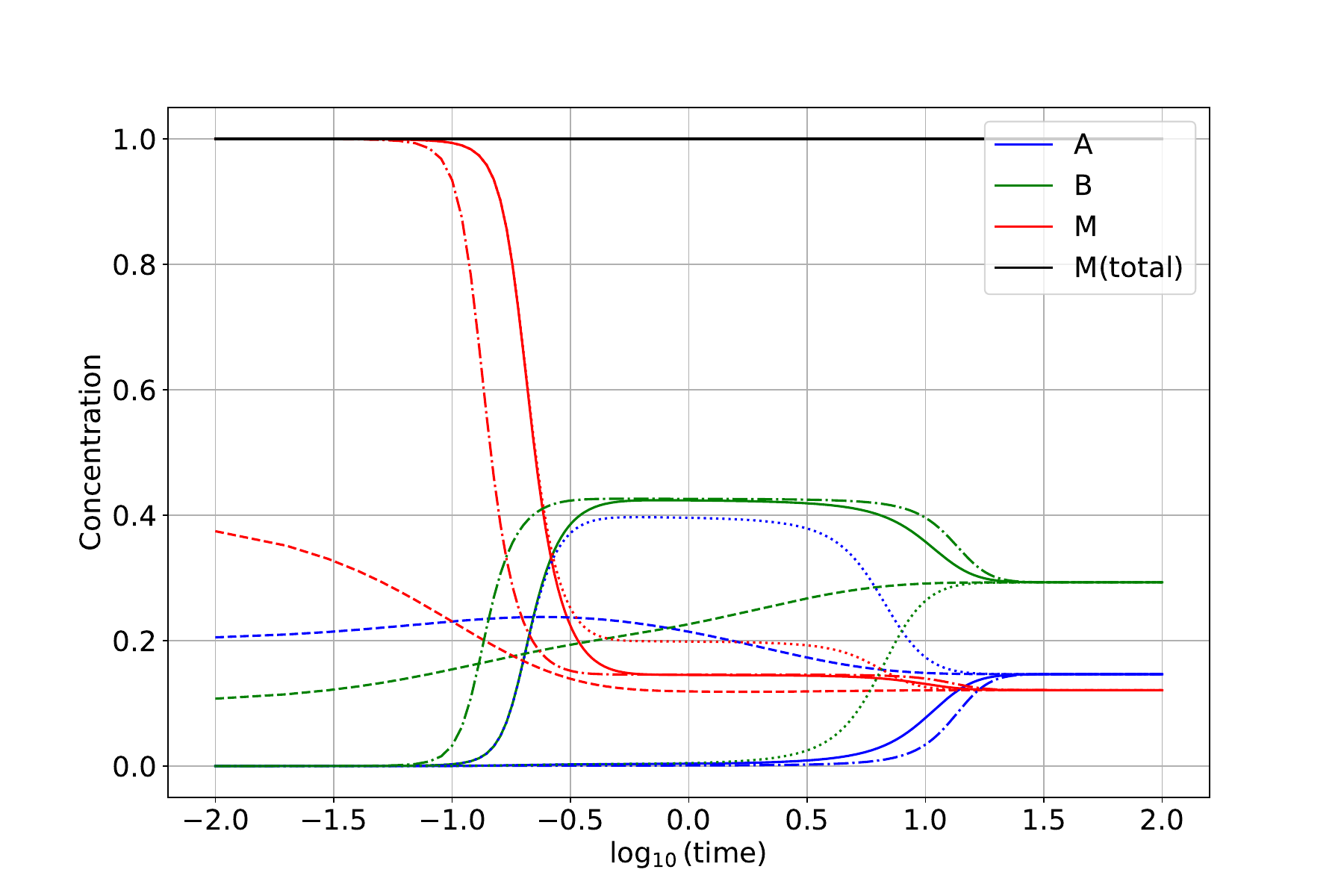}
\caption{No withdrawal/addition -- reaching the same thermodynamic
equilibrium independent of parameters}
\end{figure}

Diagram 2 shows that, as we increase the withdrawal / monomer addition
rate, this affects equilibrium concentrations: Solid (-) = default
example, dashed (--) = \((a, w)=(0.1, 0.1)\), dash-dotted (-.) =
\((a, w)=(0.5, 0.5)\), dotted (.) = \((a, w)=(10, 10)\). As one would
expect, the flow equilibrium will differ from the thermodynamic
equilibrium. For very high flow rates (dotted), reactions struggle to
build up concentrations of both reactants and less of each product gets
produced than what we would find in thermodynamic equilibrium. This way,
we can get extreme ratios of flow equilibrium product concentrations, at
the price of overall low concentrations of both products. For lower flow
rates (dashed / dash-dotted), we can be in an in-between situation where
the less effective autocatalyst struggles to build up a relevant
concentration and the more effective autocatalyst benefits from this in
absolute terms: if we turn on a small amount of flow, the equilibrium
concentration of the B-species (which is both more stable and more
effective an autocatalyst) increases, while the equilibrium
concentration of the A-species decreases. Still, we cannot fully drive
the A-species to extinction this way. One also notes that increasing the
flow rate makes the system reach its equilibrium faster, which is
plausible, given that over any given time interval, a smaller proportion
of reactants is not merely passing through. Despite the more stable
B-species benefiting for moderate flow rates, the total fraction of
unprocessed monomer always increases with increasing flow rate.

\begin{figure}
\centering
\includegraphics[width=\textwidth,height=8cm]{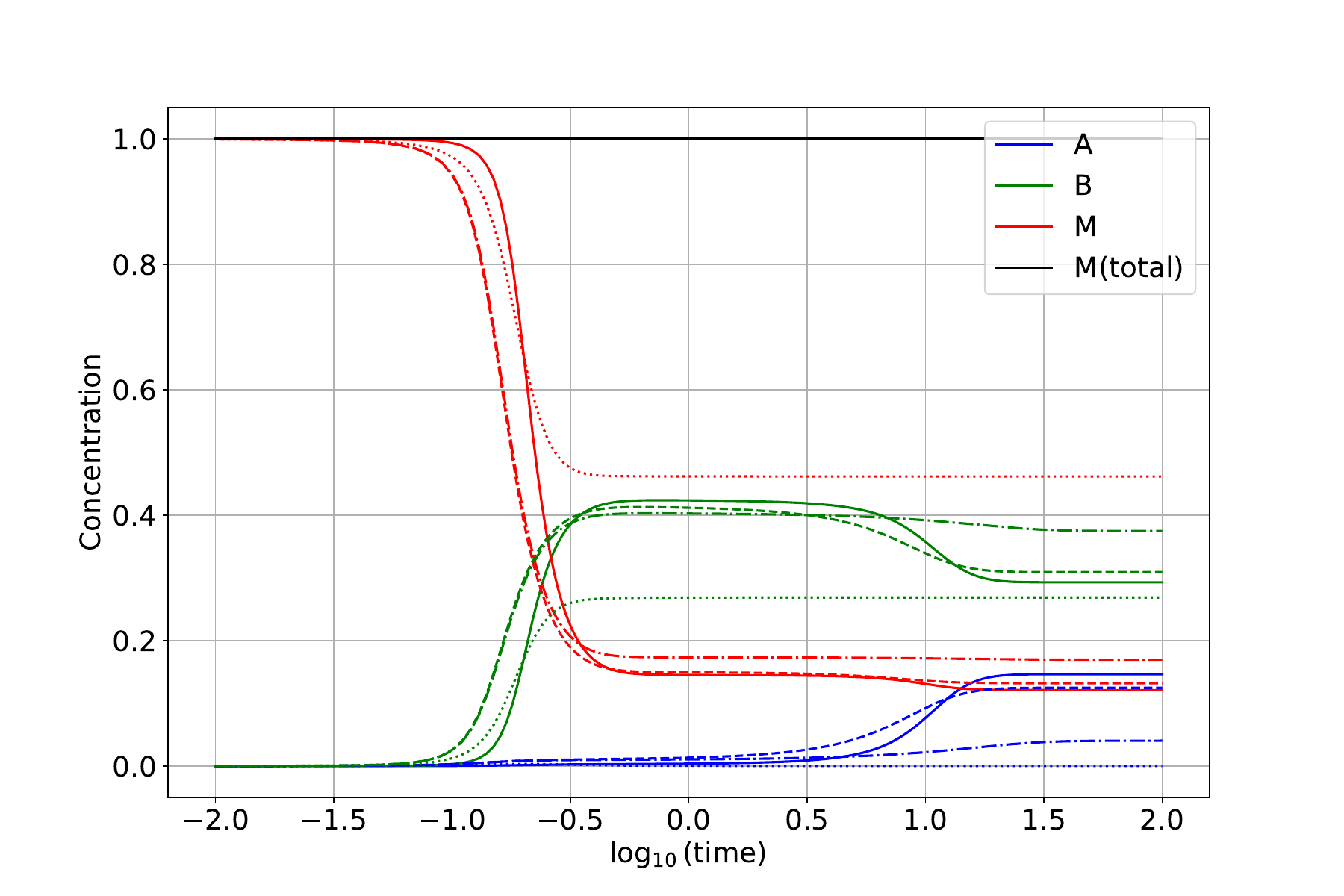}
\caption{Competing autocatalysts with changing flow rates}
\end{figure}

Also for two thermodynamically equally stable species that differ in
their autocatalytic effectiveness, the flow equilibrium
concentration-ratio changes with the flow rate, favoring the more
efficient autocatalyst the more the higher the flow rate. For this final
diagram, the reference parameters are \((A(0), B(0), M(0)) = (0, 0, 1)\)
as before, but
\((\sigma_A, \alpha_A, s_A, \sigma_B, \alpha_B, s_B, a, w)=(0.05, 20, 10, 0.05, 25, 10, r, r)\),
with \(r=0.1\) (solid (-)), \(r=1\) (dashed (--)), \(r=5\) (dash-dotted
(-.)), \(r=30\) (dotted (.)). For low flow rates, the more effective
autocatalyst benefits even in absolute terms, and at first the more the
stronger the flow rate is increased. As the flow rate increases further,
both products do not have time to build their concentration before they
are removed from the reactor, and so the concentrations of reactants
decline, with the more effective autocatalyst retaining a small
advantage from autocatalysis that in the infinite flow rate limit
becomes irrelevant. That limit is governed by the (here, equal)
spontaneous creation rates and the flow rate. In that limit, the reactor
is mostly passing through unreacted monomer, so \(M(t)\approx 1\), and
over a time interval \(dt\), we are freshly producing
\(\sigma_A M(t)^2\,dt\approx \sigma_A\, dt\) species-A (and
correspondingly species-B) from \(a\,dt\) fresh monomer that got added,
a fraction \(a\,dt\cdot M(t)\approx a\,dt\) of the total monomer. If
there were no decay, the equilibrium concentration in the effluent
(which would match that in the reactor) would approach
\(A(t)=M(t)\sigma_A/a\approx\sigma_A/a\). In this limit, decay can be
neglected vs.~spontaneous creation since creation is proportional to
\(M(t)\approx 1\), while decay is proportional to \(A(t)\ll 1\).

\begin{figure}
\centering
\includegraphics[width=\textwidth,height=8cm]{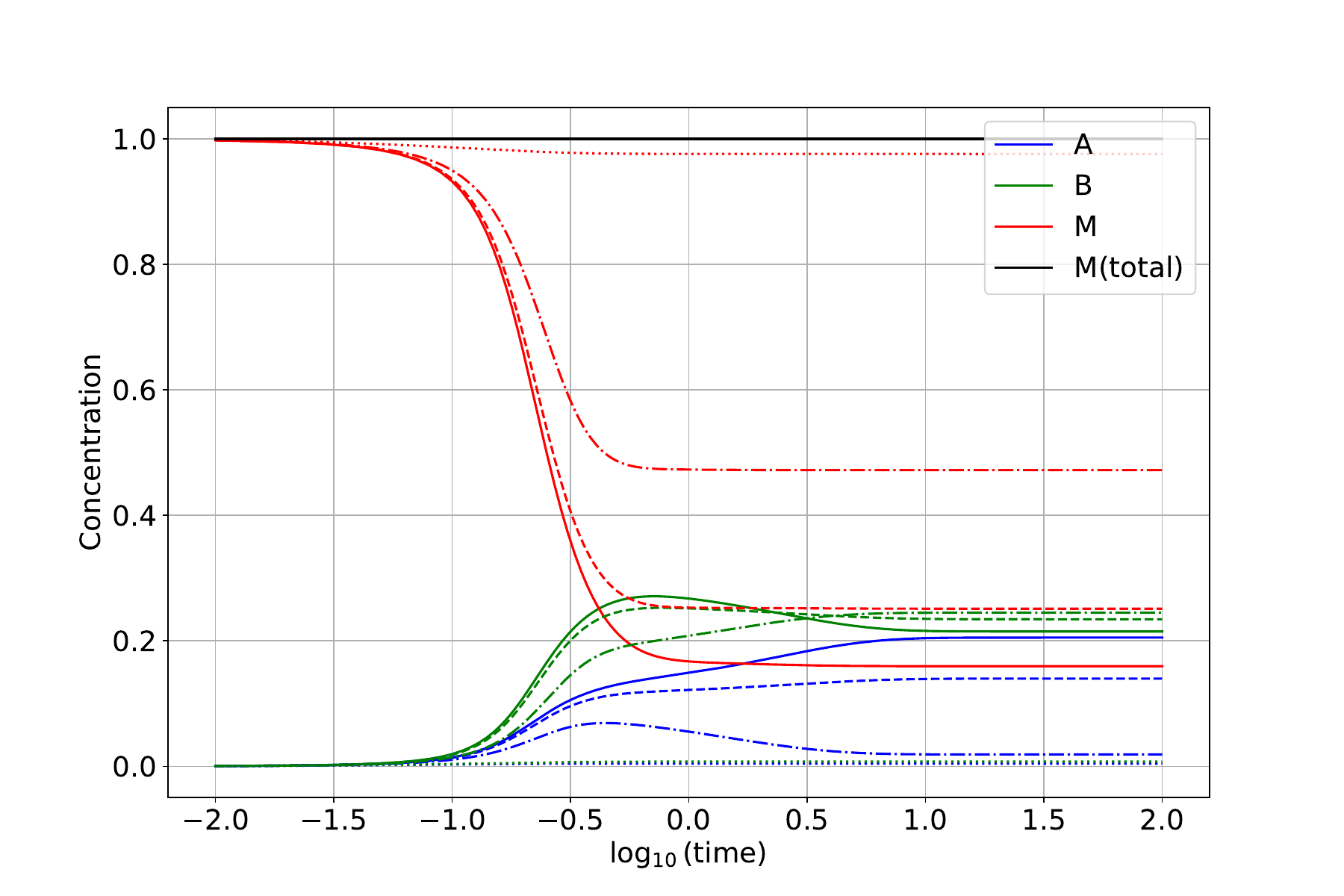}
\caption{Equal stability but differing autocatalytic power at various
flow rates}
\end{figure}

Intuitively, one would not expect concentration of the less effective
self replicator to go to zero -- even if some flow equilibrium with
high flow rate only leaves little time for reactions before reactants
are removed, looking at a single molecule, the entropy contributions
for this one molecule to be of species $A$ or $B$ are proportional to
the negative logarithms of their respective concentrations, so not
having any molecules of the more rare form becomes exceedingly
unlikely.

The code for this example can be found in the supplementary material for
this article.

\hypertarget{appendix-b-modeling-details}{%
\section{Modeling Details}\label{appendix-b-modeling-details}}

This appendix provides more detail on the algorithms underlying the
framework. As explained, at every point in time, system-state is
represented in terms of a numerical \(k\)-axis
length-\(k\)-subsequence-probability array of shape \texttt{{[}A{]}*k},
where \texttt{A} is the number of symbols in the alphabet. This array
must satisfy some implicit constraints: probabilities all must be in the
range \([0;1]\) and sum to 1. Also, these subsequence probabilities must
be a fixed point of predicting the next symbol from conditional
probabilities given the prefix and removing the leading symbol. The
framework offers functions to check these properties on the initial
state, but using ODE-integration, numerical rounding errors might lead
to slight violations.

At every point in time, the rate-of-change is then computed by making
adjustments to a rate-accumulator-array for every change that can arise
from running the user-provided code on all possible tape-contents.

This approach arises from a ``very many tape-sites'' (in the chemical
sense, such as \(N_\text{sites}>10^{20})\) limit of the construction
presented in~\cite{alakuijala2024computational} as follows: Execution
of a single (terminating) program will affect only some finite
neighborhood of a single site, so will not really change the
proportions in a composition.  Looking at possible chemical
realizations of such systems, we generally would want to consider the
situation that more than one program-executing agent is at work. So,
let us suppose we have \(C\) many ``CPUs'' that can
independently-and-concurrently operate on the tapes by running
programs.

We also want to assume that program execution is so fast that we do not
have to concern ourselves with interference between the actions of
different CPUs performing concurrent mutations on overlapping
tape-sequences. This effectively parallels the view that chemical
reactions between gases normally can be studied by looking at
2-molecule collisions only -- since molecules are typically so far apart
that it is very unlikely for a third molecule to be closeby during the
time when two molecules collide. So, we consider a ``rarefied''
situation with many more possible execution-sites \(N_\text{sites}\)
than active CPUs \(C\): \(N_\text{sites}\gg C\): If the time needed for
one instruction is \(\tau_\gamma\), and each program terminates after
executing at most \(N_\gamma^\text{max}\) instructions, and each CPU
starts programs at a rate of \(\rho\) program executions per unit of
time, the probability for any specific site to be relevant for the
execution of some running program at any point in time is
\(p_R\le(C/N_\text{sites})N_\gamma^\text{max}\tau_\gamma\rho\). The
probability for any specific site at any given point in time to be
affected by more than one running program is hence upper-bounded by
\(p_R^2\), and by making \(C\rho/N_\text{sites}\) sufficiently small, we
can suppress the impact of concurrent programs interfering with one
another relative to normal program execution arbitrarily small. (In a
chemical realization, we would generally expect concurrent overlapping
program execution that involves multiple instances of some molecular
execution machinery trying to work in the same space -- and getting into
each other's way -- to behave very differently from what we would get
when interleaving the execution of two programs on one memory region at
the per-instruction level.)

Intuitively, since program-execution will typically force some
particular state-change, and hence require entropy by making us forget
the original tape-state at the site of program-action (leaving us with
``uncertainty towards the past''), a program cannot execute by just
existing as an execution-sequence, it will need to be ``powered'' by
some entropy-creating process. In a chemical setting, one may imagine
(not to be taken too literally) program-executing sequences to be
``invoked'' by being zapped by (perhaps) a solar photon which then
powers the computational processes that release heat at environmental
heat bath temperature, effectively creating entropy by cracking up a
high energy photon. In this cartoon picture, the rate at which
incoming high-energy photons energize program-execution for any given
``CPU'' execution-engine translates to the program-starting rate
\(\rho\). One further aspect of this picture is that it makes us
naturally consider each single program invocation as having a budget
on the maximal number of instructions that can be executed in that
invocation. This is not an obstacle per se, since
finite-number-of-instructions programs can be chained by leaving
markers that are then picked up by continuation-programs that get
activated
later. Section~\ref{example-4-chemically-reversible-turing-machine}
has fully worked out examples for such a modeling approach. There, the
computational machinery is powered by energy-carrier molecules.

In the chemical ``very many sites'' limit, we get a description of time
evolution of the system in terms of an ordinary differential equation.
Since we are free to rescale our unit of time arbitrarily, we can pick a
natural scale that gives one unit of time an interpretation in terms of
how often a random site will be expected to get picked as tape
starting-position over unit time.

\emph{It is important to understand that one would not get a mathematically
consistent implementation by simply running the user-provided code on
each of the length-\(k\) sequences}. Pragmatically, this would already
fail to allow user code to probe the tape at any index relative to the
starting cursors (one for the P-tape segment, one for the D-type
segment). Instead, whenever the user's code needs to access tape-content
at position \(i\), the computational universe gets split into
sub-universes as needed, akin to how a way a quantum mechanical
measurement would split the multiverse into allowing different
experiment outcomes for observers in different realities -- here using
not quantum mechanics but the probabilistic (Markov) model for
tape-content at that point. In any such sub-universe, once the user code
finishes executing, we both know the exact probability for having landed
in such a situation, as well as the content of the P-tape and D-tape
stretches before and after execution of user code.

For each of the two stretches of tape that may have seen mutations (the
P-tape and D-tape), once user code reached its final state in some
universe, the generic situation on the tape will look as follows -- where
\texttt{V} means that tape-content has become visible, and \texttt{M}
means that additionally, a mutation has happened. If we keep track of
length-6 subsequence probabilities, then a post-execution tape-state as
shown below will require further world-splitting (and corresponding
probability-updates to cover nine different windows, as shown below:

\begin{verbatim}
Final:         ... V V M V V M V ...

         ...(? ? ? V V M)V V M V ...          #0
           ...(? ? V V M V)V M V ...          #1
             ...(? V V M V V)M V ...          #2
               ...(V V M V V M)V ...          #3
               ... V(V M V V M V)...          #4
               ... V V(M V V M V ?)...        #5
               ... V V M(V V M V ? ?)...      #6
               ... V V M V(V M V ? ? ?)...    #7
               ... V V M V V(M V ? ? ? ?)...  #8
\end{verbatim}

It may well happen that, as a consequence of multiverse-splitting,
different execution pathways uncover different parts of the tape. So, it
may happen that, depending on the content of the \texttt{V}-cells, some
other branch led to the following situation, which then would require
unfolding to cover only eight different windows:

\begin{verbatim}
Final:               ... M V M ...

          ....(? ? ? ? ? M)V M...              #0
            ....(? ? ? ? M V)M...              #1
               ...(? ? ? M V M)...             #2
                 ...(? ? M V M ?)...           #3
                   ...(? M V M ? ?)...         #4
                     ...(M V M ? ? ?)...       #5
                      ...M(V M ? ? ? ?)...     #6
                      ...M V(M ? ? ? ? ?)...   #7
\end{verbatim}

Here, care must be taken to do the extra unfolding to cover the cells
marked with \texttt{?} correctly. For \texttt{\#7} in this second
example, we could not simply start from the original content of the
cell holding the right \texttt{M}, and take all length-5 continuations
of that length-1 prefix given the known probabilities. Rather, one
possible correct approach is first to left-expand and ``measure'' two
more positions beyond the three known ones to have a full prefix of
five symbols, and then iteratively predict the subsequent symbol from
length-5 prefixes five times over.

At any endpoint of executing user code, any such further world-splitting
done to cover all subsequence-windows impacted by a change need not
utilize the costly \texttt{call-with-current-continuation} based
mechanism, but can instead use direct iteration intertwined with
recursion.

At the start of user-code evaluation, no tape-content (on the two
segments) is revealed, and the probability to be in a world like this is
1. Multiverse-splitting reduces this probability whenever
previously-unknown tape-state gets revealed. Once we know the full
initial and final P-tape and D-tape states
\(S_{P,i}, S_{D,i}; S_{P,f}, S_{D,f}\), world-probability is reduced to
\(p_w\), and for every tape-window, the rate-of-change for the
probability to encounter the observed window-content gets decreased by
\(p_w\) if this sequence-occurrence gets destroyed by the change, or
increased by \(p_w\) if it gets created.

This then implies that, if we were to start from an all-zeros binary
tape, and were to write a 1, using a subsequence length of 5, the
probability to discover a zero on the tape in this situation is 1, and
the probability rate-of-change for every 5-sequence containing a single
1 is 1, while the rate-of-change for the sequence \texttt{00000} is -5,
with five contributions \texttt{-1} from each of the
\(\mbox{\texttt{00000}}\to\mbox{\texttt{00001}}\),
\(\mbox{\texttt{00000}}\to\mbox{\texttt{00010}}\),
etc. adjustments. This sets the meaning of the time scale: over
\(\Delta t=1\), each site will be expected to get picked as a P-tape
starting cursor location once -- and likewise as a D-tape starting cursor
location.

Given that with this modeling approach, rate-changes generally will be
very smooth functions of concentrations, high-order Runge-Kutta
integration schemes such as the eighth-order DOP853 integrator
provided by SciPy~\cite{Virtanen2020} are expected to work well for
numerically integrating the time evolution.

\hypertarget{appendix-c-problem-definitions}{%
\section{Problem Definitions}\label{appendix-c-problem-definitions}}

This appendix contains the precise computational definitions of all
nontrivial example systems introduced in the main text. These
definitions not only precisely specify the system under study, but also
can serve as a starting point for exploring structurally similar
problems.

\hypertarget{c.1-radioactive-decay}{%
\subsection{Radioactive Decay}\label{c.1-radioactive-decay}}

For completeness, we repeat the definitions of the ``radioactive decay''
examples, without rate-adjustment:

{\small
\begin{Shaded}
\begin{Highlighting}[]
\CommentTok{;; Example: Radioactive Decay}
\NormalTok{(register{-}problem}
 \StringTok{"ex1{-}radioactive{-}decay"}
\NormalTok{ \#(A B)}
\NormalTok{ (}\KeywordTok{if}\NormalTok{ (}\KeywordTok{eq?}\NormalTok{ (tape{-}get{-}sym }\DecValTok{\#t} \DecValTok{0}\NormalTok{) \textquotesingle{}B)}
\NormalTok{     (tape{-}set{-}sym! }\DecValTok{\#t} \DecValTok{0}\NormalTok{ \textquotesingle{}A)))}
\end{Highlighting}
\end{Shaded}
}

as well as with rate-adjustment:

{\small
\begin{Shaded}
\begin{Highlighting}[]
\CommentTok{;; Example: Radioactive Decay (reduced{-}rate variant)}
\NormalTok{(register{-}problem}
 \StringTok{"ex1var1{-}radioactive{-}decay"}
\NormalTok{ \#(A B)}
\NormalTok{ (}\KeywordTok{if}\NormalTok{ (}\KeywordTok{and}\NormalTok{ (}\KeywordTok{eq?}\NormalTok{ (tape{-}get{-}sym }\DecValTok{\#t} \DecValTok{0}\NormalTok{) \textquotesingle{}B)}
\NormalTok{          (choose \textquotesingle{}((}\FloatTok{1.0} \DecValTok{\#t}\NormalTok{) (}\FloatTok{7.0} \DecValTok{\#f}\NormalTok{))))}
\NormalTok{     (tape{-}set{-}sym! }\DecValTok{\#t} \DecValTok{0}\NormalTok{ \textquotesingle{}A)))}
\end{Highlighting}
\end{Shaded}
}

\hypertarget{c.2-classical-ferromagnetic-spin-chain}{%
\subsection{Classical Ferromagnetic Spin Chain}\label{c.2-classical-ferromagnetic-spin-chain}}

{\small
\begin{Shaded}
\begin{Highlighting}[]
\NormalTok{(}\KeywordTok{let}\NormalTok{ ((param{-}J }\FloatTok{1.0}\NormalTok{)}
\NormalTok{      (param{-}h {-}}\FloatTok{0.25}\NormalTok{)}
\NormalTok{      (beta }\FloatTok{1.0}\NormalTok{)}
\NormalTok{      )}
\NormalTok{  (register{-}problem}
   \StringTok{"ex2{-}ferromagnetic{-}chain"}
\NormalTok{   \#(D U)}
\NormalTok{   (}\KeywordTok{let}\NormalTok{ ((p{-}mid (tape{-}get{-}sym }\DecValTok{\#t} \DecValTok{0}\NormalTok{))}
\NormalTok{         (p{-}left (tape{-}get{-}sym }\DecValTok{\#t}\NormalTok{ {-}}\DecValTok{1}\NormalTok{))}
\NormalTok{         (p{-}right (tape{-}get{-}sym }\DecValTok{\#t}\NormalTok{ +}\DecValTok{1}\NormalTok{)))}
\NormalTok{     (}\KeywordTok{let*}\NormalTok{ ((energy{-}J (}\OperatorTok{+}\NormalTok{ (}\KeywordTok{if}\NormalTok{ (}\KeywordTok{eq?}\NormalTok{ p{-}left p{-}mid) }\DecValTok{1}\NormalTok{ {-}}\DecValTok{1}\NormalTok{)}
\NormalTok{                         (}\KeywordTok{if}\NormalTok{ (}\KeywordTok{eq?}\NormalTok{ p{-}mid p{-}right) }\DecValTok{1}\NormalTok{ {-}}\DecValTok{1}\NormalTok{)))}
\NormalTok{            (factor{-}a (}\KeywordTok{exp}\NormalTok{ (}\OperatorTok{{-}}\NormalTok{ (}\OperatorTok{*}\NormalTok{ beta param{-}J (}\OperatorTok{+} \DecValTok{4}\NormalTok{ (}\OperatorTok{*} \DecValTok{2}\NormalTok{ energy{-}J))))))}
\NormalTok{            (factor{-}b}
\NormalTok{             (}\KeywordTok{if}\NormalTok{ (}\KeywordTok{eqv?}\NormalTok{ (}\OperatorTok{\textgreater{}}\NormalTok{ param{-}h }\DecValTok{0}\NormalTok{) (}\KeywordTok{eq?}\NormalTok{ p{-}mid \textquotesingle{}U))}
                 \CommentTok{;; Either field is pushing up, and we are in an}
                 \CommentTok{;; up{-}configuration, or field is pushing down,}
                 \CommentTok{;; and we are in a down{-}configuration.}
                 \CommentTok{;; In both cases, flip{-}rate has to be suppressed.}
\NormalTok{                 (}\KeywordTok{exp}\NormalTok{ (}\OperatorTok{{-}}\NormalTok{ (}\OperatorTok{*} \DecValTok{2}\NormalTok{ beta (}\KeywordTok{abs}\NormalTok{ param{-}h))))}
                 \FloatTok{1.0}\NormalTok{))}
\NormalTok{            (p{-}flip (}\OperatorTok{*}\NormalTok{ factor{-}a factor{-}b))}
\NormalTok{            (p{-}stay (}\OperatorTok{{-}} \DecValTok{1}\NormalTok{ p{-}flip)))}
\NormalTok{       (}\KeywordTok{if}\NormalTok{ (choose \textasciigrave{}((,p{-}flip }\DecValTok{\#t}\NormalTok{) (,p{-}stay }\DecValTok{\#f}\NormalTok{)))}
\NormalTok{           (tape{-}set{-}sym! }\DecValTok{\#t} \DecValTok{0}\NormalTok{ (}\KeywordTok{if}\NormalTok{ (}\KeywordTok{eq?}\NormalTok{ p{-}mid \textquotesingle{}U) \textquotesingle{}D \textquotesingle{}U))}
           \DecValTok{\#t}\NormalTok{)))))}
\end{Highlighting}
\end{Shaded}
}

\hypertarget{c.3-co-polymerization}{%
\subsection{Co-Polymerization}\label{c.3-co-polymerization}}

Basic problem:

{\small
\begin{Shaded}
\begin{Highlighting}[]
\CommentTok{;; Example: "Nylon copolymerization"}
\NormalTok{(register{-}problem}
 \StringTok{"ex3{-}copolymerization"}
\NormalTok{ \#(O A M N)}
\NormalTok{ (}\KeywordTok{let}\NormalTok{ ((p0 (tape{-}get{-}sym }\DecValTok{\#f} \DecValTok{0}\NormalTok{)))}
\NormalTok{   (}\KeywordTok{if}\NormalTok{ (}\KeywordTok{and}\NormalTok{ (}\KeywordTok{not}\NormalTok{ (}\KeywordTok{eq?}\NormalTok{ p0 \textquotesingle{}O))}
\NormalTok{            (}\KeywordTok{eq?}\NormalTok{ (tape{-}get{-}sym }\DecValTok{\#f}\NormalTok{ {-}}\DecValTok{1}\NormalTok{) \textquotesingle{}O)}
\NormalTok{            (}\KeywordTok{eq?}\NormalTok{ (tape{-}get{-}sym }\DecValTok{\#f}\NormalTok{ +}\DecValTok{1}\NormalTok{) \textquotesingle{}O))}
       \CommentTok{;; We have an isolated monomer on the P{-}tape.}
\NormalTok{       (}\KeywordTok{let}\NormalTok{ ((d0 (tape{-}get{-}sym }\DecValTok{\#t} \DecValTok{0}\NormalTok{)))}
\NormalTok{         (}\KeywordTok{if}\NormalTok{ (}\KeywordTok{or}\NormalTok{ (}\KeywordTok{and}\NormalTok{ (}\KeywordTok{eq?}\NormalTok{ p0 \textquotesingle{}A) (}\KeywordTok{or}\NormalTok{ (}\KeywordTok{eq?}\NormalTok{ d0 \textquotesingle{}M) (}\KeywordTok{eq?}\NormalTok{ d0 \textquotesingle{}N)))}
\NormalTok{                 (}\KeywordTok{and}\NormalTok{ (}\KeywordTok{eq?}\NormalTok{ d0 \textquotesingle{}A) (}\KeywordTok{or}\NormalTok{ (}\KeywordTok{eq?}\NormalTok{ p0 \textquotesingle{}M) (}\KeywordTok{eq?}\NormalTok{ p0 \textquotesingle{}N))))}
             \CommentTok{;; We have compatible monomers.}
             \CommentTok{;; Need to see if we have an available end to connect to.}
             \CommentTok{;; We try both sides with equal probability.}
             \CommentTok{;; If there is no open end on one side, we give up.}
\NormalTok{             (}\KeywordTok{let*}\NormalTok{ ((i (choose \textquotesingle{}((}\FloatTok{1.0}\NormalTok{ {-}}\DecValTok{1}\NormalTok{) (}\FloatTok{1.0}\NormalTok{ +}\DecValTok{1}\NormalTok{))))}
\NormalTok{                    (di (tape{-}get{-}sym }\DecValTok{\#t}\NormalTok{ i)))}
\NormalTok{               (}\KeywordTok{if}\NormalTok{ (}\KeywordTok{and}\NormalTok{ (}\KeywordTok{eq?}\NormalTok{ di \textquotesingle{}O)}
                        \CommentTok{;; post{-}addition, we also have an endpoint,}
                        \CommentTok{;; i.e. this is not an accidental place where}
                        \CommentTok{;; two chain{-}ends meet.}
\NormalTok{                        (}\KeywordTok{eq?}\NormalTok{ (tape{-}get{-}sym }\DecValTok{\#t}\NormalTok{ (}\OperatorTok{*} \DecValTok{2}\NormalTok{ i)) \textquotesingle{}O))}
\NormalTok{                   (}\KeywordTok{begin}
                     \CommentTok{;; Remove the monomer from the P{-}tape.}
\NormalTok{                     (tape{-}set{-}sym! }\DecValTok{\#f} \DecValTok{0}\NormalTok{ \textquotesingle{}O)}
                     \CommentTok{;; Connect the monomer on the D{-}tape.}
\NormalTok{                     (tape{-}set{-}sym! }\DecValTok{\#t}\NormalTok{ i p0)))))))))}
\end{Highlighting}
\end{Shaded}
}

Variant 1 (slight preference for \texttt{AMANAMAN} alternation):

{\small
\begin{Shaded}
\begin{Highlighting}[]
\CommentTok{;; Example: "Nylon copolymerization" (asymmetric variant)}
\NormalTok{(register{-}problem}
 \StringTok{"ex3var1{-}copolymerization"}
\NormalTok{ \#(O A M N)}
\NormalTok{ (}\KeywordTok{let}\NormalTok{ ((p0 (tape{-}get{-}sym }\DecValTok{\#f} \DecValTok{0}\NormalTok{)))}
\NormalTok{   (}\KeywordTok{if}\NormalTok{ (}\KeywordTok{and}\NormalTok{ (}\KeywordTok{not}\NormalTok{ (}\KeywordTok{eq?}\NormalTok{ p0 \textquotesingle{}O))}
\NormalTok{            (}\KeywordTok{eq?}\NormalTok{ (tape{-}get{-}sym }\DecValTok{\#f}\NormalTok{ {-}}\DecValTok{1}\NormalTok{) \textquotesingle{}O)}
\NormalTok{            (}\KeywordTok{eq?}\NormalTok{ (tape{-}get{-}sym }\DecValTok{\#f}\NormalTok{ +}\DecValTok{1}\NormalTok{) \textquotesingle{}O))}
       \CommentTok{;; We have an isolated monomer on the P{-}tape.}
\NormalTok{       (}\KeywordTok{let}\NormalTok{ ((d0 (tape{-}get{-}sym }\DecValTok{\#t} \DecValTok{0}\NormalTok{)))}
\NormalTok{         (}\KeywordTok{if}\NormalTok{ (}\KeywordTok{or}\NormalTok{ (}\KeywordTok{and}\NormalTok{ (}\KeywordTok{eq?}\NormalTok{ p0 \textquotesingle{}A) (}\KeywordTok{or}\NormalTok{ (}\KeywordTok{eq?}\NormalTok{ d0 \textquotesingle{}M) (}\KeywordTok{eq?}\NormalTok{ d0 \textquotesingle{}N)))}
\NormalTok{                 (}\KeywordTok{and}\NormalTok{ (}\KeywordTok{eq?}\NormalTok{ d0 \textquotesingle{}A) (}\KeywordTok{or}\NormalTok{ (}\KeywordTok{eq?}\NormalTok{ p0 \textquotesingle{}M) (}\KeywordTok{eq?}\NormalTok{ p0 \textquotesingle{}N))))}
             \CommentTok{;; We have compatible monomers.}
             \CommentTok{;; Need to see if we have an available end to connect to.}
             \CommentTok{;; We try both sides with equal probability.}
             \CommentTok{;; If there is no open end on one side, we give up.}
\NormalTok{             (}\KeywordTok{let*}\NormalTok{ ((i (choose \textquotesingle{}((}\FloatTok{1.0}\NormalTok{ {-}}\DecValTok{1}\NormalTok{) (}\FloatTok{1.0}\NormalTok{ +}\DecValTok{1}\NormalTok{))))}
\NormalTok{                    (di (tape{-}get{-}sym }\DecValTok{\#t}\NormalTok{ i)))}
\NormalTok{               (}\KeywordTok{if}\NormalTok{ (}\KeywordTok{and}\NormalTok{ (}\KeywordTok{eq?}\NormalTok{ di \textquotesingle{}O)}
                        \CommentTok{;; post{-}addition, we also have an endpoint,}
                        \CommentTok{;; i.e. this is not an accidental place where}
                        \CommentTok{;; two chain{-}ends meet.}
\NormalTok{                        (}\KeywordTok{eq?}\NormalTok{ (tape{-}get{-}sym }\DecValTok{\#t}\NormalTok{ (}\OperatorTok{*} \DecValTok{2}\NormalTok{ i)) \textquotesingle{}O))}
                   \CommentTok{;; if the opposite{-}side of the D{-}tape unit is of the same}
                   \CommentTok{;; type as the new unit, prevent the reaction with some}
                   \CommentTok{;; probability.}
\NormalTok{                   (}\KeywordTok{if}\NormalTok{ (}\KeywordTok{and}
                        \CommentTok{;; New unit is M or N.}
\NormalTok{                        (}\KeywordTok{not}\NormalTok{ (}\KeywordTok{eq?}\NormalTok{ p0 \textquotesingle{}A))}
                        \CommentTok{;; Same unit already on other side of A.}
\NormalTok{                        (}\KeywordTok{eq?}\NormalTok{ (tape{-}get{-}sym }\DecValTok{\#t}\NormalTok{ (}\OperatorTok{{-}}\NormalTok{ i)) p0)}
                        \CommentTok{;; High rejection rate for this case. {-} XXX transfer to paper}
\NormalTok{                        (choose \textquotesingle{}((}\FloatTok{75.0} \DecValTok{\#t}\NormalTok{) (}\FloatTok{25.0} \DecValTok{\#f}\NormalTok{))))}
                       \DecValTok{\#f}  \CommentTok{; do nothing}
\NormalTok{                       (}\KeywordTok{begin}  \CommentTok{; otherwise, react as before.}
                         \CommentTok{;; Remove the monomer from the P{-}tape.}
\NormalTok{                         (tape{-}set{-}sym! }\DecValTok{\#f} \DecValTok{0}\NormalTok{ \textquotesingle{}O)}
                         \CommentTok{;; Connect the monomer on the D{-}tape.}
\NormalTok{                         (tape{-}set{-}sym! }\DecValTok{\#t}\NormalTok{ i p0))))))))))}
\end{Highlighting}
\end{Shaded}
}

Variant 2 (reversible reactions):

{\small
\begin{Shaded}
\begin{Highlighting}[]
\CommentTok{;; Example: "Nylon copolymerization" (reversible variant)}
\NormalTok{(register{-}problem}
 \CommentTok{;; Sequences \textasciigrave{}O[AMN]?...\textasciigrave{} and \textasciigrave{}...[AMN]?O\textasciigrave{} plus \textasciigrave{}OOO\textasciigrave{} can depolymerize,}
 \CommentTok{;; but with low relative rate.}
 \CommentTok{;; Rate{-}ratio \{polymerization\} : \{depolymerization\} is related to}
 \CommentTok{;; the free enthalpy of the reaction, i.e. thermodynamic stability}
 \CommentTok{;; of the polymer relative to monomers.}
 \StringTok{"ex3var2{-}copolymerization"}
\NormalTok{ \#(O A M N)}
\NormalTok{ (}\KeywordTok{let}\NormalTok{ ((p0 (tape{-}get{-}sym }\DecValTok{\#f} \DecValTok{0}\NormalTok{)))}
\NormalTok{   (}\KeywordTok{if}
\NormalTok{    (}\KeywordTok{eq?}\NormalTok{ p0 \textquotesingle{}O)}
    \CommentTok{;; "program{-}tape" cell 0 is empty, try dissociation.}
\NormalTok{    (}\KeywordTok{if}\NormalTok{ (}\KeywordTok{and}\NormalTok{ (}\KeywordTok{eq?}\NormalTok{ (tape{-}get{-}sym }\DecValTok{\#f}\NormalTok{ {-}}\DecValTok{1}\NormalTok{) \textquotesingle{}O)}
\NormalTok{             (}\KeywordTok{eq?}\NormalTok{ (tape{-}get{-}sym }\DecValTok{\#f}\NormalTok{ +}\DecValTok{1}\NormalTok{) \textquotesingle{}O))}
        \CommentTok{;; We have free space on the P{-}tape.}
\NormalTok{        (}\KeywordTok{let}\NormalTok{ ((d0 (tape{-}get{-}sym }\DecValTok{\#t} \DecValTok{0}\NormalTok{)))}
\NormalTok{          (}\KeywordTok{if}\NormalTok{ (}\KeywordTok{not}\NormalTok{ (}\KeywordTok{eq?}\NormalTok{ d0 \textquotesingle{}O))}
\NormalTok{              (}\KeywordTok{let}\NormalTok{ ((d1{-}right (tape{-}get{-}sym }\DecValTok{\#t} \DecValTok{1}\NormalTok{))}
\NormalTok{                    (d1{-}left (tape{-}get{-}sym }\DecValTok{\#t}\NormalTok{ {-}}\DecValTok{11}\NormalTok{)))}
\NormalTok{                (}\KeywordTok{if}\NormalTok{ (}\OperatorTok{=} \DecValTok{1}\NormalTok{ (}\OperatorTok{+}\NormalTok{ (}\KeywordTok{if}\NormalTok{ (}\KeywordTok{eq?}\NormalTok{ d1{-}left \textquotesingle{}O) }\DecValTok{0} \DecValTok{1}\NormalTok{)}
\NormalTok{                            (}\KeywordTok{if}\NormalTok{ (}\KeywordTok{eq?}\NormalTok{ d1{-}right \textquotesingle{}O) }\DecValTok{0} \DecValTok{1}\NormalTok{)))}
                    \CommentTok{;; We are at the end of a chain.}
                    \CommentTok{;; Depolymerization happens at a reduced rate,}
                    \CommentTok{;; since we take the polymer to be thermodynamically}
                    \CommentTok{;; more stable than the monomers.}
\NormalTok{                    (}\KeywordTok{if}\NormalTok{ (choose \textquotesingle{}((}\FloatTok{1.0} \DecValTok{\#t}\NormalTok{) (}\FloatTok{50.0} \DecValTok{\#f}\NormalTok{)))}
\NormalTok{                        (}\KeywordTok{begin}
\NormalTok{                          (tape{-}set{-}sym! }\DecValTok{\#f} \DecValTok{0}\NormalTok{ d0)}
\NormalTok{                          (tape{-}set{-}sym! }\DecValTok{\#t} \DecValTok{0}\NormalTok{ \textquotesingle{}O))))))))}
    \CommentTok{;; else, "program{-}tape" cell 0 is not empty, try polycondensation.}
\NormalTok{    (}\KeywordTok{if}\NormalTok{ (}\KeywordTok{and}\NormalTok{ (}\KeywordTok{eq?}\NormalTok{ (tape{-}get{-}sym }\DecValTok{\#f}\NormalTok{ {-}}\DecValTok{1}\NormalTok{) \textquotesingle{}O)}
\NormalTok{             (}\KeywordTok{eq?}\NormalTok{ (tape{-}get{-}sym }\DecValTok{\#f}\NormalTok{ +}\DecValTok{1}\NormalTok{) \textquotesingle{}O))}
        \CommentTok{;; We have an isolated monomer on the P{-}tape.}
\NormalTok{        (}\KeywordTok{let}\NormalTok{ ((d0 (tape{-}get{-}sym }\DecValTok{\#t} \DecValTok{0}\NormalTok{)))}
\NormalTok{          (}\KeywordTok{if}\NormalTok{ (}\KeywordTok{or}\NormalTok{ (}\KeywordTok{and}\NormalTok{ (}\KeywordTok{eq?}\NormalTok{ p0 \textquotesingle{}A) (}\KeywordTok{or}\NormalTok{ (}\KeywordTok{eq?}\NormalTok{ d0 \textquotesingle{}M) (}\KeywordTok{eq?}\NormalTok{ d0 \textquotesingle{}N)))}
\NormalTok{                  (}\KeywordTok{and}\NormalTok{ (}\KeywordTok{eq?}\NormalTok{ d0 \textquotesingle{}A) (}\KeywordTok{or}\NormalTok{ (}\KeywordTok{eq?}\NormalTok{ p0 \textquotesingle{}M) (}\KeywordTok{eq?}\NormalTok{ p0 \textquotesingle{}N))))}
              \CommentTok{;; We have compatible monomers.}
              \CommentTok{;; Need to see if we have an available end to connect to.}
              \CommentTok{;; We try both sides with equal probability.}
              \CommentTok{;; If there is no open end on one side, we give up.}
\NormalTok{              (}\KeywordTok{let*}\NormalTok{ ((i (choose \textquotesingle{}((}\FloatTok{1.0}\NormalTok{ {-}}\DecValTok{1}\NormalTok{) (}\FloatTok{1.0}\NormalTok{ +}\DecValTok{1}\NormalTok{))))}
\NormalTok{                     (di (tape{-}get{-}sym }\DecValTok{\#t}\NormalTok{ i)))}
\NormalTok{                (}\KeywordTok{if}\NormalTok{ (}\KeywordTok{and}\NormalTok{ (}\KeywordTok{eq?}\NormalTok{ di \textquotesingle{}O)}
                         \CommentTok{;; post{-}addition, we also have an endpoint,}
                         \CommentTok{;; i.e. this is not an accidental place where}
                         \CommentTok{;; two chain{-}ends meet.}
\NormalTok{                         (}\KeywordTok{eq?}\NormalTok{ (tape{-}get{-}sym }\DecValTok{\#t}\NormalTok{ (}\OperatorTok{*} \DecValTok{2}\NormalTok{ i)) \textquotesingle{}O))}
\NormalTok{                    (}\KeywordTok{begin}
                      \CommentTok{;; Remove the monomer from the P{-}tape.}
\NormalTok{                      (tape{-}set{-}sym! }\DecValTok{\#f} \DecValTok{0}\NormalTok{ \textquotesingle{}O)}
                      \CommentTok{;; Connect the monomer on the D{-}tape.}
\NormalTok{                      (tape{-}set{-}sym! }\DecValTok{\#t}\NormalTok{ i p0))))))))))}
\end{Highlighting}
\end{Shaded}
}

\hypertarget{c.4-chemical-turing-machine}{%
\subsection{Chemical Turing Machine}\label{c.4-chemical-turing-machine}}

Basic problem:

{\small
\begin{Shaded}
\begin{Highlighting}[]
\CommentTok{;; Example: Basic Turing Machine}
\CommentTok{;; We can move some definitions out of problem{-}registration.}
\NormalTok{(}\KeywordTok{let*}\NormalTok{ ((is{-}io? (}\KeywordTok{lambda}\NormalTok{ (x) (}\KeywordTok{or}\NormalTok{ (}\KeywordTok{eq?}\NormalTok{ x \textquotesingle{}I) (}\KeywordTok{eq?}\NormalTok{ x \textquotesingle{}O))))}
\NormalTok{       (px{-}relative{-}stability{-}reverse{-}suppression{-}factor }\FloatTok{0.05}\NormalTok{)}
\NormalTok{       (px{-}reverse{-}suppression{-}choices}
\NormalTok{        \textasciigrave{}((,(}\OperatorTok{{-}} \FloatTok{1.0}\NormalTok{ px{-}relative{-}stability{-}reverse{-}suppression{-}factor) }\DecValTok{\#f}\NormalTok{)}
\NormalTok{          (,px{-}relative{-}stability{-}reverse{-}suppression{-}factor }\DecValTok{\#t}\NormalTok{))))}
\NormalTok{  (register{-}problem}
   \StringTok{"ex4{-}chemical{-}turing"}
\NormalTok{   \#(A B C D I O P X S)  }\CommentTok{; S = Solvent, P = Powered, X = De{-}Powered.}
\NormalTok{   (}\KeywordTok{let}\NormalTok{ ((p0 (tape{-}get{-}sym }\DecValTok{\#f} \DecValTok{0}\NormalTok{)))}
\NormalTok{     (}\KeywordTok{cond}
\NormalTok{      ((}\KeywordTok{and}\NormalTok{ (}\KeywordTok{eq?}\NormalTok{ p0 \textquotesingle{}P)  }\CommentTok{; powered{-}\textgreater{}de{-}powered}
            \CommentTok{;; We need to suppress this by a factor 2, since otherwise,}
            \CommentTok{;; back{-} and forward{-} reaction constants would not be the same at}
            \CommentTok{;; px{-}relative{-}stability{-}reverse{-}suppression{-}factor = 0.}
\NormalTok{            (choose \textquotesingle{}((}\FloatTok{1.0} \DecValTok{\#t}\NormalTok{) (}\FloatTok{1.0} \DecValTok{\#f}\NormalTok{)))}
\NormalTok{            )}
\NormalTok{       (}\KeywordTok{let}\NormalTok{ ((d0 (tape{-}get{-}sym }\DecValTok{\#t} \DecValTok{0}\NormalTok{)))}
\NormalTok{         (}\KeywordTok{cond}
\NormalTok{          ((}\KeywordTok{and}\NormalTok{ (}\KeywordTok{eq?}\NormalTok{ d0 \textquotesingle{}A)}
                \CommentTok{;; Can we advance?}
\NormalTok{                (is{-}io? (tape{-}get{-}sym }\DecValTok{\#t} \DecValTok{1}\NormalTok{))}
                \CommentTok{;; Post{-}advancement, we again have to be in a valid state}
                \CommentTok{;; where the cursor{-}marker is just before an I or O.}
\NormalTok{                (is{-}io? (tape{-}get{-}sym }\DecValTok{\#t} \DecValTok{2}\NormalTok{))                  }
\NormalTok{                )}
\NormalTok{           (tape{-}set{-}sym! }\DecValTok{\#f} \DecValTok{0}\NormalTok{ \textquotesingle{}X)             }
\NormalTok{           (tape{-}set{-}sym! }\DecValTok{\#t} \DecValTok{0}\NormalTok{ \textquotesingle{}I)}
\NormalTok{           (tape{-}set{-}sym! }\DecValTok{\#t} \DecValTok{1}\NormalTok{ \textquotesingle{}B))}
\NormalTok{          ((}\KeywordTok{and}\NormalTok{ (}\KeywordTok{eq?}\NormalTok{ d0 \textquotesingle{}B)}
\NormalTok{                (is{-}io? (tape{-}get{-}sym }\DecValTok{\#t} \DecValTok{1}\NormalTok{))}
\NormalTok{                (is{-}io? (tape{-}get{-}sym }\DecValTok{\#t} \DecValTok{2}\NormalTok{)))}
\NormalTok{           (tape{-}set{-}sym! }\DecValTok{\#f} \DecValTok{0}\NormalTok{ \textquotesingle{}X)                }
\NormalTok{           (tape{-}set{-}sym! }\DecValTok{\#t} \DecValTok{0}\NormalTok{ \textquotesingle{}O)}
\NormalTok{           (tape{-}set{-}sym! }\DecValTok{\#t} \DecValTok{1}\NormalTok{ \textquotesingle{}C))}
\NormalTok{          ((}\KeywordTok{and}\NormalTok{ (}\KeywordTok{eq?}\NormalTok{ d0 \textquotesingle{}C)}
\NormalTok{                (is{-}io? (tape{-}get{-}sym }\DecValTok{\#t} \DecValTok{1}\NormalTok{))}
\NormalTok{                (is{-}io? (tape{-}get{-}sym }\DecValTok{\#t} \DecValTok{2}\NormalTok{)))}
\NormalTok{           (tape{-}set{-}sym! }\DecValTok{\#f} \DecValTok{0}\NormalTok{ \textquotesingle{}X)  }
\NormalTok{           (tape{-}set{-}sym! }\DecValTok{\#t} \DecValTok{0}\NormalTok{ \textquotesingle{}I)}
\NormalTok{           (tape{-}set{-}sym! }\DecValTok{\#t} \DecValTok{1}\NormalTok{ \textquotesingle{}D)))))}
\NormalTok{      ((}\KeywordTok{eq?}\NormalTok{ p0 \textquotesingle{}X)  }\CommentTok{; de{-}powered{-}\textgreater{}powered}
\NormalTok{       (}\KeywordTok{let}\NormalTok{ ((d0 (tape{-}get{-}sym }\DecValTok{\#t} \DecValTok{0}\NormalTok{)))}
\NormalTok{         (}\KeywordTok{if}\NormalTok{ (}\KeywordTok{and}\NormalTok{ (}\KeywordTok{or}\NormalTok{ (}\KeywordTok{eq?}\NormalTok{ d0 \textquotesingle{}B) (}\KeywordTok{eq?}\NormalTok{ d0 \textquotesingle{}C) (}\KeywordTok{eq?}\NormalTok{ d0 \textquotesingle{}D))}
\NormalTok{                  (is{-}io? (tape{-}get{-}sym }\DecValTok{\#t}\NormalTok{ {-}}\DecValTok{1}\NormalTok{))  }\CommentTok{; Can move back}
\NormalTok{                  (is{-}io? (tape{-}get{-}sym }\DecValTok{\#t}\NormalTok{ {-}}\DecValTok{2}\NormalTok{))  }\CommentTok{; Won\textquotesingle{}t move next to a cursor.}
                  \CommentTok{;; Also, the previous symbol needs to be compatible with}
                  \CommentTok{;; the forward{-}reaction end{-}state.}
\NormalTok{                  (}\KeywordTok{or}\NormalTok{ (}\KeywordTok{and}\NormalTok{ (}\KeywordTok{eq?}\NormalTok{ d0 \textquotesingle{}C) (}\KeywordTok{eq?}\NormalTok{ (tape{-}get{-}sym }\DecValTok{\#t}\NormalTok{ {-}}\DecValTok{1}\NormalTok{) \textquotesingle{}O))}
\NormalTok{                      (}\KeywordTok{and}\NormalTok{ (}\KeywordTok{not}\NormalTok{ (}\KeywordTok{eq?}\NormalTok{ d0 \textquotesingle{}C)) (}\KeywordTok{eq?}\NormalTok{ (tape{-}get{-}sym }\DecValTok{\#t}\NormalTok{ {-}}\DecValTok{1}\NormalTok{) \textquotesingle{}I)))}
                  \CommentTok{;; If P is thermodynamically more stable than X, we need to}
                  \CommentTok{;; further suppress this reaction.}
\NormalTok{                  (choose px{-}reverse{-}suppression{-}choices))}
\NormalTok{             (}\KeywordTok{begin}
\NormalTok{               (tape{-}set{-}sym! }\DecValTok{\#f} \DecValTok{0}\NormalTok{ \textquotesingle{}P)}
\NormalTok{               (tape{-}set{-}sym! }\DecValTok{\#t} \DecValTok{0}\NormalTok{ (choose \textquotesingle{}((}\FloatTok{1.0}\NormalTok{ I) (}\FloatTok{1.0}\NormalTok{ O))))}
\NormalTok{               (tape{-}set{-}sym! }\DecValTok{\#t}\NormalTok{ {-}}\DecValTok{1}
\NormalTok{                              (}\KeywordTok{cond}\NormalTok{ ((}\KeywordTok{eq?}\NormalTok{ d0 \textquotesingle{}B) \textquotesingle{}A)}
\NormalTok{                                    ((}\KeywordTok{eq?}\NormalTok{ d0 \textquotesingle{}C) \textquotesingle{}B)}
\NormalTok{                                    ((}\KeywordTok{eq?}\NormalTok{ d0 \textquotesingle{}D) \textquotesingle{}C)))))))))))}
\end{Highlighting}
\end{Shaded}
}

Variant 1 (equal thermodynamic stability): The definition can be
obtained from the above one by setting
\texttt{(px-relative-stability-reverse-suppression-factor\ 0.0)} and
changing the registration name to \texttt{"ex4var1-chemical-turing"} (to
align with the Python code published alongside this article).

Variant 2 (detachable evaluators): This somewhat complex example is
expected to be instructive for defining similar models where ensuring
that reaction rates are in alignment with thermodynamics is best
realized by starting from Gibbs free energies of formation.

{\small
\begin{Shaded}
\begin{Highlighting}[]
\CommentTok{;; Variant 2: Detachable Evaluator, Symbol{-}Under{-}Cursor.}
\NormalTok{(}\KeywordTok{let*}\NormalTok{ ((is{-}io? (}\KeywordTok{lambda}\NormalTok{ (x) (}\KeywordTok{or}\NormalTok{ (}\KeywordTok{eq?}\NormalTok{ x \textquotesingle{}I) (}\KeywordTok{eq?}\NormalTok{ x \textquotesingle{}O))))}
\NormalTok{       (choice{-}IO \textquotesingle{}((}\FloatTok{1.0}\NormalTok{ I) (}\FloatTok{1.0}\NormalTok{ O)))}
\NormalTok{       (choice{-}1:1 \textquotesingle{}((}\FloatTok{1.0} \DecValTok{\#t}\NormalTok{) (}\FloatTok{1.0} \DecValTok{\#f}\NormalTok{)))}
\NormalTok{       (beta }\FloatTok{1.0}\NormalTok{)  }\CommentTok{; Adjustable 1/(k\_B T) factor.}
       \CommentTok{;; Free enthalpies of formation. A large G{-}E disfavors}
       \CommentTok{;; evaluators in solution.}
\NormalTok{       (G{-}P }\FloatTok{6.0}\NormalTok{) (G{-}X }\FloatTok{0.0}\NormalTok{) (G{-}E }\FloatTok{1.0}\NormalTok{)}
       \CommentTok{;; Final D{-}state must be thermodynamically less stable}
       \CommentTok{;; if we want D to detach more easily to E than A.}
\NormalTok{       (G{-}A {-}}\FloatTok{1.0}\NormalTok{) (G{-}B {-}}\FloatTok{1.0}\NormalTok{) (G{-}C {-}}\FloatTok{1.0}\NormalTok{) (G{-}D }\FloatTok{1.5}\NormalTok{)}
       \CommentTok{;; With these parameters, the fastest reactions are the A+P{-}\textgreater{}B+X}
       \CommentTok{;; type reactions. }
\NormalTok{       (Delta{-}G{-}fastest (}\OperatorTok{{-}}\NormalTok{ (}\OperatorTok{+}\NormalTok{ G{-}B G{-}X) (}\OperatorTok{+}\NormalTok{ G{-}A G{-}P)))}
\NormalTok{       (get{-}rate{-}factor}
\NormalTok{        (}\KeywordTok{lambda}\NormalTok{ (G{-}left G{-}right)}
\NormalTok{          (}\KeywordTok{let}\NormalTok{ ((rate{-}factor}
\NormalTok{                 (}\KeywordTok{exp}\NormalTok{ (}\OperatorTok{{-}}\NormalTok{ (}\OperatorTok{*}\NormalTok{ beta (}\OperatorTok{{-}}\NormalTok{ G{-}right G{-}left Delta{-}G{-}fastest))))))}
\NormalTok{            (}\KeywordTok{if}\NormalTok{ (}\OperatorTok{\textgreater{}}\NormalTok{ rate{-}factor }\FloatTok{1.001}\NormalTok{)}
\NormalTok{                (}\KeywordTok{error}
                 \StringTok{"Setup error: Delta{-}G{-}fastest not actually fastest."}\NormalTok{)}
\NormalTok{                (}\KeywordTok{min} \FloatTok{1.0}\NormalTok{ rate{-}factor)))))}
\NormalTok{       (rate{-}choices}
\NormalTok{        (}\KeywordTok{lambda}\NormalTok{ (G{-}left G{-}right)}
\NormalTok{          (}\KeywordTok{let}\NormalTok{ ((r (get{-}rate{-}factor G{-}left G{-}right)))}
\NormalTok{            \textasciigrave{}((,r }\DecValTok{\#t}\NormalTok{) (,(}\OperatorTok{{-}} \DecValTok{1}\NormalTok{ r) }\DecValTok{\#f}\NormalTok{)))))}
\NormalTok{       (rate{-}choices{-}A+P{-}\textgreater{}B+X (rate{-}choices (}\OperatorTok{+}\NormalTok{ G{-}A G{-}P) (}\OperatorTok{+}\NormalTok{ G{-}B G{-}X)))}
\NormalTok{       (rate{-}choices{-}B+X{-}\textgreater{}A+P (rate{-}choices (}\OperatorTok{+}\NormalTok{ G{-}B G{-}X) (}\OperatorTok{+}\NormalTok{ G{-}A G{-}P)))}
\NormalTok{       (rate{-}choices{-}B+P{-}\textgreater{}C+X (rate{-}choices (}\OperatorTok{+}\NormalTok{ G{-}B G{-}P) (}\OperatorTok{+}\NormalTok{ G{-}C G{-}X)))}
\NormalTok{       (rate{-}choices{-}C+X{-}\textgreater{}B+P (rate{-}choices (}\OperatorTok{+}\NormalTok{ G{-}C G{-}X) (}\OperatorTok{+}\NormalTok{ G{-}B G{-}P)))}
\NormalTok{       (rate{-}choices{-}C+P{-}\textgreater{}D+X (rate{-}choices (}\OperatorTok{+}\NormalTok{ G{-}C G{-}P) (}\OperatorTok{+}\NormalTok{ G{-}D G{-}X)))}
\NormalTok{       (rate{-}choices{-}D+X{-}\textgreater{}C+P (rate{-}choices (}\OperatorTok{+}\NormalTok{ G{-}D G{-}X) (}\OperatorTok{+}\NormalTok{ G{-}C G{-}P)))}
\NormalTok{       (rate{-}choices{-}A{-}\textgreater{}E (rate{-}choices G{-}A G{-}E))}
\NormalTok{       (rate{-}choices{-}D{-}\textgreater{}E (rate{-}choices G{-}D G{-}E))                  }
\NormalTok{       (rate{-}choices{-}E{-}\textgreater{}A+D}
\NormalTok{        (}\KeywordTok{let}\NormalTok{ ((r{-}A (get{-}rate{-}factor G{-}E G{-}A))}
\NormalTok{              (r{-}D (get{-}rate{-}factor G{-}E G{-}D)))}
\NormalTok{          (}\KeywordTok{if}\NormalTok{ (}\OperatorTok{\textgreater{}}\NormalTok{ (}\OperatorTok{+}\NormalTok{ r{-}A r{-}D) }\FloatTok{1.0}\NormalTok{)}
              \CommentTok{;; In order to handle this case, we would have to set}
              \CommentTok{;; Delta{-}G{-}fastest to make this rate fastest.}
\NormalTok{              (}\KeywordTok{error} \StringTok{"E{-}\textgreater{}A+D rates too high to merge, given Delta{-}G{-}fastest."}\NormalTok{)}
\NormalTok{              \textasciigrave{}((,r{-}A A) (,r{-}D D) (,(}\OperatorTok{{-}} \FloatTok{1.0}\NormalTok{ r{-}A r{-}D) }\DecValTok{\#f}\NormalTok{))))))}
  \CommentTok{;; It can be useful to show rates at problem registration time,}
  \CommentTok{;; for visual inspection. This can be done as follows:}
  \CommentTok{;;(begin}
  \CommentTok{;;  (display \textasciigrave{}(DEBUG rates}
  \CommentTok{;;                   rate{-}choices{-}A+P{-}\textgreater{}B+X ,rate{-}choices{-}A+P{-}\textgreater{}B+X}
  \CommentTok{;;                   rate{-}choices{-}B+P{-}\textgreater{}C+X ,rate{-}choices{-}B+P{-}\textgreater{}C+X               }
  \CommentTok{;;                   rate{-}choices{-}C+P{-}\textgreater{}D+X ,rate{-}choices{-}C+P{-}\textgreater{}D+X}
  \CommentTok{;;                   rate{-}choices{-}D+X{-}\textgreater{}C+P ,rate{-}choices{-}D+X{-}\textgreater{}C+P}
  \CommentTok{;;                   rate{-}choices{-}C+X{-}\textgreater{}B+P ,rate{-}choices{-}C+X{-}\textgreater{}B+P               }
  \CommentTok{;;                   rate{-}choices{-}B+X{-}\textgreater{}A+P ,rate{-}choices{-}B+X{-}\textgreater{}A+P}
  \CommentTok{;;                   rate{-}choices{-}A{-}\textgreater{}E ,rate{-}choices{-}A{-}\textgreater{}E}
  \CommentTok{;;                   rate{-}choices{-}D{-}\textgreater{}E ,rate{-}choices{-}D{-}\textgreater{}E}
  \CommentTok{;;                   rate{-}choices{-}E{-}\textgreater{}A+D ,rate{-}choices{-}E{-}\textgreater{}A+D))}
  \CommentTok{;;  (display "\textbackslash{}n"))}
\NormalTok{  (register{-}problem}
   \StringTok{"ex4var2{-}chemical{-}turing"}
   \CommentTok{;; S = Solvent, P = Powered, X = De{-}Powered, E = Detached Evaluator.     }
\NormalTok{   \#(A B C D I O P X S E) }
\NormalTok{   (}\KeywordTok{let}\NormalTok{ ((p0 (tape{-}get{-}sym }\DecValTok{\#f} \DecValTok{0}\NormalTok{)))}
\NormalTok{     (}\KeywordTok{cond}
\NormalTok{      ((}\KeywordTok{and}\NormalTok{ (}\KeywordTok{eq?}\NormalTok{ p0 \textquotesingle{}P)  }\CommentTok{; powered{-}\textgreater{}de{-}powered}
            \CommentTok{;; Data tape is "?[IO][IO]" {-} so, if "?" is a cursor,}
            \CommentTok{;; we can advance to a valid state.}
\NormalTok{            (is{-}io? (tape{-}get{-}sym }\DecValTok{\#t} \DecValTok{1}\NormalTok{))}
\NormalTok{            (is{-}io? (tape{-}get{-}sym }\DecValTok{\#t} \DecValTok{2}\NormalTok{))}
            \CommentTok{;; We need to suppress this by another factor 2, since}
            \CommentTok{;; back{-}reaction is two different reactions, depending on}
            \CommentTok{;; what bit gets written.}
\NormalTok{            (choose choice{-}1:1))}
\NormalTok{       (}\KeywordTok{let}\NormalTok{ ((d0 (tape{-}get{-}sym }\DecValTok{\#t} \DecValTok{0}\NormalTok{)))}
\NormalTok{         (}\KeywordTok{cond}
\NormalTok{          ((}\KeywordTok{and}\NormalTok{ (}\KeywordTok{eq?}\NormalTok{ d0 \textquotesingle{}A) (choose rate{-}choices{-}A+P{-}\textgreater{}B+X))}
\NormalTok{           (tape{-}set{-}sym! }\DecValTok{\#f} \DecValTok{0}\NormalTok{ \textquotesingle{}X)}
\NormalTok{           (tape{-}set{-}sym! }\DecValTok{\#t} \DecValTok{0}\NormalTok{ \textquotesingle{}I)}
\NormalTok{           (tape{-}set{-}sym! }\DecValTok{\#t} \DecValTok{1}\NormalTok{ \textquotesingle{}B))}
\NormalTok{          ((}\KeywordTok{and}\NormalTok{ (}\KeywordTok{eq?}\NormalTok{ d0 \textquotesingle{}B) (choose rate{-}choices{-}B+P{-}\textgreater{}C+X))}
\NormalTok{           (tape{-}set{-}sym! }\DecValTok{\#f} \DecValTok{0}\NormalTok{ \textquotesingle{}X)}
\NormalTok{           (tape{-}set{-}sym! }\DecValTok{\#t} \DecValTok{0}\NormalTok{ \textquotesingle{}O)}
\NormalTok{           (tape{-}set{-}sym! }\DecValTok{\#t} \DecValTok{1}\NormalTok{ \textquotesingle{}C))}
\NormalTok{          ((}\KeywordTok{and}\NormalTok{ (}\KeywordTok{eq?}\NormalTok{ d0 \textquotesingle{}C) (choose rate{-}choices{-}C+P{-}\textgreater{}D+X))}
\NormalTok{           (tape{-}set{-}sym! }\DecValTok{\#f} \DecValTok{0}\NormalTok{ \textquotesingle{}X)}
\NormalTok{           (tape{-}set{-}sym! }\DecValTok{\#t} \DecValTok{0}\NormalTok{ \textquotesingle{}I)}
\NormalTok{           (tape{-}set{-}sym! }\DecValTok{\#t} \DecValTok{1}\NormalTok{ \textquotesingle{}D)))))}
\NormalTok{      ((}\KeywordTok{and}\NormalTok{ (}\KeywordTok{eq?}\NormalTok{ p0 \textquotesingle{}X)  }\CommentTok{; de{-}powered{-}\textgreater{}powered}
            \CommentTok{;; Data tape is "[IO][IO]?" {-} so, if "?" is a cursor,}
            \CommentTok{;; we can un{-}advance to a valid state.}
\NormalTok{            (is{-}io? (tape{-}get{-}sym }\DecValTok{\#t}\NormalTok{ {-}}\DecValTok{1}\NormalTok{))}
\NormalTok{            (is{-}io? (tape{-}get{-}sym }\DecValTok{\#t}\NormalTok{ {-}}\DecValTok{2}\NormalTok{)))}
\NormalTok{       (}\KeywordTok{let}\NormalTok{ ((d0 (tape{-}get{-}sym }\DecValTok{\#t} \DecValTok{0}\NormalTok{)))}
\NormalTok{         (}\KeywordTok{cond}
\NormalTok{          ((}\KeywordTok{and}\NormalTok{ (}\KeywordTok{eq?}\NormalTok{ d0 \textquotesingle{}B) (choose rate{-}choices{-}B+X{-}\textgreater{}A+P))}
\NormalTok{           (tape{-}set{-}sym! }\DecValTok{\#f} \DecValTok{0}\NormalTok{ \textquotesingle{}P)}
\NormalTok{           (tape{-}set{-}sym! }\DecValTok{\#t} \DecValTok{0}\NormalTok{ (choose choice{-}IO))}
\NormalTok{           (tape{-}set{-}sym! }\DecValTok{\#t}\NormalTok{ {-}}\DecValTok{1}\NormalTok{ \textquotesingle{}A))}
\NormalTok{          ((}\KeywordTok{and}\NormalTok{ (}\KeywordTok{eq?}\NormalTok{ d0 \textquotesingle{}C) (choose rate{-}choices{-}C+X{-}\textgreater{}B+P))}
\NormalTok{           (tape{-}set{-}sym! }\DecValTok{\#f} \DecValTok{0}\NormalTok{ \textquotesingle{}P)}
\NormalTok{           (tape{-}set{-}sym! }\DecValTok{\#t} \DecValTok{0}\NormalTok{ (choose choice{-}IO))}
\NormalTok{           (tape{-}set{-}sym! }\DecValTok{\#t}\NormalTok{ {-}}\DecValTok{1}\NormalTok{ \textquotesingle{}B))}
\NormalTok{          ((}\KeywordTok{and}\NormalTok{ (}\KeywordTok{eq?}\NormalTok{ d0 \textquotesingle{}D) (choose rate{-}choices{-}D+X{-}\textgreater{}C+P))}
\NormalTok{           (tape{-}set{-}sym! }\DecValTok{\#f} \DecValTok{0}\NormalTok{ \textquotesingle{}P)}
\NormalTok{           (tape{-}set{-}sym! }\DecValTok{\#t} \DecValTok{0}\NormalTok{ (choose choice{-}IO))}
\NormalTok{           (tape{-}set{-}sym! }\DecValTok{\#t}\NormalTok{ {-}}\DecValTok{1}\NormalTok{ \textquotesingle{}C)))))}
\NormalTok{      ((}\KeywordTok{and}\NormalTok{ (}\KeywordTok{eq?}\NormalTok{ p0 \textquotesingle{}E)  }\CommentTok{; Detached evaluator that can attach.}
\NormalTok{            (is{-}io? (tape{-}get{-}sym }\DecValTok{\#t} \DecValTok{0}\NormalTok{))}
\NormalTok{            (is{-}io? (tape{-}get{-}sym }\DecValTok{\#t}\NormalTok{ +}\DecValTok{1}\NormalTok{))}
\NormalTok{            (is{-}io? (tape{-}get{-}sym }\DecValTok{\#t}\NormalTok{ {-}}\DecValTok{1}\NormalTok{))}
\NormalTok{            (choose choice{-}1:1))  }\CommentTok{; We overwrite one bit.}
\NormalTok{       (}\KeywordTok{let}\NormalTok{ ((A{-}D{-}f (choose rate{-}choices{-}E{-}\textgreater{}A+D)))}
\NormalTok{         (}\KeywordTok{cond}
\NormalTok{          ((}\KeywordTok{eq?}\NormalTok{ A{-}D{-}f \textquotesingle{}A)}
\NormalTok{           (tape{-}set{-}sym! }\DecValTok{\#f} \DecValTok{0}\NormalTok{ \textquotesingle{}S)}
\NormalTok{           (tape{-}set{-}sym! }\DecValTok{\#t} \DecValTok{0}\NormalTok{ \textquotesingle{}A))}
\NormalTok{          ((}\KeywordTok{eq?}\NormalTok{ A{-}D{-}f \textquotesingle{}D)}
\NormalTok{           (tape{-}set{-}sym! }\DecValTok{\#f} \DecValTok{0}\NormalTok{ \textquotesingle{}S)}
\NormalTok{           (tape{-}set{-}sym! }\DecValTok{\#t} \DecValTok{0}\NormalTok{ \textquotesingle{}D)))))}
\NormalTok{      ((}\KeywordTok{and}\NormalTok{ (}\KeywordTok{eq?}\NormalTok{ p0 \textquotesingle{}S)}
\NormalTok{            (is{-}io? (tape{-}get{-}sym }\DecValTok{\#t}\NormalTok{ +}\DecValTok{1}\NormalTok{))}
\NormalTok{            (is{-}io? (tape{-}get{-}sym }\DecValTok{\#t}\NormalTok{ {-}}\DecValTok{1}\NormalTok{)))}
\NormalTok{       (}\KeywordTok{let}\NormalTok{ ((d0 (tape{-}get{-}sym }\DecValTok{\#t} \DecValTok{0}\NormalTok{)))}
\NormalTok{         (}\KeywordTok{cond}
\NormalTok{          ((}\KeywordTok{and}\NormalTok{ (}\KeywordTok{eq?}\NormalTok{ d0 \textquotesingle{}A) (choose rate{-}choices{-}A{-}\textgreater{}E))}
\NormalTok{           (tape{-}set{-}sym! }\DecValTok{\#f} \DecValTok{0}\NormalTok{ \textquotesingle{}E)}
\NormalTok{           (tape{-}set{-}sym! }\DecValTok{\#t} \DecValTok{0}\NormalTok{ (choose choice{-}IO)))}
\NormalTok{          ((}\KeywordTok{and}\NormalTok{ (}\KeywordTok{eq?}\NormalTok{ d0 \textquotesingle{}D) (choose rate{-}choices{-}D{-}\textgreater{}E))}
\NormalTok{           (tape{-}set{-}sym! }\DecValTok{\#f} \DecValTok{0}\NormalTok{ \textquotesingle{}E)}
\NormalTok{           (tape{-}set{-}sym! }\DecValTok{\#t} \DecValTok{0}\NormalTok{ (choose choice{-}IO)))}
\NormalTok{          )))))))}
\end{Highlighting}
\end{Shaded}
}

\hypertarget{c.5-simple-machine-language}{%
\subsection{Simple Machine Language}\label{c.5-simple-machine-language}}

For problems of this type, it is especially important to remember that
the framework requires the implementation to be purely-functional,
i.e.~\emph{not perform any state-mutation}. Side effects related to
e.g.~assigning values to variables or updating the entries of a list or
vector would violate the requirements of the ``multiverse'' evaluator
mechanism, making adjustments in one ``universe'' cause changes to other
``universes'' that must remain independent.

{\small
\begin{Shaded}
\begin{Highlighting}[]
\NormalTok{(}\KeywordTok{let}\NormalTok{ ((single{-}R{-}can{-}execute }\DecValTok{\#f}\NormalTok{))}
\NormalTok{  (register{-}problem}
   \StringTok{"ex5{-}msrtf{-}machine"}
\NormalTok{   \#(M S R T F)}
\NormalTok{   (}\KeywordTok{let}\NormalTok{ loop ((Q }\DecValTok{4}\NormalTok{) (Is }\DecValTok{0}\NormalTok{) (Ip }\DecValTok{0}\NormalTok{) (Id }\DecValTok{0}\NormalTok{) (Op }\DecValTok{\#f}\NormalTok{) (NT }\DecValTok{0}\NormalTok{) (NR }\DecValTok{0}\NormalTok{) (NF }\DecValTok{0}\NormalTok{))}
\NormalTok{     (}\KeywordTok{let}\NormalTok{ ((op{-}todo (}\KeywordTok{if}\NormalTok{ (}\OperatorTok{\textgreater{}}\NormalTok{ Q }\DecValTok{0}\NormalTok{) (tape{-}get{-}sym }\DecValTok{\#f}\NormalTok{ Ip) Op)))}
\NormalTok{       (}\KeywordTok{if}\NormalTok{ (}\OperatorTok{=}\NormalTok{ Q }\DecValTok{4}\NormalTok{)}
\NormalTok{           (}\KeywordTok{cond}
\NormalTok{            ((}\KeywordTok{eq?}\NormalTok{ op{-}todo \textquotesingle{}S)}
\NormalTok{             (loop (}\OperatorTok{{-}}\NormalTok{ Q }\DecValTok{1}\NormalTok{) Is (}\OperatorTok{+} \DecValTok{1}\NormalTok{ Ip) Id op{-}todo }\DecValTok{0} \DecValTok{0} \DecValTok{0}\NormalTok{))}
\NormalTok{            ((}\KeywordTok{and}\NormalTok{ (}\KeywordTok{eq?}\NormalTok{ op{-}todo \textquotesingle{}R) single{-}R{-}can{-}execute)}
\NormalTok{             (tape{-}set! }\DecValTok{\#t}\NormalTok{ Id (}\KeywordTok{modulo}\NormalTok{ (}\OperatorTok{+} \DecValTok{1}\NormalTok{ (tape{-}get }\DecValTok{\#t}\NormalTok{ Id)) }\DecValTok{5}\NormalTok{))))}
           \CommentTok{;; Otherwise, not{-}first{-}op.}
\NormalTok{           (}\KeywordTok{case}\NormalTok{ op{-}todo}
\NormalTok{             ((T)}
\NormalTok{              (}\KeywordTok{let}\NormalTok{ ((activated? (}\KeywordTok{and}\NormalTok{ (}\OperatorTok{\textgreater{}}\NormalTok{ NT }\DecValTok{0}\NormalTok{) (}\OperatorTok{\textgreater{}}\NormalTok{ NF }\DecValTok{0}\NormalTok{))))}
\NormalTok{                (}\KeywordTok{if}\NormalTok{ activated? (tape{-}set! }\DecValTok{\#t}\NormalTok{ Id (tape{-}get }\DecValTok{\#f}\NormalTok{ Is)))}
\NormalTok{                (}\KeywordTok{if}\NormalTok{ (}\KeywordTok{not}\NormalTok{ (}\KeywordTok{or}\NormalTok{ (}\OperatorTok{=}\NormalTok{ Q }\DecValTok{1}\NormalTok{) (}\OperatorTok{=}\NormalTok{ Q {-}}\DecValTok{3}\NormalTok{)))}
\NormalTok{                    (loop (}\OperatorTok{{-}}\NormalTok{ Q }\DecValTok{1}\NormalTok{)}
\NormalTok{                          (}\KeywordTok{if}\NormalTok{ activated? (}\OperatorTok{+} \DecValTok{1}\NormalTok{ Is) Is)}
\NormalTok{                          (}\KeywordTok{if}\NormalTok{ (}\OperatorTok{\textgreater{}}\NormalTok{ Q }\DecValTok{0}\NormalTok{) (}\OperatorTok{+} \DecValTok{1}\NormalTok{ Ip) Ip)}
\NormalTok{                          (}\KeywordTok{if}\NormalTok{ activated? (}\OperatorTok{+} \DecValTok{1}\NormalTok{ Id) Id)}
\NormalTok{                          op{-}todo}
                          \DecValTok{1}\NormalTok{ NR NF))))}
\NormalTok{             ((R)}
\NormalTok{              (}\KeywordTok{if}\NormalTok{ (}\OperatorTok{\textgreater{}}\NormalTok{ NR }\DecValTok{0}\NormalTok{)}
\NormalTok{                  (tape{-}set! }\DecValTok{\#t}\NormalTok{ Id (}\KeywordTok{modulo}\NormalTok{ (}\OperatorTok{+} \DecValTok{1}\NormalTok{ (tape{-}get }\DecValTok{\#t}\NormalTok{ Id)) }\DecValTok{5}\NormalTok{)))}
\NormalTok{              (}\KeywordTok{if}\NormalTok{ (}\KeywordTok{not}\NormalTok{ (}\KeywordTok{or}\NormalTok{ (}\OperatorTok{=}\NormalTok{ Q }\DecValTok{1}\NormalTok{) (}\OperatorTok{=}\NormalTok{ Q {-}}\DecValTok{3}\NormalTok{)))}
\NormalTok{                  (loop (}\OperatorTok{{-}}\NormalTok{ Q }\DecValTok{1}\NormalTok{) Is}
\NormalTok{                        (}\KeywordTok{if}\NormalTok{ (}\OperatorTok{\textgreater{}}\NormalTok{ Q }\DecValTok{0}\NormalTok{) (}\OperatorTok{+} \DecValTok{1}\NormalTok{ Ip) Ip)}
\NormalTok{                        Id}
\NormalTok{                        op{-}todo}
\NormalTok{                        NT }\DecValTok{1}\NormalTok{ NF)))}
\NormalTok{             ((F)}
\NormalTok{              (}\KeywordTok{if}\NormalTok{ (}\KeywordTok{not}\NormalTok{ (}\KeywordTok{or}\NormalTok{ (}\OperatorTok{=}\NormalTok{ Q }\DecValTok{1}\NormalTok{) (}\OperatorTok{=}\NormalTok{ Q {-}}\DecValTok{3}\NormalTok{)))}
\NormalTok{                  (loop (}\OperatorTok{{-}}\NormalTok{ Q }\DecValTok{1}\NormalTok{) Is }
\NormalTok{                        (}\KeywordTok{if}\NormalTok{ (}\OperatorTok{\textgreater{}}\NormalTok{ Q }\DecValTok{0}\NormalTok{) (}\OperatorTok{+} \DecValTok{1}\NormalTok{ Ip) Ip)}
\NormalTok{                        Id}
\NormalTok{                        op{-}todo}
\NormalTok{                        NT NR }\DecValTok{1}\NormalTok{)))}
\NormalTok{             ((M)}
\NormalTok{              (}\KeywordTok{if}\NormalTok{ (}\KeywordTok{or}\NormalTok{ (}\KeywordTok{eq?}\NormalTok{ Op \textquotesingle{}R) (}\KeywordTok{eq?}\NormalTok{ Op \textquotesingle{}T))}
\NormalTok{                  (loop {-}}\DecValTok{1}\NormalTok{ Is Ip Id Op NT NR NF)))))))))}
\end{Highlighting}
\end{Shaded}
}

\hypertarget{c.6-mini-bff}{%
\subsection{``Mini-BFF''}\label{c.6-mini-bff}}

{\small
\begin{Shaded}
\begin{Highlighting}[]
\CommentTok{;; Example: "Mini{-}BFF"}
\NormalTok{(}\KeywordTok{let}\NormalTok{ ((alphabet}
\NormalTok{       \#(sym\textless{} sym\textgreater{} sym{-}cl sym{-}cr  }\CommentTok{; cl/cr = curly left/right \{\}}
\NormalTok{         sym{-} sym+ sym{-}dot sym{-}comma}
\NormalTok{         sym{-}bl sym{-}br sym0 sym{-}nop)  }\CommentTok{; bl/br = bracket left/right []}
\NormalTok{       ))}
\NormalTok{  (register{-}problem}
   \StringTok{"ex6{-}mini{-}bff"}
\NormalTok{   alphabet}
\NormalTok{   (}\KeywordTok{let}\NormalTok{ loop ((max{-}num{-}syms{-}to{-}still{-}read }\DecValTok{10}\NormalTok{)}
\NormalTok{              (p{-}offset }\DecValTok{0}\NormalTok{)}
\NormalTok{              (d0{-}offset }\DecValTok{0}\NormalTok{)  }\CommentTok{; "head 0" offset}
\NormalTok{              (d1{-}offset }\DecValTok{12}\NormalTok{)  }\CommentTok{; "head 1" offset}
              \CommentTok{;; Scan{-}mode N\textless{}0: Look for ({-}N){-}th [{-}bracket on the left.}
              \CommentTok{;; Scan mode N\textgreater{}0: Look for N{-}th ]{-}bracket on the right.}
              \CommentTok{;; Scan mode 0: Execute local command.}
\NormalTok{              (scan{-}mode }\DecValTok{0}\NormalTok{))}
\NormalTok{     (}\KeywordTok{if}\NormalTok{ (}\OperatorTok{=}\NormalTok{ max{-}num{-}syms{-}to{-}still{-}read }\DecValTok{0}\NormalTok{)}
         \DecValTok{\#f}  \CommentTok{; Done.}
\NormalTok{         (}\KeywordTok{let}\NormalTok{ ((op (tape{-}get{-}sym }\DecValTok{\#f}\NormalTok{ p{-}offset)))}
\NormalTok{           (}\KeywordTok{cond}
\NormalTok{            ((}\OperatorTok{\textless{}}\NormalTok{ scan{-}mode }\DecValTok{0}\NormalTok{)}
\NormalTok{             (}\KeywordTok{cond}
\NormalTok{              ((}\KeywordTok{eq?}\NormalTok{ op \textquotesingle{}sym{-}bl)}
\NormalTok{               (}\KeywordTok{if}\NormalTok{ (}\OperatorTok{=}\NormalTok{ scan{-}mode {-}}\DecValTok{1}\NormalTok{)}
\NormalTok{                   (loop (}\OperatorTok{{-}}\NormalTok{ max{-}num{-}syms{-}to{-}still{-}read }\DecValTok{1}\NormalTok{)}
\NormalTok{                         (}\OperatorTok{+}\NormalTok{ p{-}offset }\DecValTok{1}\NormalTok{)  }\CommentTok{; right after [}
\NormalTok{                         d0{-}offset d1{-}offset }\DecValTok{0}\NormalTok{)}
\NormalTok{                   (loop (}\OperatorTok{{-}}\NormalTok{ max{-}num{-}syms{-}to{-}still{-}read }\DecValTok{1}\NormalTok{)}
\NormalTok{                         (}\OperatorTok{+}\NormalTok{ p{-}offset {-}}\DecValTok{1}\NormalTok{)}
\NormalTok{                         d0{-}offset d1{-}offset (}\OperatorTok{+}\NormalTok{ scan{-}mode }\DecValTok{1}\NormalTok{))))}
\NormalTok{              ((}\KeywordTok{eq?}\NormalTok{ op \textquotesingle{}sym{-}br)}
\NormalTok{               (loop (}\OperatorTok{{-}}\NormalTok{ max{-}num{-}syms{-}to{-}still{-}read }\DecValTok{1}\NormalTok{)}
\NormalTok{                     (}\OperatorTok{+}\NormalTok{ p{-}offset {-}}\DecValTok{1}\NormalTok{)}
\NormalTok{                     d0{-}offset d1{-}offset (}\OperatorTok{+}\NormalTok{ scan{-}mode {-}}\DecValTok{1}\NormalTok{)))}
\NormalTok{              (}\DecValTok{\#t}
\NormalTok{               (loop (}\OperatorTok{{-}}\NormalTok{ max{-}num{-}syms{-}to{-}still{-}read }\DecValTok{1}\NormalTok{)}
\NormalTok{                     (}\OperatorTok{+}\NormalTok{ p{-}offset {-}}\DecValTok{1}\NormalTok{) d0{-}offset d1{-}offset scan{-}mode))))}
\NormalTok{            ((}\OperatorTok{\textgreater{}}\NormalTok{ scan{-}mode }\DecValTok{0}\NormalTok{)}
\NormalTok{             (}\KeywordTok{cond}
\NormalTok{              ((}\KeywordTok{eq?}\NormalTok{ op \textquotesingle{}sym{-}br)}
\NormalTok{               (}\KeywordTok{if}\NormalTok{ (}\OperatorTok{=}\NormalTok{ scan{-}mode }\DecValTok{1}\NormalTok{)}
\NormalTok{                   (loop (}\OperatorTok{{-}}\NormalTok{ max{-}num{-}syms{-}to{-}still{-}read }\DecValTok{1}\NormalTok{)}
\NormalTok{                         (}\OperatorTok{+}\NormalTok{ p{-}offset }\DecValTok{1}\NormalTok{)  }\CommentTok{; right after ]}
\NormalTok{                         d0{-}offset d1{-}offset }\DecValTok{0}\NormalTok{)}
\NormalTok{                   (loop (}\OperatorTok{{-}}\NormalTok{ max{-}num{-}syms{-}to{-}still{-}read }\DecValTok{1}\NormalTok{)}
\NormalTok{                         (}\OperatorTok{+}\NormalTok{ p{-}offset }\DecValTok{1}\NormalTok{)}
\NormalTok{                         d0{-}offset d1{-}offset (}\OperatorTok{+}\NormalTok{ scan{-}mode {-}}\DecValTok{1}\NormalTok{))))}
\NormalTok{              ((}\KeywordTok{eq?}\NormalTok{ op \textquotesingle{}sym{-}bl)}
\NormalTok{               (loop (}\OperatorTok{{-}}\NormalTok{ max{-}num{-}syms{-}to{-}still{-}read }\DecValTok{1}\NormalTok{)}
\NormalTok{                     (}\OperatorTok{+}\NormalTok{ p{-}offset }\DecValTok{1}\NormalTok{)}
\NormalTok{                     d0{-}offset d1{-}offset (}\OperatorTok{+}\NormalTok{ scan{-}mode }\DecValTok{1}\NormalTok{)))}
\NormalTok{              (}\DecValTok{\#t}
\NormalTok{               (loop (}\OperatorTok{{-}}\NormalTok{ max{-}num{-}syms{-}to{-}still{-}read }\DecValTok{1}\NormalTok{)}
\NormalTok{                     (}\OperatorTok{+}\NormalTok{ p{-}offset }\DecValTok{1}\NormalTok{) d0{-}offset d1{-}offset scan{-}mode))))}
\NormalTok{            (}\DecValTok{\#t}
\NormalTok{             (}\KeywordTok{cond}
\NormalTok{              ((}\KeywordTok{or}\NormalTok{ (}\KeywordTok{eq?}\NormalTok{ op sym\textless{}) (}\KeywordTok{eq?}\NormalTok{ op sym\textgreater{}))}
\NormalTok{               (loop (}\OperatorTok{{-}}\NormalTok{ max{-}num{-}syms{-}to{-}still{-}read }\DecValTok{1}\NormalTok{)}
\NormalTok{                     (}\OperatorTok{+}\NormalTok{ p{-}offset }\DecValTok{1}\NormalTok{)}
\NormalTok{                     (}\OperatorTok{+}\NormalTok{ d0{-}offset (}\KeywordTok{if}\NormalTok{ (}\KeywordTok{eq?}\NormalTok{ op sym\textless{}) {-}}\DecValTok{1}\NormalTok{ +}\DecValTok{1}\NormalTok{))}
\NormalTok{                     d1{-}offset }\DecValTok{0}\NormalTok{))}
\NormalTok{              ((}\KeywordTok{or}\NormalTok{ (}\KeywordTok{eq?}\NormalTok{ op sym{-}cl) (}\KeywordTok{eq?}\NormalTok{ op sym{-}cr))}
\NormalTok{               (loop (}\OperatorTok{{-}}\NormalTok{ max{-}num{-}syms{-}to{-}still{-}read }\DecValTok{1}\NormalTok{)}
\NormalTok{                     (}\OperatorTok{+}\NormalTok{ p{-}offset }\DecValTok{1}\NormalTok{)}
\NormalTok{                     d0{-}offset}
\NormalTok{                     (}\OperatorTok{+}\NormalTok{ d1{-}offset (}\KeywordTok{if}\NormalTok{ (}\KeywordTok{eq?}\NormalTok{ op sym\textless{}) {-}}\DecValTok{1}\NormalTok{ +}\DecValTok{1}\NormalTok{))}
                     \DecValTok{0}\NormalTok{))}
\NormalTok{              ((}\KeywordTok{or}\NormalTok{ (}\KeywordTok{eq?}\NormalTok{ op sym+) (}\KeywordTok{eq?}\NormalTok{ op sym{-}))}
\NormalTok{               (tape{-}set! }\DecValTok{\#t}\NormalTok{ d0{-}offset}
\NormalTok{                          (}\KeywordTok{modulo}
\NormalTok{                           (}\OperatorTok{+}\NormalTok{ (tape{-}get }\DecValTok{\#t}\NormalTok{ d0{-}offset }\DecValTok{1}\NormalTok{) (}\KeywordTok{if}\NormalTok{ (}\KeywordTok{eq?}\NormalTok{ op sym+) +}\DecValTok{1}\NormalTok{ {-}}\DecValTok{1}\NormalTok{))}
\NormalTok{                           (}\KeywordTok{vector{-}length}\NormalTok{ alphabet)))}
\NormalTok{               (loop (}\OperatorTok{{-}}\NormalTok{ max{-}num{-}syms{-}to{-}still{-}read }\DecValTok{1}\NormalTok{)}
\NormalTok{                     (}\OperatorTok{+}\NormalTok{ p{-}offset }\DecValTok{1}\NormalTok{) d0{-}offset d1{-}offset }\DecValTok{0}\NormalTok{))}
\NormalTok{              ((}\KeywordTok{eq?}\NormalTok{ op sym{-}dot)}
\NormalTok{               (tape{-}set! }\DecValTok{\#t}\NormalTok{ d1{-}offset (tape{-}get }\DecValTok{\#t}\NormalTok{ d0{-}offset))}
\NormalTok{               (loop (}\OperatorTok{{-}}\NormalTok{ max{-}num{-}syms{-}to{-}still{-}read }\DecValTok{1}\NormalTok{)}
\NormalTok{                     (}\OperatorTok{+}\NormalTok{ p{-}offset }\DecValTok{1}\NormalTok{) d0{-}offset d1{-}offset }\DecValTok{0}\NormalTok{))}
\NormalTok{              ((}\KeywordTok{eq?}\NormalTok{ op sym{-}comma)}
\NormalTok{               (tape{-}set! }\DecValTok{\#t}\NormalTok{ d0{-}offset (tape{-}get }\DecValTok{\#t}\NormalTok{ d1{-}offset))}
\NormalTok{               (loop (}\OperatorTok{{-}}\NormalTok{ max{-}num{-}syms{-}to{-}still{-}read }\DecValTok{1}\NormalTok{)}
\NormalTok{                     (}\OperatorTok{+}\NormalTok{ p{-}offset }\DecValTok{1}\NormalTok{) d0{-}offset d1{-}offset }\DecValTok{0}\NormalTok{))}
\NormalTok{              ((}\KeywordTok{eq?}\NormalTok{ op sym{-}bl)}
\NormalTok{               (loop (}\OperatorTok{{-}}\NormalTok{ max{-}num{-}syms{-}to{-}still{-}read }\DecValTok{1}\NormalTok{)}
\NormalTok{                     (}\OperatorTok{+}\NormalTok{ p{-}offset }\DecValTok{1}\NormalTok{) d0{-}offset d1{-}offset}
\NormalTok{                     (}\KeywordTok{if}\NormalTok{ (}\KeywordTok{eq?}\NormalTok{ sym0 (tape{-}get{-}sym }\DecValTok{\#t}\NormalTok{ d0{-}offset))}
                         \CommentTok{;; Either enter scan{-}mode or make this a no{-}op.}
\NormalTok{                         +}\DecValTok{1} \DecValTok{0}\NormalTok{)))}
\NormalTok{              ((}\KeywordTok{eq?}\NormalTok{ op sym{-}br)}
\NormalTok{               (}\KeywordTok{if}\NormalTok{ (}\KeywordTok{eq?}\NormalTok{ sym0 (tape{-}get{-}sym }\DecValTok{\#t}\NormalTok{ d0{-}offset))}
                   \CommentTok{;; no{-}op.}
\NormalTok{                   (loop (}\OperatorTok{{-}}\NormalTok{ max{-}num{-}syms{-}to{-}still{-}read }\DecValTok{1}\NormalTok{)}
\NormalTok{                         (}\OperatorTok{+}\NormalTok{ p{-}offset }\DecValTok{1}\NormalTok{) d0{-}offset d1{-}offset }\DecValTok{0}\NormalTok{)}
                   \CommentTok{;; otherwise, scan backwards.}
\NormalTok{                   (loop (}\OperatorTok{{-}}\NormalTok{ max{-}num{-}syms{-}to{-}still{-}read }\DecValTok{1}\NormalTok{)}
\NormalTok{                         (}\OperatorTok{+}\NormalTok{ p{-}offset {-}}\DecValTok{1}\NormalTok{) d0{-}offset d1{-}offset {-}}\DecValTok{1}\NormalTok{)))}
\NormalTok{              (}\DecValTok{\#t}  \CommentTok{; no{-}op.}
\NormalTok{               (loop (}\OperatorTok{{-}}\NormalTok{ max{-}num{-}syms{-}to{-}still{-}read }\DecValTok{1}\NormalTok{)}
\NormalTok{                     (}\OperatorTok{+}\NormalTok{ p{-}offset }\DecValTok{1}\NormalTok{) d0{-}offset d1{-}offset }\DecValTok{0}\NormalTok{))))))))))}
\end{Highlighting}
\end{Shaded}
}

\end{appendices}

\end{document}